\documentclass[twocolumn, usenatbib]{mnras}
\usepackage{savesym}
\usepackage{graphicx}
\usepackage{longtable}
\usepackage{changepage}

\expandafter\let\csname equation*\endcsname\relax
  \expandafter\let\csname endequation*\endcsname\relax 
\usepackage{subfig}
\usepackage{amsmath}
\usepackage{amssymb}
\usepackage{verbatim}
\usepackage[yyyymmdd,hhmmss]{datetime}
\usepackage{array}
\usepackage{times}
\usepackage{xcolor}

\DeclareRobustCommand{\DE}[3]{#2}
\let\DEthebibliography\thebibliography
\def\thebibliography{\DeclareRobustCommand{\DE}[3]{##3}\DEthebibliography}

\newcommand{\beq}{\begin{equation}}
\newcommand{\eeq}{\end{equation}}

\title[ Hills masses of Kerr black holes ]{ The maximum mass of a black hole which can tidally disrupt a star: \\ measuring black hole spins with tidal disruption events  }
\author [Andrew Mummery]{Andrew Mummery$^1$\thanks{E-mail:
andrew.mummery@physics.ox.ac.uk}
\\
$^1$Oxford Theoretical Physics, Beecroft Building,  Clarendon Laboratory, Parks Road, Oxford, OX1 3PU, United Kingdom}

\date{}

\begin{document}

\pagerange{\pageref{firstpage}--\pageref{lastpage}} \pubyear{2023}

\maketitle

\label{firstpage}

\begin{abstract} 
The tidal acceleration experienced by an object at the event horizon of a black hole decreases as one over the square of the black hole's mass. As such there is a maximum mass at which a black hole can tidally disrupt an object outside of its event horizon and potentially produce observable emission.  This maximum mass is known as the ``Hills mass'', and in full general relativity is a function of both the black hole's spin $a_\bullet$ and the inclination angle of the incoming object's orbit with respect to the black hole's spin axis $\psi$. In this paper we demonstrate that the Hills mass can be represented by a simple analytical function of $a_\bullet$ and $\psi$, the first general solution of this problem. This general solution is found by utilising the symmetries of a class of critical Kerr metric orbits known as the innermost bound spherical orbits. Interestingly, at fixed black hole spin the maximum Hills mass can lie at incoming orbital inclinations outside of the black hole's equatorial plane $\psi \neq \pi/2$. When compared to previous results in the literature this effect can lead to an increase in the maximum Hills mass (at fixed spin) by as much as a factor of $\sqrt{11/5} \simeq 1.48$ for a maximally rotating black hole. We then demonstrate how Bayesian inference, {coupled with} an estimate of the mass of a black hole in a tidal disruption event, can be used to place conservative constraints on that black hole's spin. We provide a publicly available code {\tt tidalspin} which computes these spin distributions. 
\end{abstract}

\noindent

\section{Introduction} 
The supermassive black holes which reside in galactic centres are surrounded by a dense population of stars.   On rare occasion \citep[typically once every $\sim 10^4-10^5$ years in a given galaxy;][]{Magorrian99} a star may be perturbed by N-body gravitational interactions onto a near radial orbit about the central black hole.   If the tidal force experienced by the star  overcomes the star's own self gravity, then the star will become tidally unbound and subsequently accreted -- a process which produces bright electromagnetic signatures \citep{Rees88}.  A large number ($\sim 100$) of these so-called `tidal disruption events' have been discovered in recent years, at observed wavelengths across the entire electromagnetic spectrum, from X-rays \citep[e.g.,][]{Greiner00, Cenko12}, and optical \citep[e.g.,][]{Gezari08, vanVelzen20} to radio frequencies \citep[e.g.,][]{Alexander16}. 

A simple Newtonian analysis shows that the dimensional scale of tidal accelerations is 
\begin{equation}
    a_T \sim GM_\bullet R_\star /r^3 ,
\end{equation} 
with $M_\bullet$ the black hole mass, $R_\star$ is the stellar radius, and $r$ the radial separation. For those orbits with radial scale  of order the event horizon of the black hole $r_S \propto M_\bullet$, the tidal acceleration scales as the one over the square of the black holes mass
\begin{equation}
    a_T \sim 1/M_\bullet^2 , \quad r \sim {\rm event \,  horizon},
\end{equation} 
and therefore decreases for more massive black holes. This well known result implies that there is a maximum mass at which a black hole can tidally disrupt a star outside of its event horizon, a mass scale known as the \cite{Hills75} mass. Importantly, this Hills mass sets the upper mass scale for black holes which can produce electromagnetically observable tidal disruption events.   The precise value of this Hills mass is therefore of real interest to current and future studies of tidal disruption events, as it sets the maximum observable black hole mass scale which should be detected in transient surveys. 

A proper calculation of the Hills mass requires an analysis of tidal accelerations in the Kerr metric.  This field of study has a long history, \cite{Marck83} first derived the relevant tidal tensor experienced by particles following geodesic orbits, which we will make use of in this paper. Later \cite{Kesden12}, following simulations by \cite{Beloborodov92} and \cite{Ivanov06}, highlighted the effects of the Kerr spin parameter on the rates of tidal disruption events for black holes of different masses. \cite{Kesden12} demonstrated that the black hole spin enhances the tidal disruption rate substantially for large mass black holes.  In other words, rotating black holes have a substantially larger Hills mass than non-rotating black holes. 

The physical reason for the positive spin dependence of the Hills mass is two-fold. The first is relatively trivial: the event horizon of rapidly rotating black holes is smaller than their more slowly rotating counterparts. The second is that the tidal force itself is an increasing function of the Kerr spin parameter; in effect there is a component of the tidal force which results from the azimuthal sheer intrinsic to the Kerr metric. While the dominant trend of Hills mass with black hole spin was demonstrated in \cite{Kesden12}, these results were purely numerical, and based on Monte Carlo simulations of incoming stars.  In \cite{Mummery_et_al_23} an analytic expression for the Hills mass was presented, in the limit of stars confined to the equatorial plane of the Kerr spacetime. The Hills mass however depends on both the black hole spin {\it and} the inclination of the incoming star's orbit. No general solution for the Hills mass as a function of both of these parameters has previously been derived.   

In this paper we derive this general expression for the Hills mass as a function of black hole spin and incoming orbital inclination. Finding this general solution is possible because the orbits of the last stars which can possibly be tidally disrupted posses a number of simplifying symmetries. They are known as the `innermost bound spherical orbits' and have been studied extensively in the literature \citep[e.g.,][]{Grossman12, Will12, Hod13, Stein20, Teo21}. 

The fact that there is a limiting black hole mass at which a tidal disruption event of a given star can occur, which depends strongly on the black hole's spin, means that posterior constraints can be placed on the black hole's spin if a tidal disruption event is observed to occur.  In the latter parts of this paper we demonstrate how Bayesian inference allows constraints on black hole spins to be determined if an estimate for the black hole's mass at the centre of the tidal disruption event is known. These constraints only rely on the properties of tidal forces in the Kerr metric, and we provide spin estimates for 9 tidal disruption events from the literature with large inferred masses.  These types of constraints will be of interest to the large populations of tidal disruption events discovered by future optical surveys \citep[e.g.,][]{Bricman20}. 

The layout of this paper is as follows. In section \ref{dynamic_sec} we discuss the properties of test particle motion in the Kerr metric, and introduce the properties of the innermost bound spherical orbits. In section \ref{tidal_sec} we discuss the properties of tidal forces in the Kerr metric, and derive an analytical expression for the maximum tidal acceleration experienced by a test particle on the critical innermost bound spherical orbit. In section \ref{hills_mass_sec} we present the general solution for the maximum black hole mass which can tidally disrupt a star, and analyse its properties. In section \ref{spin_sec} we discuss how these results can be used to place constraints on black hole spins in individual tidal disruption events, an approach we expand to populations of tidal disruption events in section \ref{sec:population}, before concluding in section \ref{conc}. Some technical results are presented in three Appendices.

\section{Orbital dynamics in the Kerr metric} \label{dynamic_sec}
\subsection{The metric}
The Kerr metric in Boyer-Lindquist coordinates takes the following form, presented in terms of its invariant line element  ${\rm d}s$
\begin{multline}\label{metric}
    {\rm d}s^2 = - \left(1 - {2r_g r \over r^2 + a^2 \cos^2\theta }\right) c^2 {\rm d}t^2 - {4 r_g r a c \sin^2 \theta \over r^2 + a^2 \cos^2\theta} \, {\rm d}t \, {\rm d}\phi \\ + {r^2 - 2r_gr + a^2 \over r^2 + a^2 \cos^2\theta }\, {\rm d}r^2 + (r^2 + a^2 \cos^2 \theta ) \, {\rm d}\theta^2  \\ + \left(r^2 + a^2 + {2 r_g a^2 r \sin^2 \theta \over r^2 + a^2 \cos^2 \theta } \right)\sin^2 \theta \, {\rm d}\phi^2 ,
\end{multline}
where $r_g \equiv GM_\bullet / c^2$ is the gravitational radius, $M_\bullet$ is the black hole mass, and $a$ is the angular momentum constant (with dimensions of length) of the black hole $a = J/M_\bullet c$, where $J$ is the total angular momentum of the black hole. We shall also work with the dimensionless black hole spin $a_\bullet$ in this paper 
\begin{equation}
    a_\bullet \equiv a/r_g, \quad -1 < a_\bullet < 1. 
\end{equation}
The coordinates are $t$, the time as measured at infinity, and the three spatial coordinates $(r, \theta, \phi)$ which have their usual quasi-spherical interpretations. 

\subsection{Equations of motion}
In this paper we are interested with the dynamical evolution of a stellar orbit around a supermassive Kerr black hole. Fortunately, in the limit of a supermassive black hole we can treat the orbiting star as a test particle, which substantially simplifies the dynamics.  The reason this limit is valid is because both the mass and radius of the star are significantly smaller than the mass and radial scales set by the black hole 
\begin{equation}
    \left({M_\star \over M_\bullet}\right) \ll 1 , \quad \left({R_\star \over r_g}\right) = \left({R_\star c^2 \over GM_\star}\right) \left({M_\star \over M_\bullet}\right) \sim 10^{-3} \left({10^8M_\odot \over M_\bullet}\right) ,
\end{equation}
where we have substituted solar values for the second equality, and $10^8M_\odot$ is the rough scale of the maximum black hole masses of which we are interested in. 

The equations of motion for a test particle evolving in the Kerr metric are the following \citep[see e.g.,][presented in units where $c=1$]{Bardeen72, MTW}
\begin{align}
    (r^2 + a^2 \cos^2\theta)^2 & \left({{\rm d} r \over {\rm d}\tau}\right)^2 = \left[\epsilon (r^2 + a^2) - a l_z \right]^2 \nonumber \\ - &(r^2 - 2r_gr + a^2) \left[r^2 + q + (l_z - a \epsilon)^2 \right] ,\label{rad_mot} \\
    (r^2 + a^2 \cos^2\theta)^2 & \left({{\rm d} \theta \over {\rm d}\tau}\right)^2 = q - l_z^2 \cot^2\theta \nonumber \\ &\quad\quad\quad\quad \quad\quad  - a^2(1-\epsilon^2)\cos^2\theta , \label{thet_mot}\\ 
    (r^2 + a^2 \cos^2\theta) & \left({{\rm d} \phi \over {\rm d}\tau}\right) = {l_z \over \sin^2\theta } + {2r_g a r \epsilon - a^2l_z \over r^2 - 2r_g r + a^2} \label{phi_mot}.
\end{align}
These expressions are written in terms of the three constants of motion of the Kerr spacetime, the specific energy of the star ($p_\mu$ is the star's 4-momenta, $M_\star$ is the star's rest mass)
\begin{equation}
    \epsilon = -p_t/M_\star ,
\end{equation}
the specific axial angular momentum of the star 
\begin{equation}
    l_z = p_\phi/M_\star , 
\end{equation}
and $q$, the specific \cite{Carter68} constant of the star 
\begin{equation}\label{carter}
    q \equiv {Q \over M_\star^2} = {p_\theta^2 \over M_\star^2} + \cos^2\theta \left( a^2 (1 -\epsilon^2) + {l_z^2 \over \sin^2\theta } \right) ,
\end{equation}
here $Q$ is the original Carter constant. 
In these units $\epsilon$ is dimensionless, while $l_z$ has dimensions of length and $q$ has dimensions of length squared. Another constant of interest is the modified \cite{Carter68} constant 
\begin{equation}
    k = q + (l_z - a\epsilon)^2 ,
\end{equation}
which has the useful property $k \geq 0$. Note that for stars orbiting about the galactic centre (as is relevant for the tidal disruption event problem), the specific energy constant 
\begin{equation}
    \epsilon \sim 1 + {\sigma^2 \over c^2} + \dots \sim 1, 
\end{equation}
where $\sigma$ is the velocity dispersion of the galaxy, with typical value $ \sigma \sim 100$ km/s  $\sim 10^{-4}c$. 

\subsection{The innermost bound spherical orbits}
Clearly, the orbital equations of motion in the full Kerr metric are incredibly algebraically complex. Fortunately, there is a class of special Kerr metric orbits which posses a number of simplifying symmetries which are relevant for the study of the limiting orbits which produce observable tidal disruptions. These orbits are known as the spherical orbits, and of most relevance are the innermost bound spherical orbits. 

The spherical orbits are characterised by test particle evolution with constant radial coordinate $r(\tau) = r_{\rm sp} = {\rm cst}$. Unlike Newtonian, or indeed Schwarzschild, dynamics, in the general Kerr metric these orbits are not confined to a plane, and have an azimuthal angle {which evolves in a non-trivial manner} $\theta(\tau)$. 

Remarkably, not all of these spherical orbits are formally bound to the black hole, in the sense that spherical orbits exist in the Kerr spacetime with both $\epsilon < 1$, and $\epsilon > 1$. Test particles evolving with $\epsilon > 1$ are unbound. The critical value $\epsilon = 1$ corresponds to the ``innermost bound'' spherical orbit. This orbit corresponds physically to a particle which approaches the black hole on a parabolic flyby from infinity, before asymptoting as $\tau \to \infty$ to a  spherical orbit with $r(\tau\to\infty) \to r_{\rm sp} = {\rm cst}$. While these orbits remain somewhat algebraically complex in the full Kerr metric, things simplify for the Schwarzschild $(a = 0)$ limit where the orbit is planar and asymptotes to a circular orbit. The full, exact, solution of the orbital equations is given in this limit by 
\begin{equation}
    r(\phi) = {4r_g \over \tanh^2\left(\phi/2\sqrt{2}\right)}, \quad \theta(\tau) = \pi/2, 
\end{equation}
where $\phi$ varies from 0 to $+\infty$ \citep{MummeryBalbus2023PRD}. This solution, while simplified by the Schwarzschild limit, highlights the key physical properties of these orbits. Initially the particle is on a parabolic flyby $\phi \to 0, r\to \infty$, but then asymptotes to $r = 4r_g$ and undergoes infinite rotations of the black hole. 

This last bound spherical orbit is of critical importance for the study of the Hills mass, and last disrupt-able orbits.   Formally it represents a {\it separatrix}: all orbits which start at $r \to \infty$ with axial angular momentum  $l_z$ smaller than the innermost bound spherical orbit are guaranteed to terminate at $r = 0$, while all orbits with larger angular momentum will escape again to $r\to \infty$ as $\tau \to \infty$ \citep[e.g.,][]{Chandrasekhar83}. Even if a star is tidally disrupted outside of the event horizon of a black hole on an orbit with smaller axial angular momentum than the innermost bound spherical orbit, all of the tidally disrupted debris will rapidly cross the event horizon on their plunge towards the singularity. Therefore, the relevant orbit for determining the limiting black hole mass at  which a star can be tidally disrupted and still produce observable emission is the innermost bound spherical orbit.  

This realisation simplifies the problem of determining the evolving properties of the stars orbit substantially.   Firstly, the specific energy of every innermost bound spherical orbit is unity, i.e., 
\begin{equation}
    \epsilon = 1. 
\end{equation}
This means that the stellar orbits are entirely determined by just two constants of motion, the axial angular momentum $l_z$ and the Carter constant $q$.  However, it will be more convenient to express the orbital dynamics in terms of a different conserved quantity $\psi$, defined in the following manner
\begin{equation}
    \sin \psi = {l_z \over \sqrt{l_z^2 + q}} .
\end{equation}

\begin{figure}
    \centering
    \includegraphics[width=0.45\textwidth]{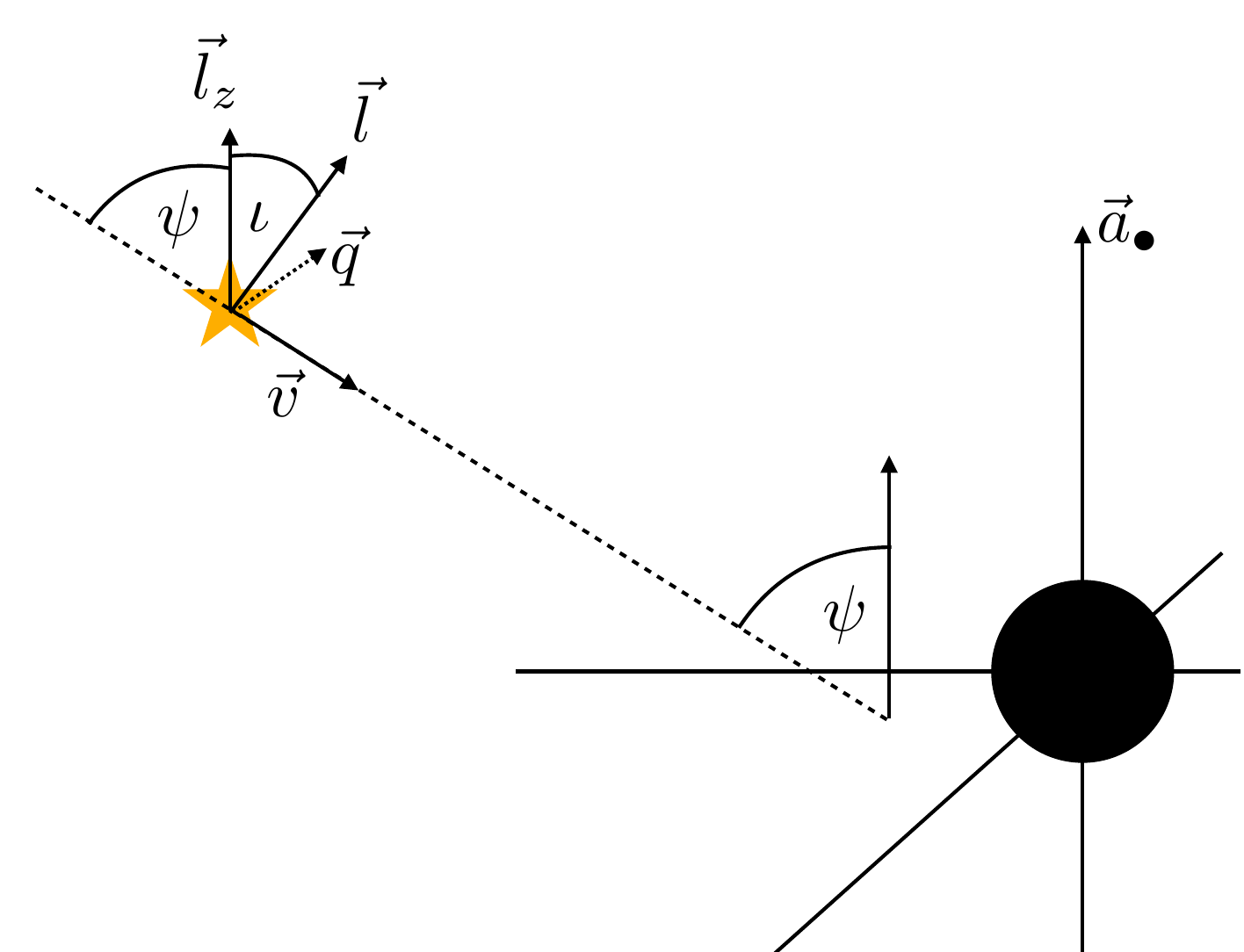}
    \caption{Geometrical interpretation of the constants of motion for the innermost bound spherical orbits relevant to this paper. The black hole is at the centre of the coordinate system and the spin axis of the black hole sets the vertical $z$ direction. The constant of motion $l_z$ represents the axial angular momentum, while Carter's constant $q = p_\theta^2 + l_z^2 \cot^2\theta = l_x^2 + l_y^2$ is the square of the angular momentum vector in the $x-y$ plane. The angle $\psi$ then represents the asymptotic ($r\to \infty$) inclination angle between the star's initial orbital velocity and the black hole's spin axis.   }
    \label{fig:geometric}
\end{figure}

\begin{figure*}
    \centering
    \includegraphics[width=.49\textwidth]{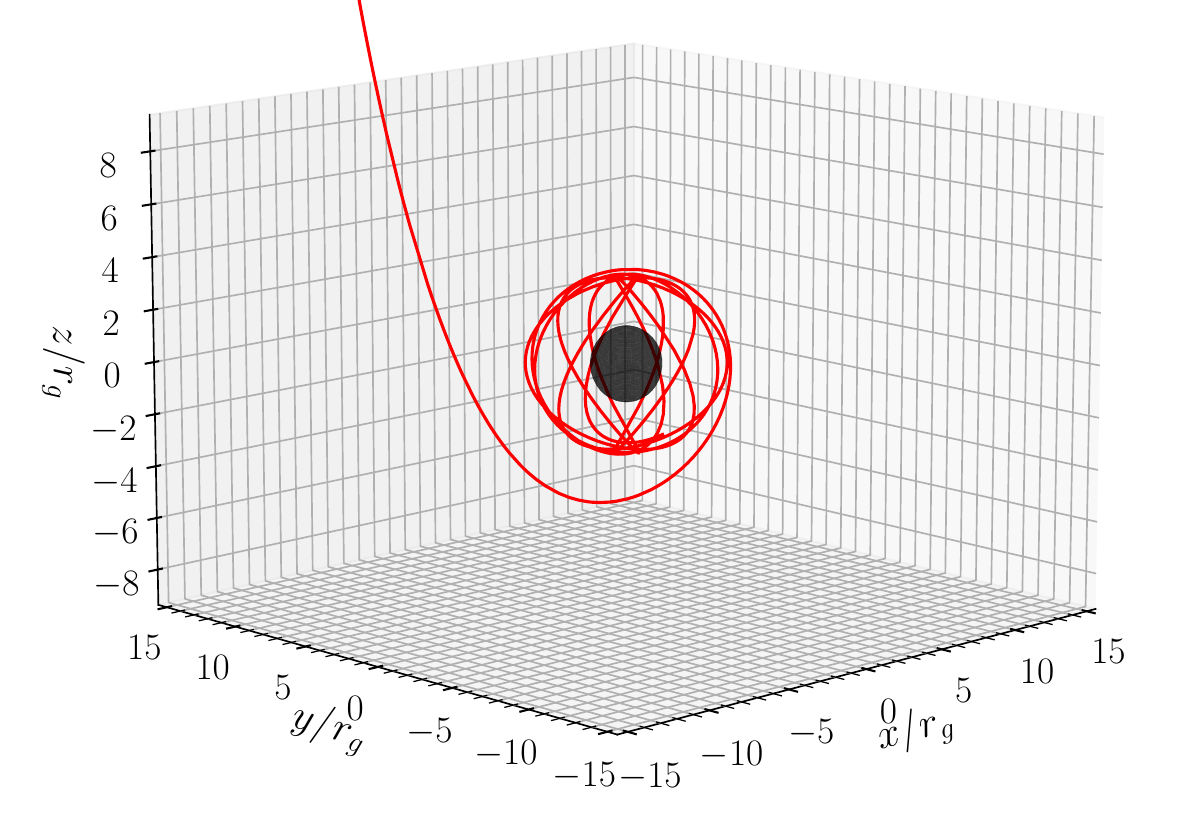}
    \includegraphics[width=.49\textwidth]{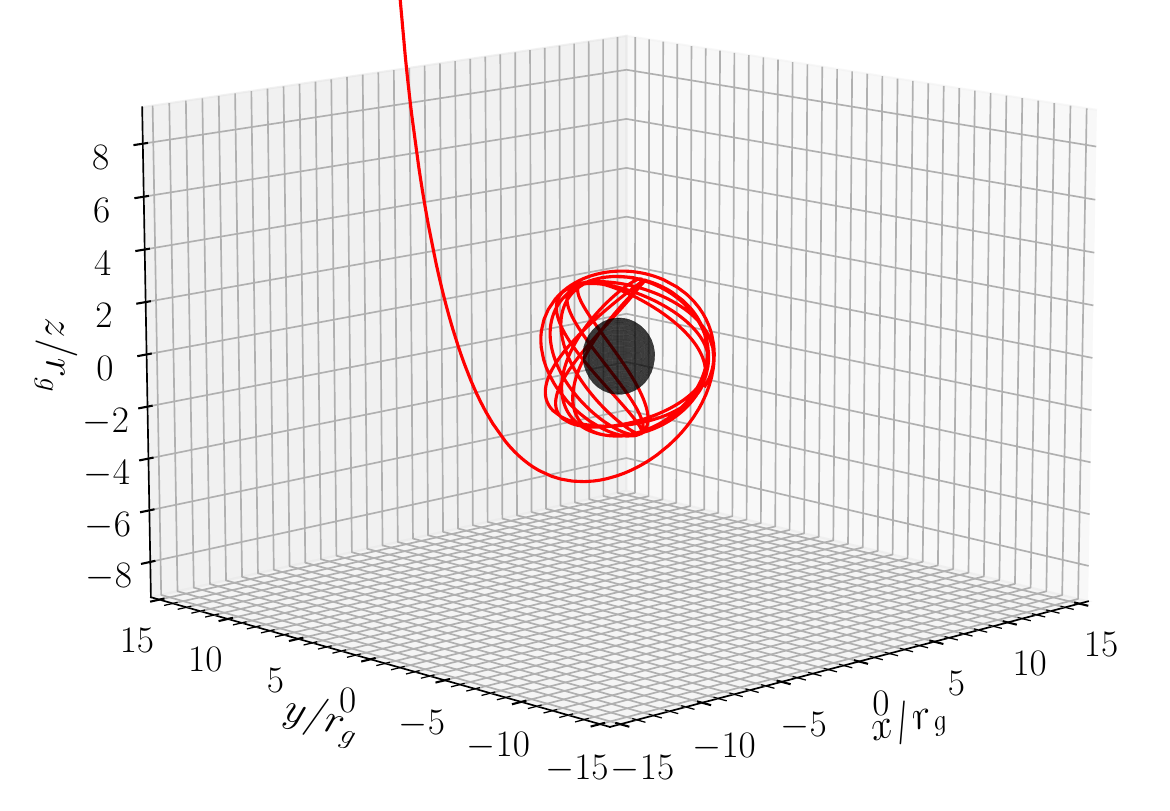}
    \includegraphics[width=.49\textwidth]{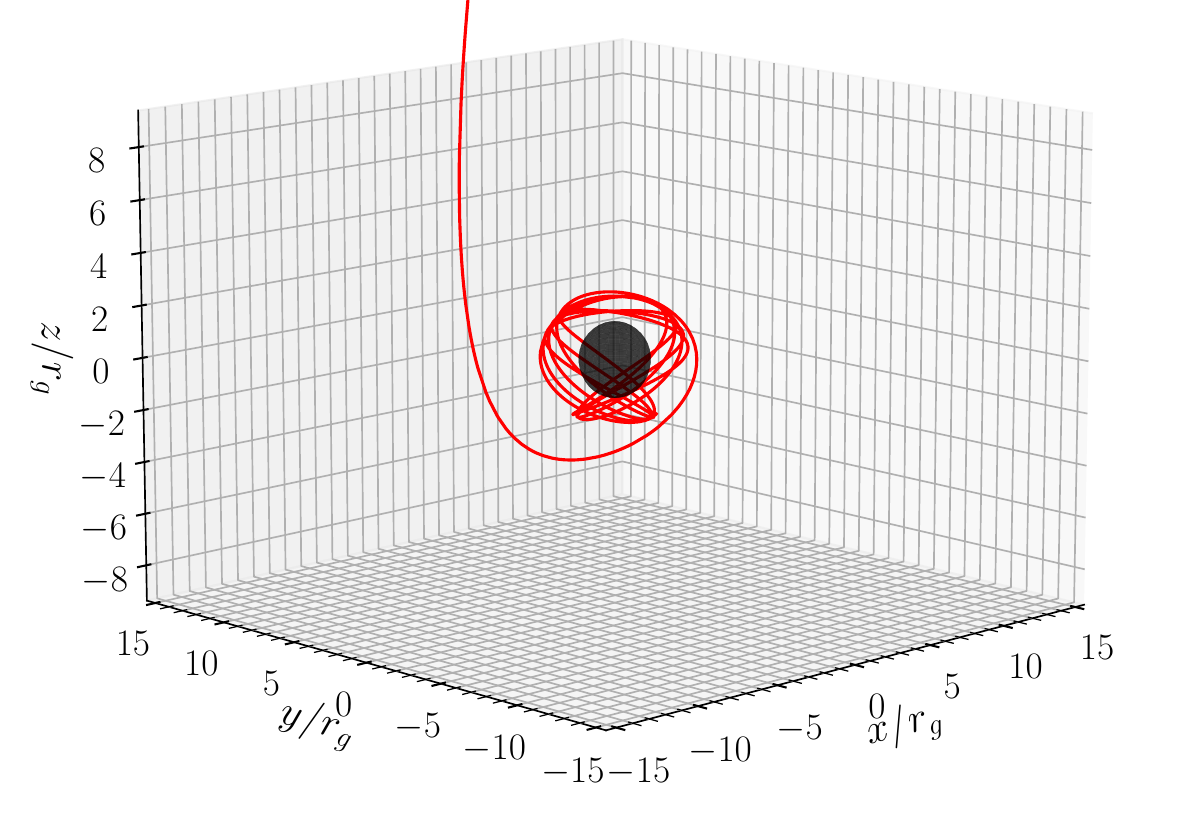}
    \includegraphics[width=.49\textwidth]{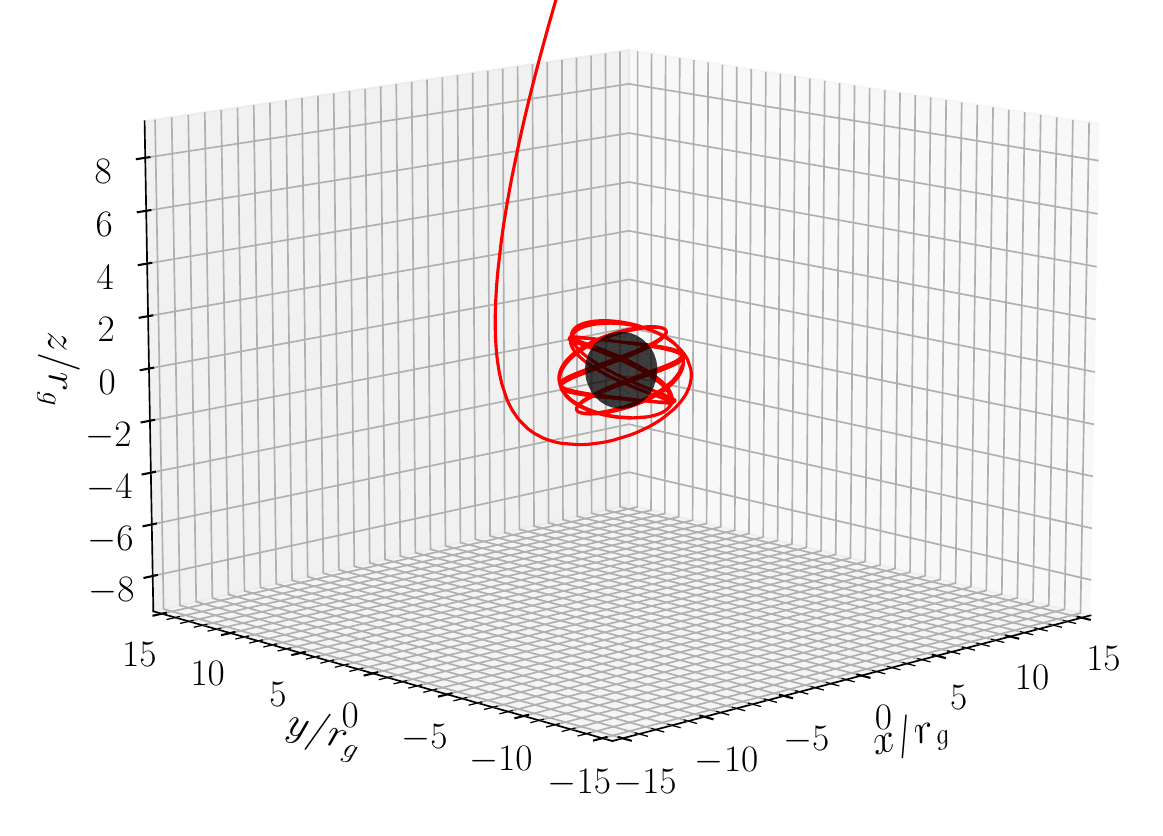}
    \includegraphics[width=.49\textwidth]{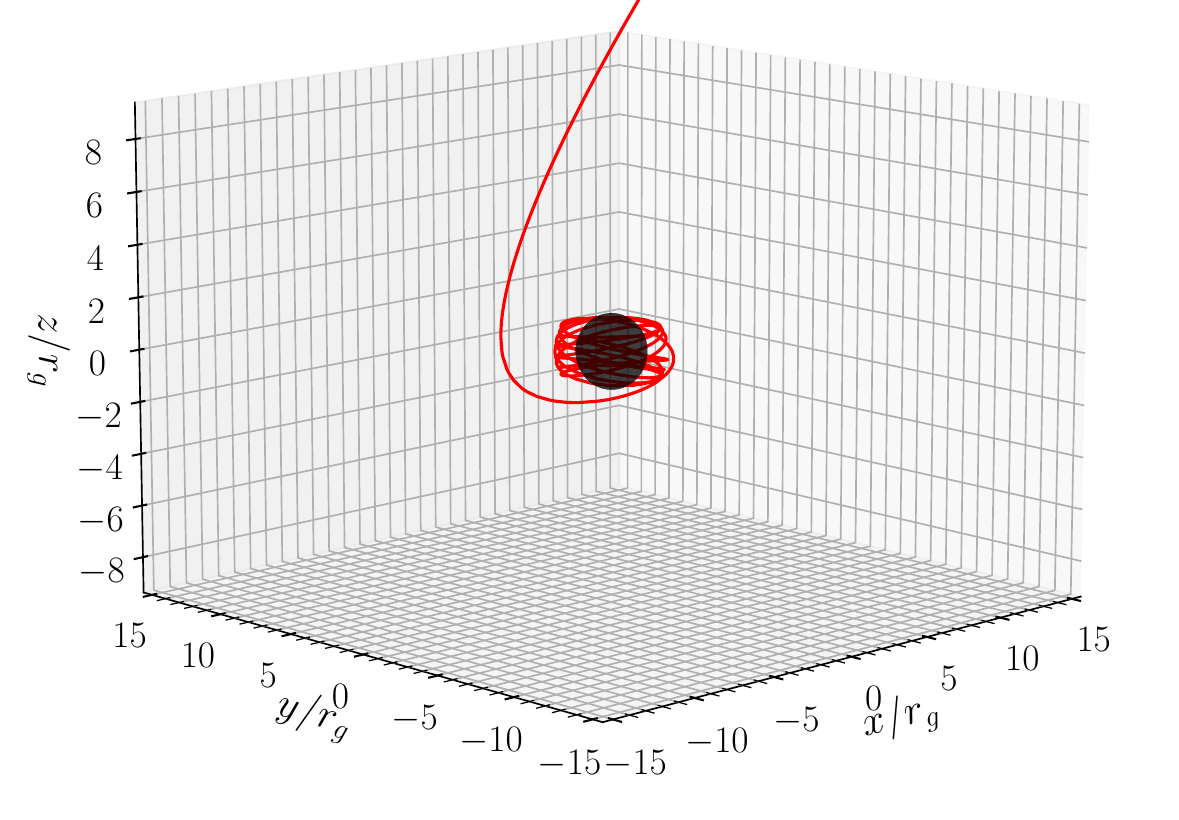}
    \includegraphics[width=.49\textwidth]{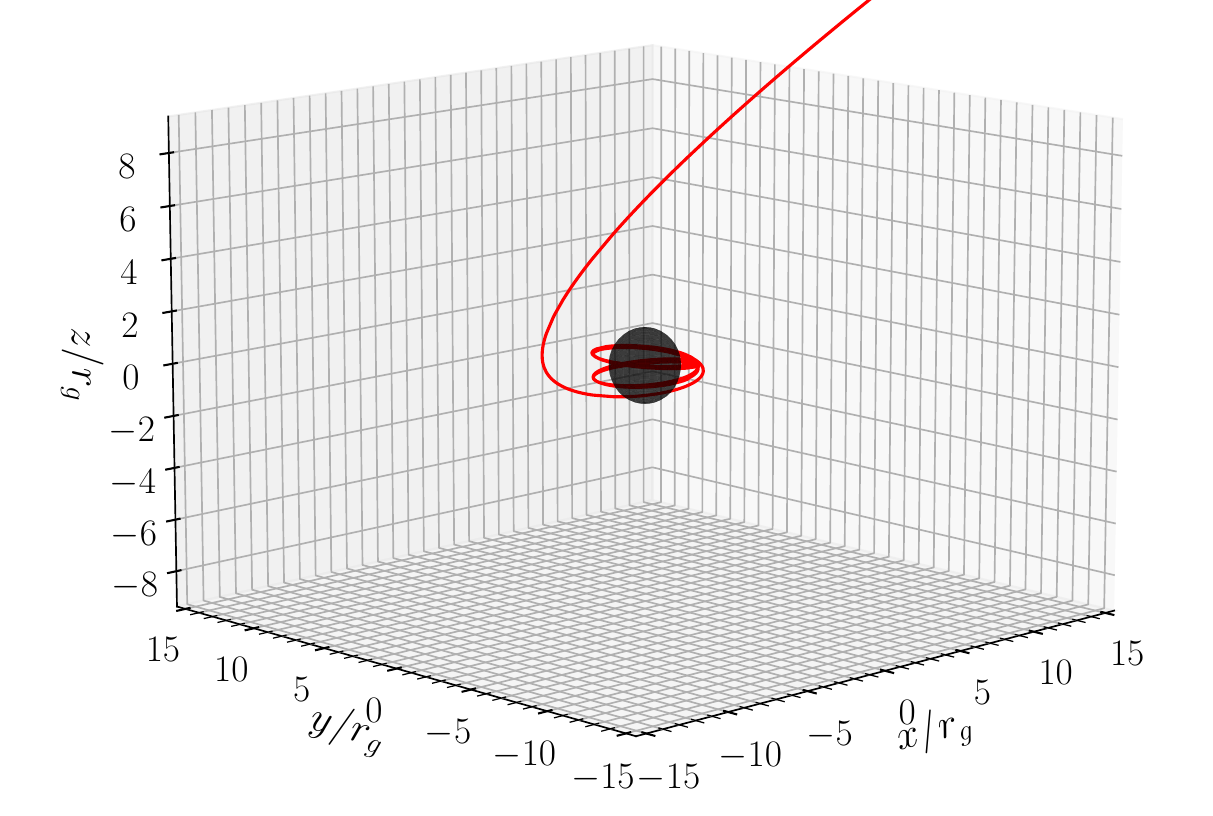}
    \caption{Example innermost bound spherical orbits for an $a_\bullet = 0.9$ Kerr black hole and six different  asymptotic inclinations $\psi$. The different systems have $\psi = 5^\circ$ (upper left), $\psi = 15^\circ$ (upper right), $\psi = 30^\circ$ (centre left), $\psi = 45^\circ$ (centre right), $\psi = 60^\circ$ (lower left) and $\psi = 75^\circ$ (lower right).  The test particle trajectory is displayed by a solid red curve, while the black holes radial extent is displayed by the shaded sphere. The 3 axes are the quasi-cartesian spheroidal representation of the Boyer-Lindquist coordinate system (see text).  }
    \label{fig:ibso_orbits}
\end{figure*}

 Naturally, as a function of only conserved quantities, $\psi$ itself is conserved over the course of an orbit, and all innermost bound spherical orbits can be parameterised in terms of the pair ($a_\bullet, \psi$). A geometric interpretation of $\psi$ is demonstrated in Fig. \ref{fig:geometric}, which highlights why $\psi$ is a much more natural parameter to use than $q$. In the limit of $\epsilon = 1$ the Carter constant is given by (see eq. \ref{carter})
 \begin{equation}
     q = p_\theta^2 + l_z^2 \cot^2\theta = l_x^2 + l_y^2 , 
 \end{equation}
and represents the square of the stars orbital angular momentum in the plane of the black holes equator. Therefore $l = \sqrt{q + l_z^2}$ corresponds to the total orbital angular momentum of the star, meaning that the angle $\iota$ 
\begin{equation}
    \cos \iota = {l_z \over l} ,
\end{equation}
is the angle between the particles orbital angular momentum vector and the black holes spin axis.   For the study of tidal disruption events it is often more relevant to consider the angle between the particles asymptotic {\it velocity} and the vertical axis. These two angles are trivially related by $\psi = \pi/2 - \iota$. 

The properties of the innermost bound spherical orbits in the Kerr metric have been studied extensively in the literature. Their properties are entirely specified by $a_\bullet$ and $\psi$, as we now demonstrate.  We write the right hand side of the radial equation of motion (eq. \ref{rad_mot}) as a pseudo-potential denoted $\Psi_R$, which is equal to 
\begin{multline}
    \Psi_R(r) = \left[ (r^2 + a^2) - a l_z \right]^2  \\ - (r^2 - 2r_gr + a^2) \left[r^2 + l_z^2\cot^2\psi + (l_z - a )^2 \right] ,
\end{multline}
where we have used $\epsilon = 1$ and $q = l_z^2 \cot^2\psi$ appropriate for the innermost bound spherical orbit. The radius of the innermost bound spherical orbit must represent a zero of this potential (as the radial velocity of the particle is zero in this limit). Not only must the potential be zero, but so must its radial gradient (or else the particle would not stay at this radius). The simultaneous equations 
\begin{equation}
    \Psi_R(r_{\rm ibso}) = 0, \quad \left. {{\partial \Psi_R} \over \partial r}\right|_{r_{\rm ibso}} = 0, 
\end{equation}
are sufficient to solve for both $l_z(a_\bullet, \psi)$ and the innermost bound spherical orbit radius $r_{\rm ibso}(a_\bullet, \psi)$. With these parameters determined the full orbit is specified. 

A particularly useful presentation of these solutions is due to \cite{Hod13} \citep[see also][]{Will12}, who derived the axial angular momentum of the innermost bound spherical orbits as a function of $a_\bullet$ and $\psi$ \citep[note that][actually used $\iota$ in place of $\psi$ as the conserved variable]{Hod13} 
\begin{equation}
    l_z = \sqrt{4 G M_\bullet r_g \chi^3    \over \chi^2 - a_\bullet^2 \cos^2\psi } \sin \psi .  
\end{equation}
This allows us to calculate the Carter constant $q$ as a function of $a_\bullet$ and $\psi$, which will be of use later 
\begin{equation}
    q = { 4 G M_\bullet r_g \chi^3 \cos^2\psi    \over \chi^2 - a_\bullet^2 \cos^2\psi } . 
\end{equation}
In these two expression we have defined $\chi$, which is the ratio of the innermost bound spherical orbit to the gravitational radius 
\begin{equation}
    \chi \equiv {r_{\rm ibso} \over r_g} . 
\end{equation}
Finally, the location of the innermost bound spherical orbit is given by the solution of the following equation \citep{Hod13} 
\begin{multline} \label{chi_eq_1}
    \chi^4 - 4\chi^3 - a_\bullet^2(1 - 3 \cos^2\psi)\chi^2 + a_\bullet^4\cos^2\psi \\ + 4a_\bullet \sin \psi \sqrt{\chi^5 - a_\bullet^2\chi^3\cos^2\psi} = 0 .
\end{multline}
Note that there is an intrinsic degeneracy in this expression between $a_\bullet < 0$ and inclination angles $-\pi/2 < \psi < 0$. This degeneracy is physical (inclinations $\psi < 0$ correspond to counter rotating orbits), and we for simplicity restrict the domain of interest to $0 \leq \psi \leq \pi/2$ and allow spin values $-1 \leq a_\bullet \leq 1$. An important result to note is that all angular terms (with dependence on $\psi$) are also functions of $a_\bullet$. The Schwarzschild solution $\chi = 4$ is independent of inclination, as expected. 

No closed form analytical solution for the root of this equation is known, but it is simple to solve numerically (it can be rearranged slightly to be more numerically stable, see Appendix \ref{poly_app}). Once a dimensionless black hole spin parameter and asymptotic inclination angle are chosen, the above expression is solved numerically for $\chi(a_\bullet, \psi)$, from which the axial angular momentum and Carter constant are uniquely determined.   It will turn out that this will be sufficient to compute the maximum tidal acceleration of the star's orbit, and therefore the Hills mass, without having to solve the geodesic motion in full.  This is discussed further in the following section.

\begin{figure}
    \centering
    \includegraphics[width=0.45\textwidth]{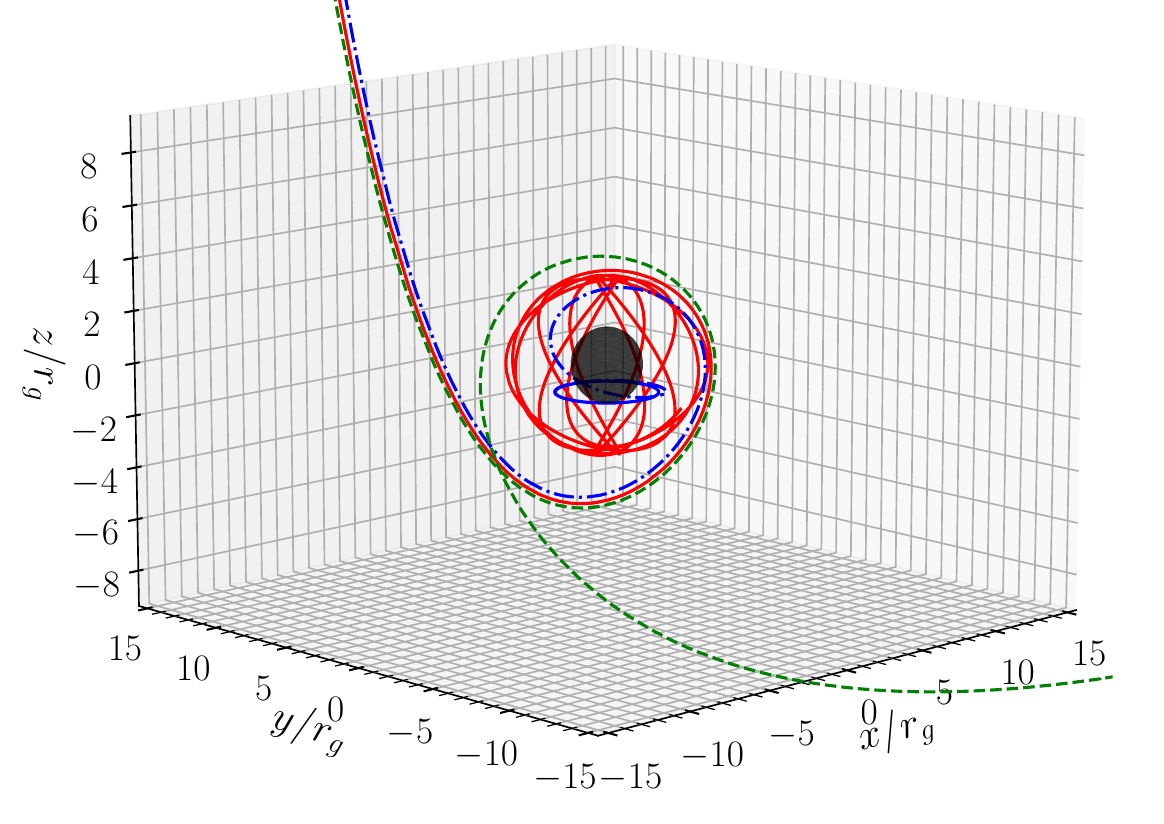}
    \caption{The seperatrix nature of the innermost bound spherical orbit. The innermost bound spherical orbit test particle trajectory is displayed by a solid red curve ($a_\bullet = 0.9, \psi= 5^\circ$). Any orbit with slightly less axial angular momentum (blue dot dashed) will terminate at the black hole singularity, while any more axial angular momentum (green dashed curve) will escape to infinity.   }
    \label{fig:seperatrix}
\end{figure}

Some example innermost bound spherical orbit trajectories are displayed in Figure \ref{fig:ibso_orbits} (see Appendix \ref{integrator_app} for a discussion of the numerical algorithm used to integrate the orbital equations). Each orbit was about an $a_\bullet = 0.9$ Kerr black hole, and six different  asymptotic inclinations $\psi$ where used: $\psi = 5^\circ$ (upper left), $\psi = 15^\circ$ (upper right), $\psi = 30^\circ$ (centre left), $\psi = 45^\circ$ (centre right), $\psi = 60^\circ$ (lower left) and $\psi = 75^\circ$ (lower right).  The test particle trajectory is displayed by a solid red curve, while the black hole's radial extent is displayed by the shaded sphere. The 3 axes are the quasi-cartesian spheroidal representation of the Boyer-Lindquist coordinate system 
\begin{align}
    x &= \sqrt{r^2 + a^2} \sin\theta \cos\phi, \\
    y &= \sqrt{r^2 + a^2} \sin\theta \sin\phi, \\
    z &= r \cos\theta. 
\end{align}

The seperatrix nature of these orbits is highlighted in Fig. \ref{fig:seperatrix}.  In red we display the critical innermost bound spherical orbit ($a_\bullet = 0.9, \psi= 15$). We also display two different orbits with subtly different axial angular momenta. Any orbit with slightly less axial angular momentum (blue dot dashed) will terminate at the black hole singularity, while any more axial angular momentum (green dashed curve) will escape to infinity.

\section{The tidal tensor and acceleration } \label{tidal_sec}
In this section we derive the relevant properties of the tidal tensor in the Kerr geometry for our calculation.  We begin with a recap of Newtonian tidal forces (section \ref{newt_tides}), and a general derivation of the relativistic tidal tensor (section \ref{gr_tides}), before specialising to the Kerr geometry in section \ref{kerr_tides}. Those readers familiar with a relativistic formulation of tidal forces may wish to skip directly to section \ref{kerr_tides}. 
\subsection{Newtonian tidal forces} \label{newt_tides}
Before we discuss tidal forces in full general relativity it will be worth recapping a Newtonian description of their origin. Consider a gravitational potential $\Phi$, and a test particle at location $X^j$ which experiences a force per unit rest mass $f_i$ 
\begin{equation}
    f_i = - \left.{\partial \Phi \over \partial x^i}\right|_X .
\end{equation}
A second particle at location $X^{'j} = X^j + \delta x^j$ experiences a force per unit rest mass
\begin{equation}
    f_i' = - \left.{\partial \Phi \over \partial x^i}\right|_{X+\delta x} = - \left.{\partial \Phi \over \partial x^i}\right|_X - \left.{\partial^2 \Phi \over \partial x^i \partial x^j}\right|_X \delta x^j + {\cal O}(\delta x^2),
\end{equation}
where we shall consider deviations $\delta x$ small enough so that the quadratic terms are negligible. 
The differential acceleration experienced by the two particles is 
\begin{equation}
    {{\rm d}^2 \over {\rm d}t^2} \delta x_i = f'_i - f_i = - \left.{\partial^2 \Phi \over \partial x^i \partial x^j}\right|_X \delta x^j \equiv - C_{ij} \delta x^j ,
\end{equation}
which defines the tidal tensor 
\begin{equation}
    C_{ij}(X) \equiv \left.{\partial^2 \Phi \over \partial x^i \partial x^j}\right|_X .
\end{equation}
The Newtonian tidal tensor has a number of properties which are shared by its relativistic counterpart.  The first is that in the vacuum external to a gravitational source it is traceless, as can be seen by using the Poisson equation 
\begin{equation}
    \sum_i C_{ii} = \left.\nabla^2 \Phi\right|_X = 4\pi G \rho(X) = 0 .
\end{equation}
Secondly, of principle physical interest are the eigenvalues and eigenvectors of the this tidal tensor, as they correspond to both the direction (eigenvectors) and magnitude (eigenvalues) of the principle axis of the tidal acceleration. This can be verified with 
\begin{equation}
    {{\rm d}^2  x_i \over {\rm d}t^2} = - C_{ij}  x^j = - \lambda  x_i,
\end{equation}
where $\vec x$ is the eigenvector and $\lambda$ its corresponding eigenvalue. We see that $\lambda < 0$ corresponds to positive tidal accelerations (stretching) and $\lambda > 0$ corresponds to negative tidal accelerations (squeezing).   With this groundwork set we now proceed to a relativistic description of tidal accelerations.

\subsection{The relativistic tidal tensor} \label{gr_tides}
The computation of the relativistic tidal tensor proceeds along much the same lines as its Newtonian counterpart.   We begin by considering the differential acceleration of two test particles following geodesics  $X^\mu(\tau)$ and $X^{'\mu}(\tau) = X^\mu(\tau) + \delta x^\mu(\tau)$, where $\tau$ is the proper time. We shall again only consider excursions $\delta x^\mu$ sufficiently small so that terms at quadratic and higher orders in $\delta x$ can be neglected.   We also work, for convenience, in locally free-falling coordinates where the affine connection $\Gamma^\mu_{\alpha \beta} = 0$.  The gradients of $\Gamma$ are crucially not zero in these coordinates. 

We wish to compute the ``differential acceleration'' of the two particles, or explicitly 
\begin{equation}
    {{\rm D}^2 \over {{\rm D}\tau}^2} \delta x^\mu = U^\beta \nabla_\beta \left[ U^\alpha \nabla_\alpha \delta x^\mu \right],
\end{equation}
where 
\begin{equation} 
    U^\mu \equiv {{\rm d}X^\mu \over {\rm d} \tau} , 
\end{equation}
and $\nabla_\mu$ is the covariant derivative, and 
\begin{equation}
    {{\rm d} \over {\rm d}\tau } \equiv U^\gamma {\partial \over \partial x^\gamma}.
\end{equation} 
The derivative ${\rm D}/{\rm D}\tau$ corresponds physically to a derivative with respect to proper time along the geodesic. Expanding out the first derivative leaves 
\begin{equation}
    {{\rm D}^2 \over {{\rm D}\tau}^2} \delta x^\mu = U^\beta \nabla_\beta \left[{{\rm d} \over {\rm d}\tau} \delta x^\mu + \Gamma^\mu_{\nu \alpha } \delta x^\alpha U^\nu \right] .
\end{equation}
Upon further expansion we find  
\begin{equation}\label{accel1}
    {{\rm D}^2 \over {{\rm D}\tau}^2} \delta x^\mu ={{\rm d}^2 \over {\rm d}\tau^2} \delta x^\mu + \left.{\partial \Gamma^\mu_{\nu \alpha}\over \partial x^\lambda} \right|_X U^\lambda U^\nu \delta x^\alpha  ,
\end{equation}
where various terms proportional to $\Gamma^{a}_{bc}$ have been set to zero, a result of us working in freely falling coordinates. To continue we must now compute the first term on the right hand side. This can be accomplished with the geodesic equation. 

The equation of motion of each test particle is given by the geodesic equation, for the first particle (geodesic $X^\mu(\tau)$) this is  
\begin{equation}
    {{\rm d}^2 X^\mu \over {{\rm d}\tau^2}} = - \left.\Gamma^\mu_{\alpha \beta}\right|_X U^\alpha U^\beta = 0 ,
\end{equation}
where the $= 0$ follows from our choice of locally free-falling coordinates. 
For the second particle we have 
\begin{equation}
    {{\rm d}^2 X'^\mu \over {{\rm d}\tau^2}} = - \left.\Gamma^\mu_{\alpha \beta}\right|_{X+\delta x} U'^\alpha U'^\beta .
\end{equation}
Taking the difference between these two expressions, and expanding to linear order in $\delta x$, we find 
\begin{equation}
     {{\rm d}^2 \delta x^\mu \over {{\rm d}\tau^2}} = - \left. {\partial\Gamma^\mu_{\alpha \beta} \over \partial x^\lambda} \right|_X U^\alpha U^\beta \delta x^\lambda + {\cal O}(\delta x^2) . 
\end{equation}
Substituting this linear order result back into equation \ref{accel1}, we find after some index relabelling  
\begin{equation}\label{accel}
    {{\rm D}^2 \over {{\rm D}\tau}^2} \delta x^\mu = \left[\left.{\partial \Gamma^\mu_{\alpha \lambda}\over \partial x^\beta} \right|_X - \left.{\partial \Gamma^\mu_{\alpha \beta}\over \partial x^\lambda} \right|_X\right] U^\alpha U^\beta  \delta x^\lambda  .
\end{equation}
In locally free falling coordinates the term in the square brackets is equal to the Riemannian tensor $R^\mu_{\alpha \lambda \beta}$, i.e.,
\begin{equation}\label{accel}
    {{\rm D}^2 \over {{\rm D}\tau}^2} \delta x^\mu = R^\mu_{\alpha \lambda \beta} U^\alpha U^\beta  \delta x^\lambda \equiv - C^\mu_\lambda \delta x^\lambda  ,
\end{equation}
which defines our relativistic tidal tensor $C^\mu_\lambda$. We are of course only interested in spatial curvature, and we define our neighbouring geodesics so that $\delta x^0 = \delta t = 0$, leading to 
\begin{equation}
    C_{ij} \equiv - R_{i \alpha j \beta} U^\alpha U^\beta . 
\end{equation}
Where spacetime indices $i, j$ span 1, 2 and 3.  Much like its Newtonian analogue this tensor is 3x3, symmetric (which follows from the symmetry properties of the Riemann tensor) and traceless in the absence of matter. To prove this final point note 
\begin{equation}
    C = - R_{\alpha \beta} U^\alpha U^\beta = -{8\pi G\over c^2 } \left[ T_{\alpha \beta} - {1\over 2} g_{\alpha \beta }   T \right] U^\alpha U^\beta = 0 ,
\end{equation}
where $C$ is the trace of $C_{ij}$, $R_{\alpha \beta}$ is the Ricci tensor and we have used Einstein's field equations in going to the second to last equality. The final equality follows from $T_{\alpha\beta}=0$ in the vacuum. 

There is one final subtlety to computing relativistic tidal forces. The physical tidal acceleration we are interested in is the one experienced by the particle in its rest frame as it follows its geodesic, and therefore we wish to compute the above tensor {\it in the local rest frame of the orbiting particle}  \citep[e.g.,][for a discussion]{Pirani56}. One must therefore construct this locally free-falling frame, and project the tensor $C_{ij}$ into this coordinate system. Only then can the tidal acceleration experienced by the particle (the eigenvalues of $C_{ij}$ in this coordinate system) be computed.  Fortunately, this coordinate projection has been solved in general for the Kerr metric by \cite{Marck83}, which we now discuss. 

\subsection{ Tidal forces in the Kerr metric } \label{kerr_tides}
Although Boyer-Lindquist coordinates (e.g., equation \ref{metric}) reduce to the Minkowski metric in spherical coordinates in the formal limit $r\to \infty$, and are convenient for expressing the fundamental dynamical equations of test particles, they do not constitute an orthogonal locally free-falling reference frame at any finite radius. The relevant locally free-falling  coordinate system experienced by a moving test particle are known as ``Fermi normal'' coordinates \citep[e.g.][]{Manasse63}, which we now outline.  

The Fermi normal coordinates consist of an orthonormal tetrad which are denoted $\xi^\mu_{(\nu)}$, where $(\nu)$ labels the four different vectors in the tetrad (the brackets here are used to avoid confusion with spacetime indices), and $\mu$ labels the components of each vector.  The timelike member of this tetrad ($\xi^\mu_{(0)}$) is the tangent vector pointing along the central geodesic (i.e., it is proportional to the particles 4-velocity $\xi_{(0)}^\mu \propto U^\mu$). The spacelike vectors $\xi^\mu_{(i)}$ then span the surface orthogonal to $\xi^\mu_{(0)}$, and are centered on the location of the star at all points along its orbit. The spacelike vectors of the tetrad are chosen so that the metric in this frame is locally Minkowski,  and that each of $\xi_{(i)}$ are parallel transported with the particle as it evolves along its geodesic. 

It was shown by \cite{Marck83} that the orthonormal tetrad $\xi_{(\nu)}^\mu$ can be written down explicitly for a general geodesic orbit in the full Kerr metric.  This tetrad are algebraically complex \citep[see][for full expressions]{Marck83, vandeMeent20} but nonetheless the Riemann tensor can then be projected into this Fermi-normal coordinate system, at which point the relevant tidal tensor $C_{ij}$ can be computed, and then represented by standard Boyer-Lindquist coordinates.    The resulting tidal tensor is algebraically complex, and we present it in full in Appendix \ref{tidal_tensor_app}.

Fortunately our problem is simplified significantly, owing to the fact that we are considering the special class of spherical geodesic orbits.  A key orbital property of these solutions for the tidal problem is the following: every spherical orbit crosses the equatorial $\theta = \pi/2$ plane a formally infinite number of times (e.g., Fig. \ref{fig:ibso_orbits}). To prove this rigorously note that the $\theta$ evolution equation (eq. \ref{thet_mot}) becomes, in the $\epsilon = 1$, $q = l_z^2 \cot^2\psi$ innermost bound spherical orbit limit,  
\begin{equation}
    (r^2 + a^2 \cos^2\theta)^2 \left( {{\rm d}\theta \over {\rm d}\tau}\right)^2 = l_z^2 \left[\cot^2\psi - \cot^2\theta \right] ,
\end{equation}
and so $\theta(\tau)$ oscillates between the two angles $\theta^\pm$ 
\begin{equation}
    \theta^+ = \psi , \quad \theta^- = \pi - \psi  ,
\end{equation}
as can be seen in Fig. \ref{fig:ibso_orbits}. 

Tidal forces are maximal in the Kerr equatorial plane, and so this is the relevant part of the trajectory for our calculation. Denoting the tidal tensor in the equatorial plane as $\widetilde C_{ij}(r) = C_{ij}(r, \theta=\pi/2)$, we find that $\widetilde C_{12} = 0 = \widetilde C_{23}$ \citep[see also][]{Marck83, Kesden12}, and therefore 
\begin{equation}
    \widetilde C_{ij} = \begin{pmatrix}
        \widetilde C_{11} & 0 & \widetilde C_{13} \\
        0  & \widetilde C_{22} & 0 \\
        \widetilde C_{13} & 0 & \widetilde C_{33}  
    \end{pmatrix}
    .
\end{equation}
The remaining tensor components (see Appendix \ref{tidal_tensor_app} for a derivation) are
\begin{align}
    \widetilde C_{11} &= \left[1 - {3 (r^2 + k) \over r^2} \cos^2\gamma\right] {GM_\bullet  \over r^3} , \\ 
    \widetilde C_{22} &= \left[1 + {3 k \over r^2} \right] {GM_\bullet \over r^3} , \\ 
    \widetilde C_{33} &= \left[1 - {3 (r^2 + k) \over r^2} \sin^2\gamma\right] {GM_\bullet  \over r^3} , \\ 
    \widetilde C_{13} &= - \left[{3 (r^2 + k) \over r^2} \cos\gamma\sin\gamma\right] {GM_\bullet  \over r^3} ,
\end{align}
where we remind the reader that $k$ is the non-negative \cite{Carter68} constant $k = q + (l_z - a\epsilon)^2$ (equation \ref{carter}). The angle $\gamma$ is a dynamic quantity in the Fermi-normal coordinate system, and represents a time dependent angle introduced so that the spacelike tetrad vectors $\xi^\mu_{(i)}$ are parallel propagated over the geodesic. The angle $\gamma$ does not, as we prove below, enter the observable tidal accelerations of the particle in the equatorial plane.  For a more detailed discussion about $\gamma$ and its equation of motion see \cite{Marck83}.

Standard methods show that this tensor has eigenvalues 
\begin{align}
    \lambda_1 &= \widetilde C_{22} , \\
    \lambda_\pm &= {1\over 2} \left[ \widetilde C_{11} + \widetilde C_{33} \pm \sqrt{\left(\widetilde C_{11}  - \widetilde C_{33} \right)^2 + 4 \left(\widetilde C_{13} \right)^2 } \right] ,
\end{align}
and that, as expected $\lambda_1 + \lambda_+ + \lambda_- = {\rm trace}\left(\widetilde C_{ij}\right) = 0.$

We are only interested in the negative eigenvalue of the tidal tensor, as this corresponds physically to the tidal stretching of the star.  Upon subsitution of the above tensor components we see that the only negative eigenvalue is $\lambda_-$, with value 
\begin{equation}
    \lambda_- = -{2 G M_\bullet  \over r^3} \left[1 + {3 k \over 2 r^2} \right] .
\end{equation}
This eigenvalue corresponds physically to the tidal acceleration (per unit length) experienced by two particles separated by an infinitesimal rest frame displacement $\delta x^i$, as the particles cross the equatorial plane of a Kerr black hole at radial Boyer-Lindquist coordinate $r$. In the asymptotic limit of our innermost bound spherical orbits when $r \to r_{\rm sp} = \chi r_g$, this will represent the maximum tidal acceleration (per unit length) experienced by the particle on the special orbit which separates infalling orbits and unbound orbits.  This is precisely the acceleration we desire for our calculation of the Hills mass. 

The magnitude of the maximum tidal acceleration is therefore given approximately by the following expression 
\begin{equation}
    a_T(a_\bullet, \psi) = R_\star \times |\lambda_-(r = \chi r_g)| , 
\end{equation}
where we have made the approximation that the tidal acceleration per unit length remains constant across the radius of the star $R_\star$. This is a good approximation for the large mass black holes (where $r_g \gg R_\star$) we are considering in this paper. 

When expanded and simplified this maximum acceleration is given explicitly by 
\begin{multline}
    a_T(a_\bullet, \psi) =   {2 c^6  R_\star \over G^2 M_\bullet^2} {1\over \chi^{3}}   
     \Bigg[1 + {6\chi \over \chi^2 - a_\bullet^2 \cos^2\psi} \\ + {3a_\bullet^2 \over 2\chi^2} - {6a_\bullet \sin \psi \over \sqrt{\chi^3 - a_\bullet^2 \chi \cos^2\psi}} \Bigg] .
\end{multline}
where we have used that results presented in section \ref{dynamic_sec}
\begin{align}
    k &= q + (l_z - a)^2 , \\
    q &= {4 G M_\bullet r_g \chi^3  \cos^2\psi  \over \chi^2 - a_\bullet^2 \cos^2\psi } , \\
    l_z &= \sqrt{4 G M_\bullet r_g \chi^3    \over \chi^2 - a_\bullet^2 \cos^2\psi } \sin \psi ,  
\end{align}
and we remind the reader that $\chi$ is specified entirely by the black hole spin $a_\bullet$ and the asymptotic inclination $\psi$ (equation \ref{chi_eq_1}). We remind the reader that there exists a degeneracy between orbits with  $a_\bullet < 0, \psi > 0$ and $a_\bullet > 0, \psi < 0$. 

\section{The general solution for the Hills mass} \label{hills_mass_sec}

\begin{figure}
    \centering
    \includegraphics[width=0.49\textwidth]{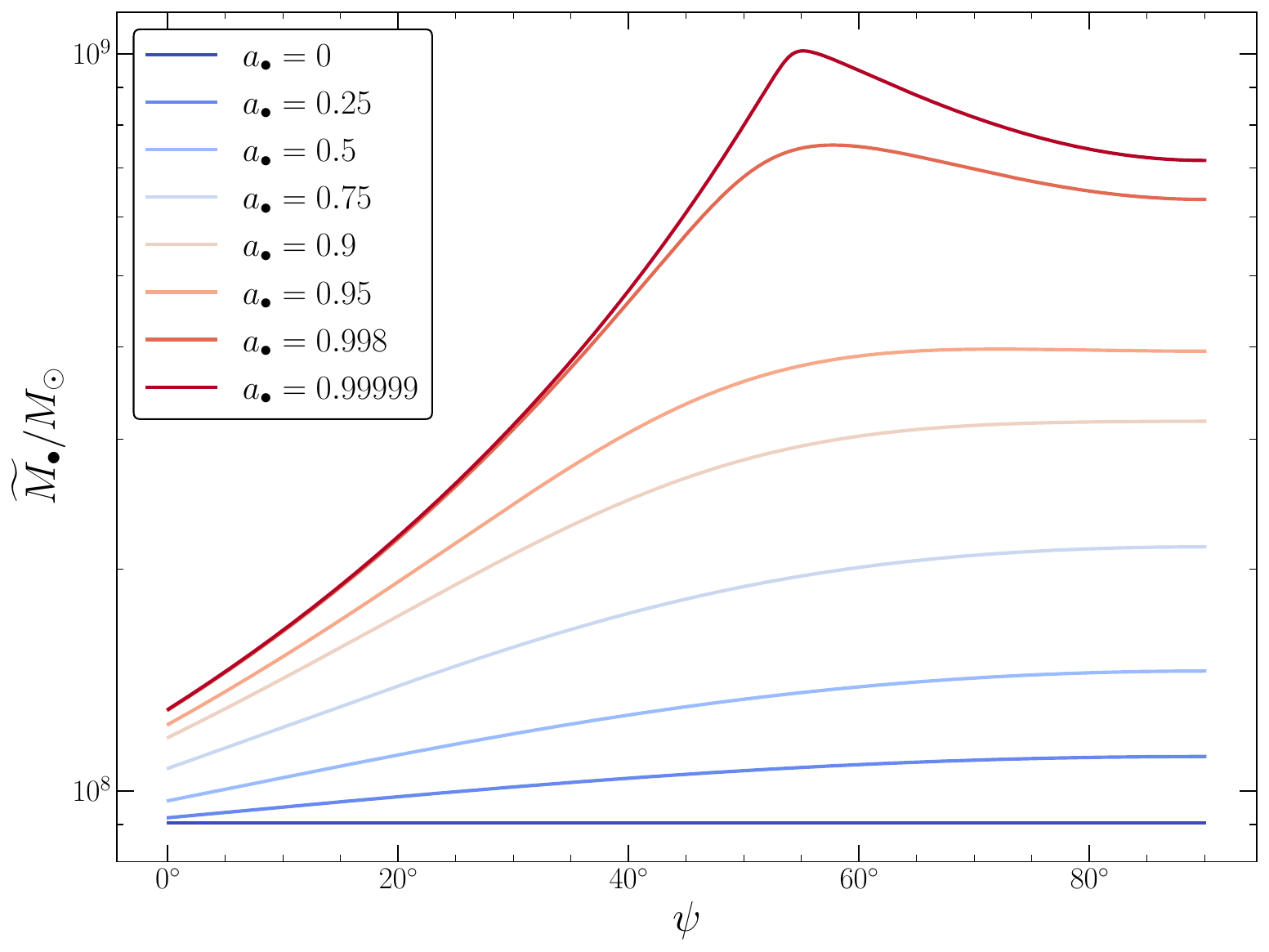}
    \includegraphics[width=0.49\textwidth]{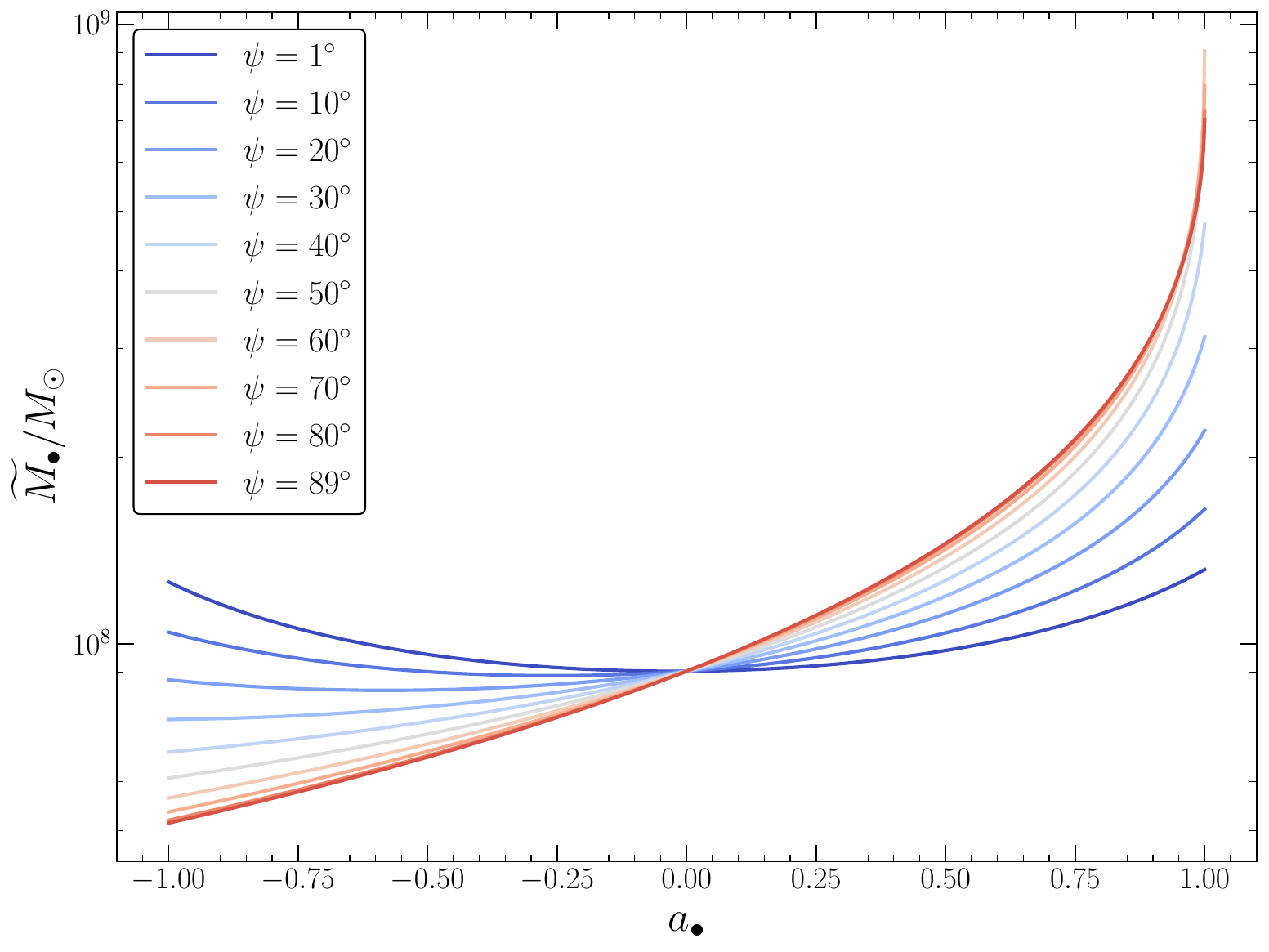}
    \caption{The Hills mass, assuming a solar-type star $M_\star = M_\odot$, $R_\star = R_\odot$, and $\eta = 1$, as a function of black hole spin $a_\bullet$ and asymptotic inclination $\psi$. At fixed inclination the Hills mass is a strictly increasing function of prograde black hole spin, but the same is not true in reverse. For high spins $a_\bullet \gtrsim 0.92$ the Hills mass peaks at asymptotic inclinations out of the equatorial plane. Recall that spins $a_\bullet < 0$ are degenerate physically with inclinations $\psi < 0$. }
    \label{fig:mass_spin_psi}
\end{figure}

\begin{figure}
    \centering
    \includegraphics[width=0.49\textwidth]{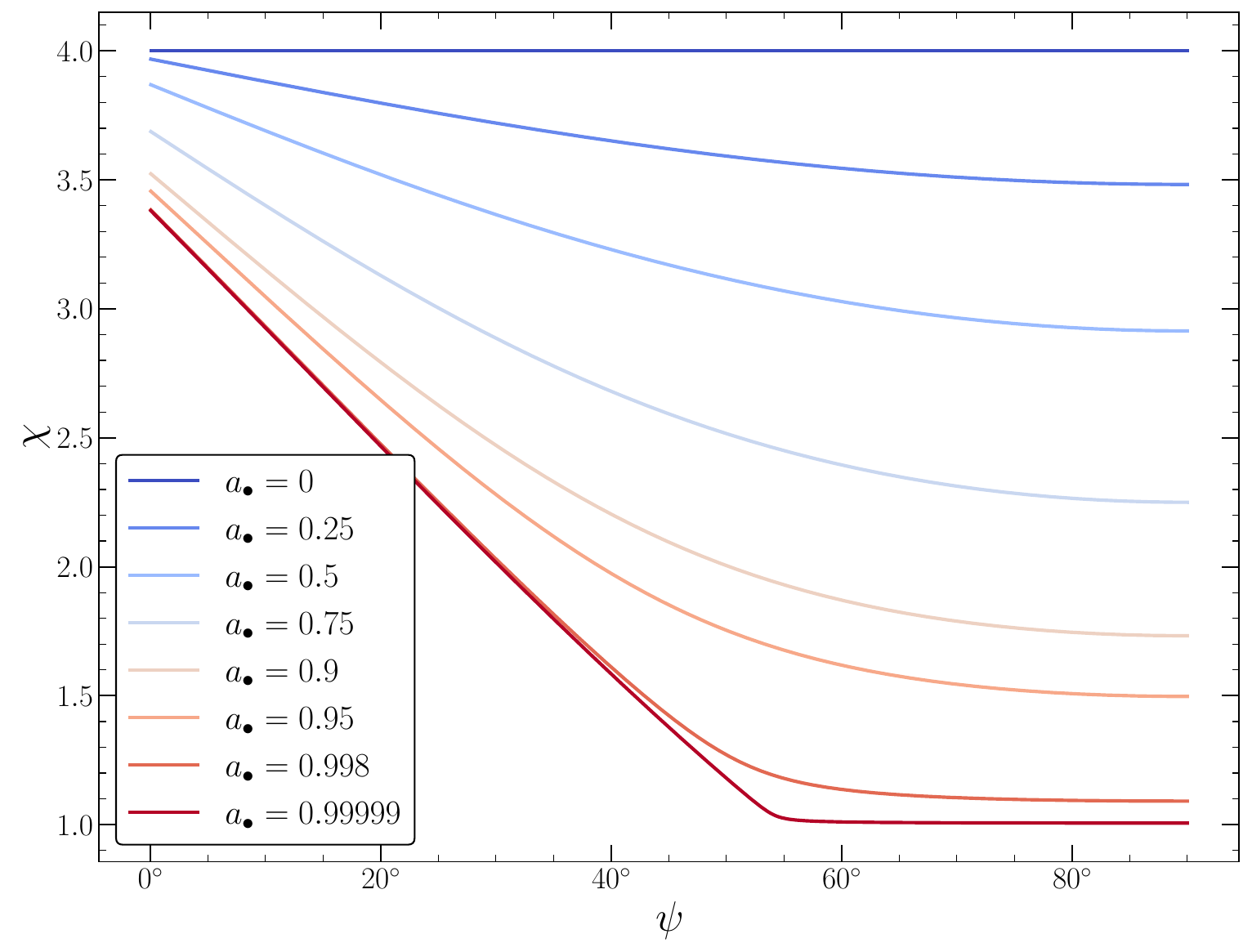}
    \caption{The radius of the innermost bound spherical orbit as a function of asymptotic inclination for a number of black hole spins $a_\bullet$. For extremal spins the $\chi$ parameter undergoes a ``phase transition'' at a critical inclination $\sin \psi^\dagger = \sqrt{2/3}$, and is equal to 1 for all $\psi \geq \psi^\dagger$.  }
    \label{fig:chi_spin_psi}
\end{figure}

The Hills mass is defined as the maximum mass at which a black hole can tidally disrupt an incoming star and produce debris streams which are in principle observable. For a star to be tidally disrupted the tidal acceleration (discussed above) must overcome the acceleration experienced by the outer edges of the star towards its centre. This self gravity is given by 
\begin{equation}
    a_S = \eta {G M_\star \over R_\star^2} ,
\end{equation}
where the scaling with stellar parameters is the natural dimensionfull scale of the acceleration, and $\eta$ parameterises our ignorance of the detailed stellar structure \citep[see e.g.,][for an attempt to parameterise $\eta$ in terms of the incoming stellar structure]{Guillochon13}.  Most analyses assume that $\eta = 1$ \citep[e.g.][]{Kesden12, MumBalb20b}, but this is likely a simplification. If the star is rotating, for example, it is less bound and more easily disrupted $\eta < 1$. On the other hand, $\eta = 1$  might overestimate the ease at which stars are fully disrupted since it is possible that a star may lose its less bound outer layers while maintaining its more dense core \citep{Ryu20}, in which case $\eta > 1$ is relevant for a full disruption \citep[e.g.,][]{Phinney89}.

Nevertheless, we may now derive the general expression for the Hills mass, which we denote $\widetilde M_\bullet$. The tidal acceleration derived in the previous section, which corresponds physically to the maximum tidal acceleration experienced by the star on the critical innermost bound spherical orbit, is equal to the stellar self gravity at the innermost bound spherical radius when 
\begin{multline}\label{HillsMass}
    \widetilde M_\bullet(a_\bullet, \psi) = \left[{2 R_\star^3 c^6 \over \eta G^3 M_\star } \right]^{1/2} {1\over \chi^{3/2}} \\ 
     \Bigg[1 + {6\chi \over \chi^2 - a_\bullet^2 \cos^2\psi} + {3a_\bullet^2 \over 2\chi^2} - {6a_\bullet \sin \psi \over \sqrt{\chi^3 - a_\bullet^2 \chi \cos^2\psi}} \Bigg]^{1/2}, 
\end{multline}
where we remind the reader that $\chi(a_\bullet, \psi)$ is the root of 
\begin{multline}
    \chi^4 - 4\chi^3 - a_\bullet^2(1 - 3 \cos^2\psi)\chi^2 + a_\bullet^4\cos^2\psi \\ + 4a_\bullet \sin \psi \sqrt{\chi^5 - a_\bullet^2\chi^3\cos^2\psi} = 0 .
\end{multline}
Any star incident at an angle $\psi$  upon a Kerr black hole with mass $M_\bullet$ and spin $a_\bullet$ where   $M_\bullet > \widetilde M_\bullet$ can not produce observable emission.   As noted in \cite{Mummery_et_al_23}, this solution is particularly simple in the equatorial plane ($\psi = \pi/2$)
\begin{equation}
    \widetilde M_\bullet(a_\bullet, \pi/2) = \left[{5 R_\star^3 c^6 \over \eta G^3 M_\star } \right]^{1/2} {1\over \left(1 + \sqrt{1-a_\bullet}\right)^{3}}.
\end{equation}

We plot the properties of $\widetilde M_\bullet$, for a canonical solar-type star $M_\star = M_\odot, R_\star = R_\odot$, $\eta=1$, as a function of $a_\bullet$ and $\psi$ in Fig. \ref{fig:mass_spin_psi}.  For a Schwarzschild black hole ($a_\bullet = 0$) the stars motion is confined to a plane and the incident inclination has no effect. For spins restricted to $0 < a_\bullet \lesssim 0.92$ increasing the incident angle $\psi$ increases the Hills mass, which is maximum in the equatorial plane. However, and potentially unexpectedly, for spins $a_\bullet \gtrsim 0.92$ the maximum Hills mass in fact peaks at an incoming inclination which is outside of the equatorial plane $\psi \neq \pi/2$. This is an important result. 

It is possible to understand the physical origin of this result by studying the location of the innermost bound spherical orbit radius as a function of inclination at high spins.  The location of the innermost bound spherical orbit (in units of $r_g$) is displayed in Fig. \ref{fig:chi_spin_psi} as a function of $a_\bullet$ and $\psi$. It is clear that, for the very highest spins, $\chi(\psi)$ shows an interesting behaviour.  It was first noted by \cite{Will12} that for extremal black hole spins $a_\bullet = 1$ the $\chi$ parameter undergoes a ``phase transition'' \citep{Hod13} at a critical inclination 
\begin{equation}
    \sin\psi^\dagger = \sqrt{2/3} , \quad \psi^\dagger \approx 54.7^\circ ,
\end{equation} 
and is equal to $\chi = 1$ for all $\psi \geq \psi^\dagger$. Taking the limit $a_\bullet \to 1$, $\chi \to 1$, the factor
\begin{equation}
    \Delta(\psi) = \Bigg[{5\over 2} + {6 \over 1 -  \cos^2\psi}  - {6\sin \psi \over \sqrt{1 -  \cos^2\psi}} \Bigg]^{1/2} ,
\end{equation}
is a strictly decreasing function of $\psi$ in the range 
\begin{equation}
    \psi^\dagger \leq \psi \leq \pi/2 . 
\end{equation}
This means that the maximum tidal acceleration (and Hills mass) occurs for $\psi^\dagger$, with corresponding Hills mass value  
\begin{equation}
    \widetilde M_\bullet(a_\bullet=1, \psi=\psi^\dagger) = \left[{11 R_\star^3 c^6 \over \eta G^3 M_\star } \right]^{1/2} , 
\end{equation}
which compares with the equatorial plane result 
\begin{equation}
    \widetilde M_\bullet(a_\bullet=1, \psi=\pi/2) = \left[{5 R_\star^3 c^6 \over \eta G^3 M_\star } \right]^{1/2} , 
\end{equation}
an increase by a factor $\sqrt{11/5}\simeq 1.48$. The Schwarzschild Hills mass is 
\begin{equation}\label{schwarz_hills}
    \widetilde M_\bullet(a_\bullet=0, \psi) = \left[{5 R_\star^3 c^6 \over 64 \eta G^3 M_\star } \right]^{1/2} , 
\end{equation}
a factor $8\sqrt{11/5} \simeq 11.87$ times lower than the theoretical maximum. 
\begin{figure}
    \centering
    \includegraphics[width=0.49\textwidth]{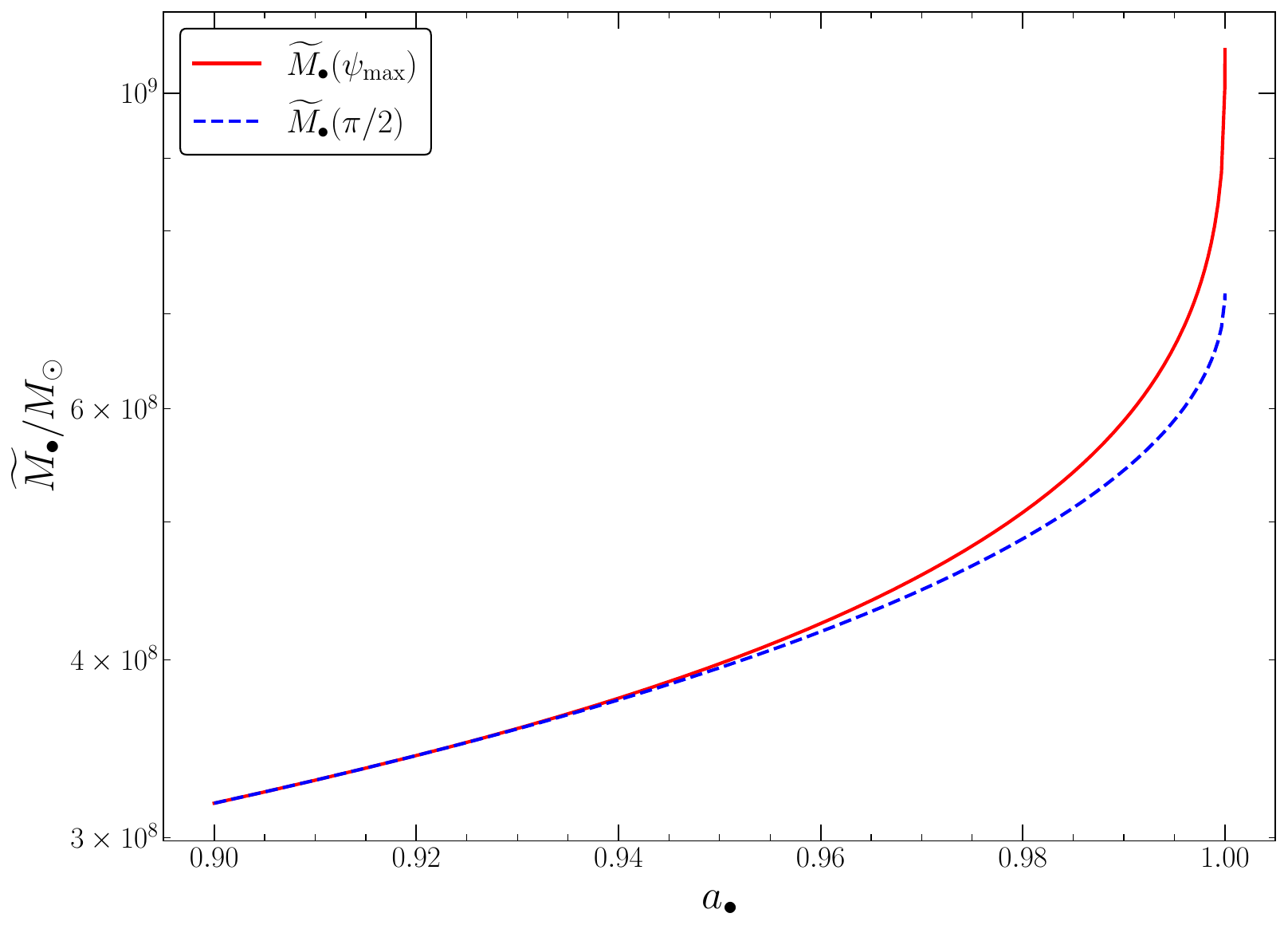}
    \includegraphics[width=0.49\textwidth]{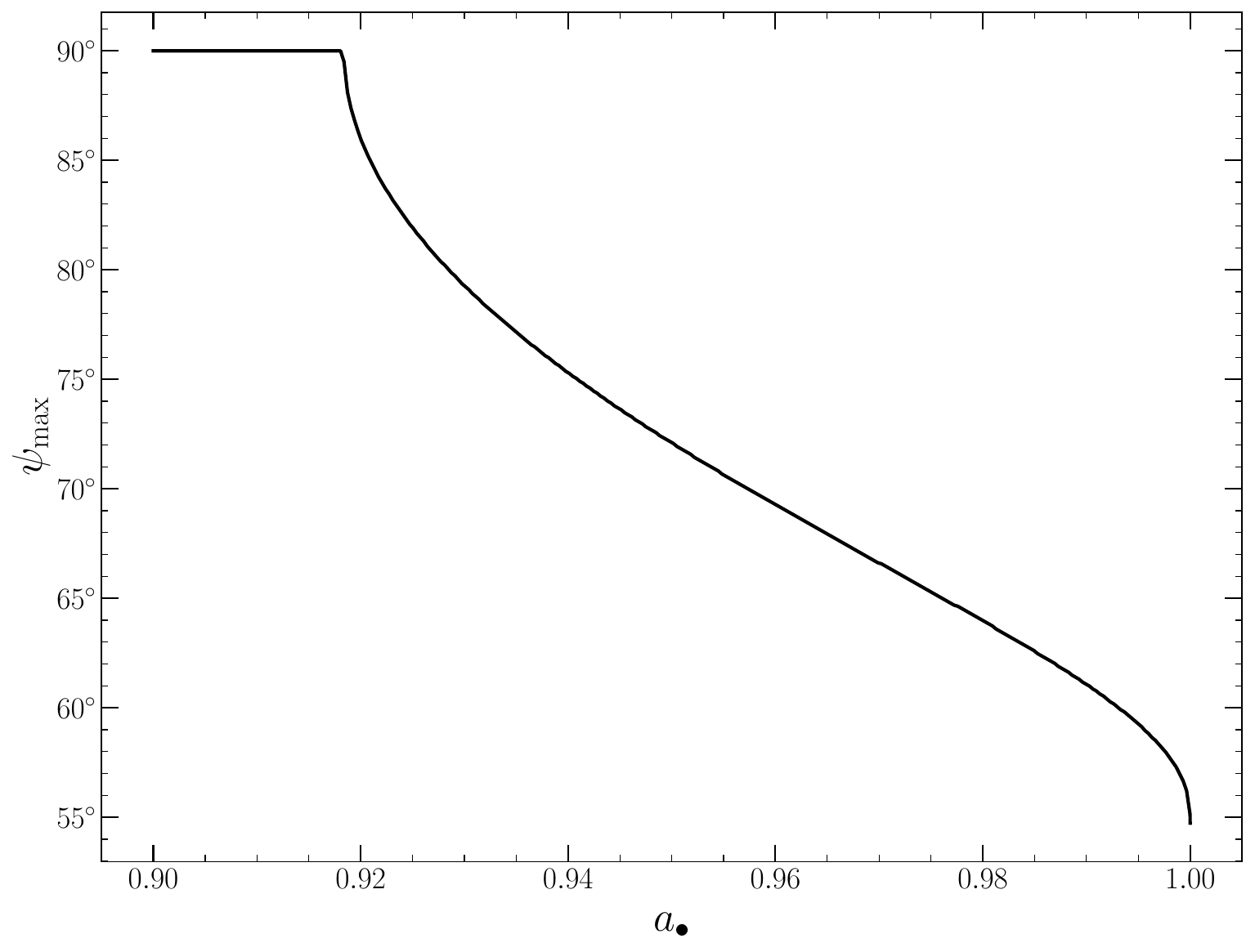}
    \caption{Upper: the maximum Hills mass of all inclinations (red solid curve) and the Hills mass in the equatorial plane (blue dashed curve), as a function of black hole spin. For spins $a_\bullet \gtrsim 0.95$ the Hills mass at an inclined angle $\psi_{\rm max}$ is notably larger than the equatorial plane value.  The value of $\psi_{\rm max}$ as a function of black hole spin is shown in the lower panel.  }
    \label{fig:new_max_mass}
\end{figure}

The deviation from the equatorial Hills mass value starts to become apparent at spins $a_\bullet \approx 0.95$ (Fig. \ref{fig:new_max_mass}), and grows increasingly worse as the extremal spin limit is reached.  The inclination of maximum Hills mass smoothly evolves from the equatorial plane $\psi_{\rm max} = \pi/2$ to the limiting value $\psi^\dagger$ as the black hole spin is increased to $a_\bullet = 1$ (lower panel, Fig. \ref{fig:new_max_mass}). 

With a general expression for the Hills mass as a function of inclination and black hole spin it is possible, given a black hole mass constraint of a tidal disruption event, to place conservative spin constraints on the black hole which produced that event.

\section{Constraining black hole spins with tidal disruption event black hole masses } \label{spin_sec}
{In this section we develop a Bayesian framework for inferring posterior distributions of the black hole parameters  (mass and spin) in an observed tidal disruption event, given a prior estimate of the black hole's mass (e.g., from a galactic scaling relationship, or the tidal disruption event's observed properties). These posterior distributions will only utilise the properties of tidal forces in their inference.   }

Given a black hole mass $M_\bullet$, it is only possible for a tidal disruption event to occur if $M_\bullet \leq \widetilde M_\bullet(a_\bullet, \psi, M_\star, R_\star)$, where the Hills mass $\widetilde M_\bullet$ is given by equation \ref{HillsMass}. {Therefore, if we observe a tidal disruption event from a given galaxy we know for certain that the black hole in that galaxy has mass $M_\bullet < \widetilde M_\bullet$.} {In a Bayesian framework the observation of a tidal disruption event represents an update to the information we have about that system, and once a likelihood function for the observation (or not) of a tidal disruption event is specified, posterior distributions of the system parameters can be computed from our prior assumptions about the system parameters. } Let us define the tidal disruption event {likelihood function} in the following manner 
\begin{equation}\label{pTDE}
    {\cal L}_{\rm TDE}(a_\bullet, \psi, M_\star, R_\star, M_\bullet) = \Theta\left[\widetilde M_\bullet(a_\bullet, \psi, M_\star, R_\star) - M_\bullet\right],
\end{equation}
where $\Theta(z)$ is the Heaviside theta function defined by 
\begin{equation}
    \Theta(z\geq0) = 1, \quad \Theta(z<0) = 0.
\end{equation}
This {likelihood function} represents a significant simplification as it assumes that every stellar orbit approaching the black hole has the precise axial angular momentum, energy and Carter constant of the innermost bound spherical orbit. {This in general is a bad assumption, as it does not utilise our knowledge of the physics of loss-cone scattering in galaxies. One result from which, for example, shows that the square of the angular momenta of those stellar orbits relevant for tidal disruption events is approximately uniformly distributed \citep{Coughlin22}, and the special orbits considered here therefore take up a small region of the relevant parameter space of scattered stellar orbits. A proper calculation of differential tidal disruption event probabilities involves performing integrals of the phase-space distribution function over the loss cone, a formalism which is discussed in detail in the  references \citep[][see \citealt{Singh23} for a more recent analysis]{Magorrian99, Wang04}.  } 

{A result found in these more detailed analyses is that it is always less likely that a tidal disruption event will occur at fixed $a_\bullet$ and $M_\bullet$ than specified by the above likelihood, as many scattered orbits will instead result in direct capture by the black hole, not a tidal disruption event. Black hole masses only slightly lower than the Hills mass will therefore be more disfavoured (compared to masses much less than the Hills mass) than the above likelihood suggests (see \cite{Coughlin22} and \cite{Singh23} for more details). A more realistic likelihood function would therefore look something like }
\begin{equation}\label{pTDEp}
    {\cal L}_{\rm TDE} = \Theta\left[\widetilde M_\bullet - M_\bullet\right] \, f\left(M_\bullet / \widetilde M_\bullet \right) ,
\end{equation}
{where $f(x)$ is some function which satisfies both $f(x) \leq 1$ for all $x \leq 1$, and the limit $\lim_{x\to 0} f(x) \to 1$. In principle $f$ should be a function of all relevant variables, in addition to the ratio $M_\bullet/\widetilde M_\bullet$. We will show in Appendix \ref{app:like} that various choices of the functional form of $f$ do not produce results which differ dramatically from the results computed using the simple Heaviside likelihood above, and in fact generally produce higher spin estimates {(the mathematical basis of this potentially surprising statement is outlined in Appendix \ref{app:like})}.  The spin constraints derived from the simplest likelihood defined above can therefore be considered conservative.    }

\begin{figure}
    \centering
    \includegraphics[width=0.49\textwidth]{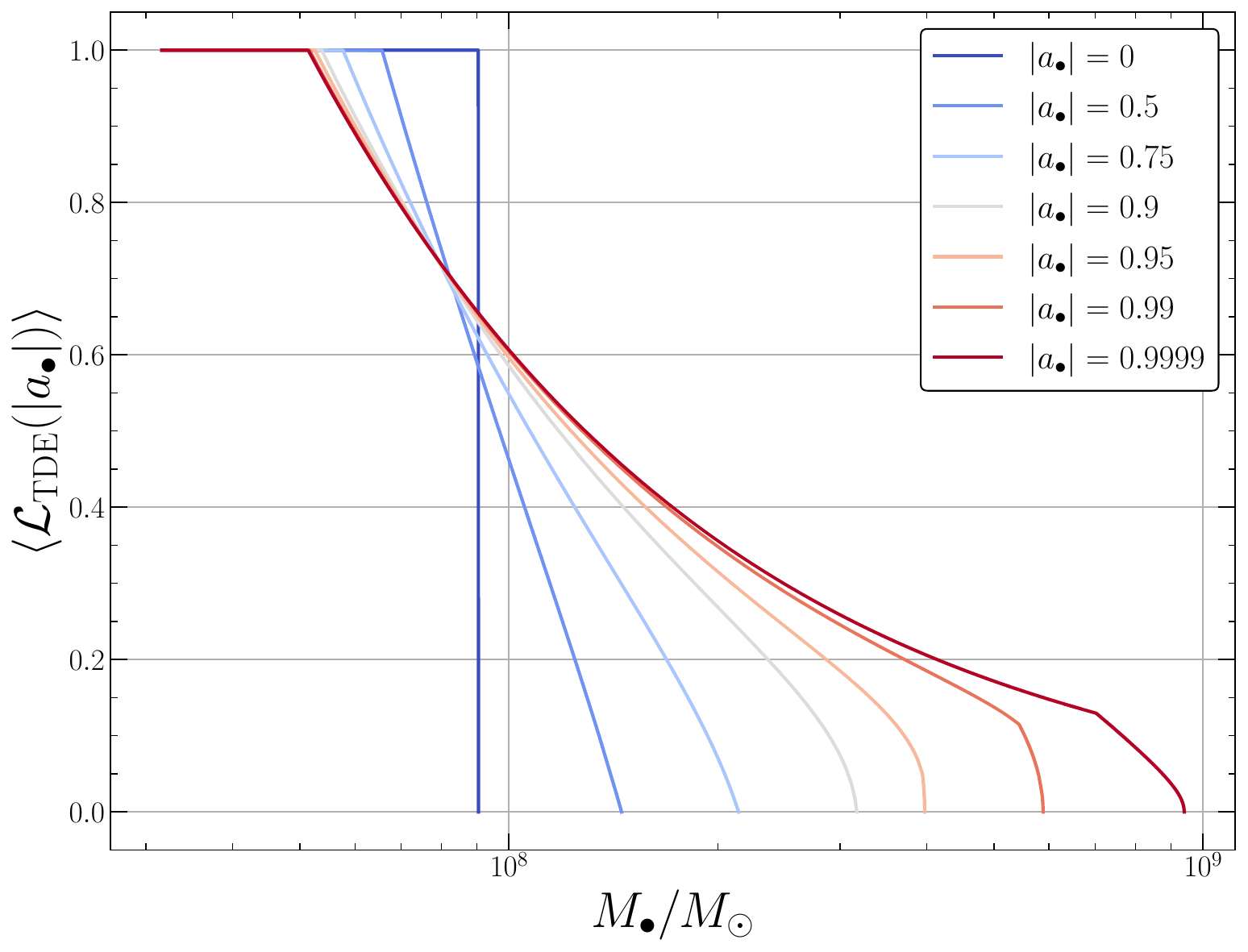}
    \caption{The likelihood of a tidal disruption event occurring, averaged over all inclinations {(prograde and retrograde)}, for a solar type star about black holes of different masses and {magnitudes of} spin. This figure was made with $M_\star = M_\odot$, $R_\star = R_\odot, \eta=1$. For other stellar parameters the shape of $\left\langle {\cal L}_{\rm TDE} \right \rangle$ is unchanged, but the values of the horizontal axis should be shifted by a factor $f=(R_\star/R_\odot)^{3/2} (M_\odot/M_\star)^{1/2} \eta^{-1/2}$ (equation \ref{HillsMass}). {This likelihood should be interpreted as an upper bound on the true likelihood.}  }
    \label{fig:p_tde}
\end{figure}

The first relevant quantity one can compute is the integrated tidal disruption event {likelihood} over all incoming inclinations. We shall compute this for a solar-type star, as the following symmetry property of the Heaviside function 
\begin{equation}
    \Theta(\alpha z - \beta) = \Theta\left(z - {\beta / \alpha}\right), \quad \alpha > 0, 
\end{equation}
means that the stellar parameter dependence can be simply scaled out. The angle integrated likelihood is therefore given by   
\begin{align}
    \left\langle {\cal L}_{\rm TDE} (a_\bullet)\right\rangle &\equiv \int_0^{\pi/2} p(\psi) \,{\cal L}_{\rm TDE}(a_\bullet, \psi, M_\odot, R_\odot, M_\bullet) \, {\rm d}\psi ,\nonumber \\
    &=  \int_0^{\pi/2}  \Theta\left(\widetilde M_\bullet(a_\bullet, \psi) - M_\bullet\right) \, \cos \psi \, {\rm d}\psi , \label{p_TDE}
\end{align}
where we have assumed that the {stellar population are isotropically distributed on the black hole's sphere}, and all non displayed arguments in $\widetilde M_\bullet$ are set equal to solar values. {Of course, what we are really interested in is the likelihood that a black hole with a given  {\it magnitude} of spin can have a tidal disruption event, which involves including both positive and negative spin parameters using the trivial relation  }
\begin{equation}
    \left\langle {\cal L}_{\rm TDE}(|a_\bullet|) \right\rangle = {1\over 2} \Big[\left\langle {\cal L}_{\rm TDE}(a_\bullet) \right\rangle  + \left\langle {\cal L}_{\rm TDE} (-a_\bullet) \right\rangle\Big].
\end{equation}
{This is consistent with the sign convention used in this manuscript (introduced after equation \ref{chi_eq_1}), namely that spins can take both signs $-1 \leq a_\bullet \leq 1$, and angles are restricted to $0 \leq \psi \leq \pi/2$. Every equation in the remainder of this manuscript should be interpreted in terms of black hole spins $a_\bullet$ which can have either sign. When the absolute magnitude of the spin is being referred to this will be denoted $|a_\bullet|$.    }

{The $\cos\psi$ in equation (\ref{p_TDE}) comes from noting that in the Newtonian limit (which is of course the relevant limit at the point at which tidally disrupted stars are scattered onto their disrupt-able orbits) $\sin \psi = (\vec r \times \vec v) . \hat z / |\vec r \times \vec v|$ (Fig \ref{fig:geometric}), where $\vec r$ and $\vec v$ are the velocity and radial vectors of the star. By definition, in an isotropic system there can be no preference for the direction between $(\vec r \times \vec v)$ and $\hat z$, and so $\sin \psi$ must be uniformly distributed and therefore $p(\psi) \propto \cos \psi$. It is important to note that the details of loss cone filling may modify the distribution of inclinations of the {\it particular} stars scattered onto near radial orbits from isotropic \citep[see][for details]{Singh23}.  }

We plot $\left\langle {\cal L}_{\rm TDE} (|a_\bullet|) \right\rangle$ as a function of $M_\bullet$ for different {magnitudes of the black hole} spin $|a_\bullet|$ in Fig. \ref{fig:p_tde}. Note that in this limit the Schwarzschild result is simply a Heaviside theta function at the value of equation \ref{schwarz_hills}, as the Schwarzschild Hills mass is inclination independent. All Kerr black holes with non-zero spin undergo some region of smoothly evolving $\left\langle {\cal L}_{\rm TDE} (|a_\bullet|)\right\rangle$, before dropping to zero for masses above the maximal value displayed in Fig. \ref{fig:new_max_mass}.  

An interesting result to note is that the gradient of $\left\langle {\cal L}_{\rm TDE} \right \rangle$ is discontinuous at some critical black hole mass for very large spins $|a_\bullet| \gtrsim 0.95$. The reason for this can be understood with reference to Fig. \ref{fig:mass_spin_psi}. While for low spins increasing the black hole mass only ever results in the loss of possible tidal disruption events at low inclinations (low $\psi$), for high spins where the Hills mass peaks outside of the equatorial plane (Fig. \ref{fig:mass_spin_psi}) above a certain mass value tidal disruption events become impossible at both low {\it and} high inclinations.  This new region where stellar orbits are lost leads to the discontinuous gradient in $\left\langle {\cal L}_{\rm TDE} \right \rangle$.

It is possible to go further than this result however, provided that one believes one knows the intrinsic probability distribution of the stellar parameters of the galactic centre, and the relative rates at which different stars are deferentially scattered onto near-radial orbits about the central black hole.  

Let us assume for simplicity that the masses and radii of stars are coupled, and so one only needs to know the stellar mass to fully specify the star's properties.  This is a good approximation for most main sequence stars, where we have the following mass-radius relationship \citep{Kippenhahn90}
\begin{equation}
    R_\star = 
    \begin{cases}
    &R_\odot \left(M_\star / M_\odot \right)^{0.56}, \quad M_\star \leq M_\odot ,\\
    \\
    &R_\odot \left(M_\star / M_\odot \right)^{0.79}, \quad M_\star > M_\odot .\\
    \end{cases}
\end{equation}
Assuming we have measured, via some independent means, a black hole mass $M_\bullet$ in an event we believe has been caused by the tidal disruption of a star, the posterior probability distribution of that black hole's spin can be constrained via Bayesian inference. 
If we have prior assumptions about the distribution of $a_\bullet, \psi$ and $M_\star$ (which we shall denote $\hat p(x)$ for variable $x$), then the posterior probability is found by marginalising over the other system parameters (we henceforth drop the subscript ``TDE'' on the likelihood function) 
\begin{align}
    &p(a_\bullet| M_\bullet) \propto \hat p(a_\bullet)  \iint \hat p(\psi) \hat p(M_\star) \,{\cal L}(a_\bullet, \psi, M_\star, M_\bullet) \, {\rm d}\psi \, {\rm d}M_\star  , \nonumber \\
    &\propto \hat p(a_\bullet) \left[\int\limits_{M_{\star, {\rm min}}}^{M_{\star, {\rm max}}}  \hat p(M_\star) \Bigg( \int\limits_0^{\pi/2}  \,{\cal L}(a_\bullet, \psi, M_\star, M_\bullet) \cos \psi \, {\rm d}\psi \Bigg) {\rm d}M_\star \right] ,
\end{align}
where in going to the second line we have again assumed an isotropic stellar distribution. In this expression ${M_{\star, {\rm min}}}/{M_{\star, {\rm max}}}$ are the minimum/maximum stellar masses in the galactic centre population. We use the notation $p(A|B)$ to represent the probability of $A$ given $B$. This is a direct application of Bayes theorem. {As we discussed earlier, the absolute magnitude of the black hole spin is the primary parameter of interest, and can be determined from the trivial relationship}
\begin{equation}\label{conversion}
    p(|a_\bullet|) = p(a_\bullet) + p(-a_\bullet), \quad {\textrm{where}} \quad 0 \leq a_\bullet \leq 1. 
\end{equation}
{which holds for every posterior distribution $p(a_\bullet)$ derived in the remainder of this paper. }
\begin{figure}
    \centering
    \includegraphics[width=0.49\textwidth]{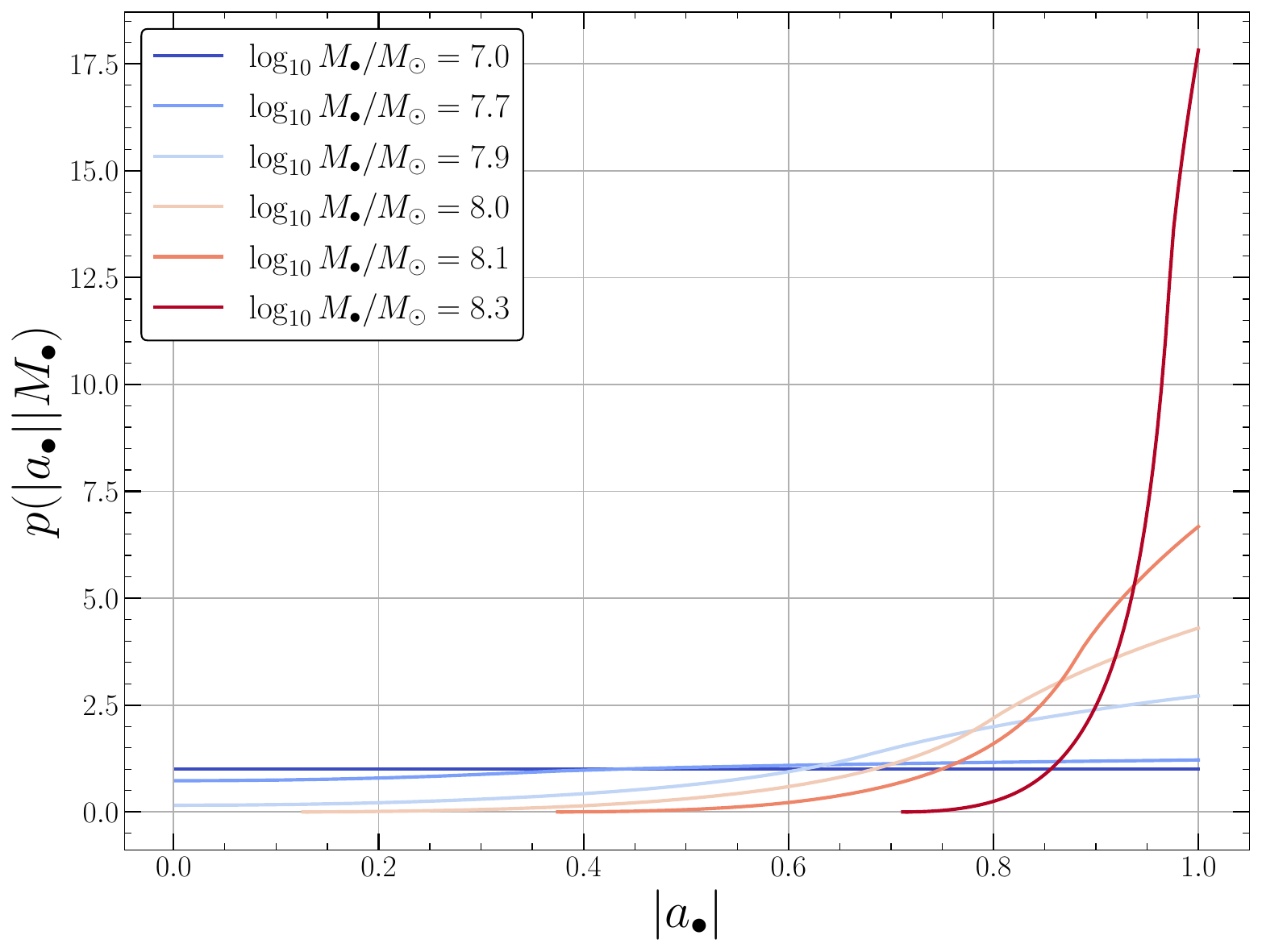}
    \caption{The inferred spin distributions for tidal disruption events about black holes with masses denoted in the figure legend {(absolute magnitude $|a_\bullet|$)}. We assume a flat prior on the black hole spin $\hat p(a_\bullet) \propto 1$. For high black hole masses a lower bound can be placed on the black hole spin at the centre of the event.  }
    \label{fig:spin_dists}
\end{figure}

The prior probability function $\hat p(M_\star)$ can be estimated in the following manner, and depends on two factors. Stars of different  masses are both formed at intrinsically different rates, but are also tidally disrupted at intrinsically different rates, due to their differing structures.  The probability we are looking for is therefore given by the product of the stellar mass function and the intrinsic dependence of tidal disruption event rate on stellar parameters.  

We use the Kroupa initial mass function \citep{Kroupa01}  to determine the intrinsic rate at which stars of different masses are formed. We assume that this equals the probability of a given star existing in the galactic centre.   The Kroupa IMF takes the form of a multiply broken power-law, with each power-law section taking the form
\beq
p_{\rm IMF}(M_\star) \propto M_\star^{k_i}.
\eeq
The values of $k_i$ are the following: $k_1 = -1.8$ for $M_\star < 0.5 M_\odot$; $k_2 = -2.7$ for $0.5 M_\odot < M_\star < M_\odot$; and $k_3 = -2.3$ for $M_\star > M_\odot$. The intrinsic rate at which tidal disruption events occur for different stellar parameters, for a given black hole mass and spin,  is a quantity which may be calculated theoretically. We use the rate calculation of \citet[][see also \citet{Magorrian99, Rees88}]{Wang04}  whereby the intrinsic rate of tidal disruptions scales as 
\beq
p_{\rm rate}(M_\star, R_\star) \propto M_\star^{-1/3} R_\star^{1/4} .
\eeq
This result encapsulates the intuitive result that more massive stars are harder to disrupt, but stars with larger radii are easier to disrupt. For our calculation we therefore define the stellar mass prior distribution as 
\begin{equation}
    \hat p(M_\star) \propto p_{\rm IMF}(M_\star) \times p_{\rm rate}(M_\star) . 
\end{equation}
We also take $M_{\star, {\rm min}} = 0.1 M_\odot$, $M_{\star, {\rm max}} = M_\odot$, relevant for older stellar populations \citep[][]{Stone16, Magorrian99}. In Fig. \ref{fig:spin_dists} we plot the inferred posterior spin {magnitude} distributions for a number of different black hole masses. We assume a flat prior on the black hole spin $\hat p(a_\bullet) \propto 1$. For high black hole masses a conservative lower bound can be placed on the black hole spin at the centre of the event. This lower bound on the spin is a strong function of black hole mass for masses above $\log_{10} M_\bullet/M_\odot > 8$.

More realistically, one expects to have some uncertainty on the black hole mass measurement. Assuming we have some mass measurement $\mu_{M_\bullet}$ with uncertainty $\sigma_{M_\bullet}$, then our spin posterior becomes 
\begin{align}
    &p(a_\bullet| \mu_{M_\bullet}, \sigma_{M_\bullet}) \propto \hat p(a_\bullet)  \left\{ \int_0^\infty \hat p(M_\bullet) \, \left[ \int\limits_{M_{\star, {\rm min}}}^{M_{\star, {\rm max}}} \hat p(M_\star) \nonumber \right.\right. \\ & \left.\left. \Bigg( \int\limits_0^{\pi/2}  \,{\cal L}(a_\bullet, \psi, M_\star, M_\bullet) \cos \psi \, {\rm d}\psi \Bigg) {\rm d}M_\star \right] {\rm d} M_\bullet \right\} ,
\end{align}
where $\hat p(M_\bullet)$ is the assumed prior probability distribution for the black hole mass. 

Typically black hole mass measurements are assumed to be described by a log-normal distribution with some dimensionless scatter $\sigma_{M_\bullet}$ expressed in dex, which corresponds to the uncertainty in the logarithm of the black hole mass. {We must also, much like for the stellar properties, take into account the differing rates at which tidal disruption events occur for different black hole masses (which we shall denote ${\cal R}(M_\bullet)$), the differing numbers of black holes in different mass bins in the local universe (which we shall denote ${\rm d}N_\bullet/{\rm d}M_\bullet$), and the differing observing volumes available to tidal disruption events around black holes of differing masses (which we shall denote ${\cal V}(M_\bullet)$). If we have an estimate from (for example) a galactic scaling relationship, with some associated probability density (denoted $\widetilde p(\mu_{M_\bullet}, \sigma_{M_\bullet})$), then our black hole mass prior is  }

\begin{figure}
    \centering
    \includegraphics[width=0.49\textwidth]{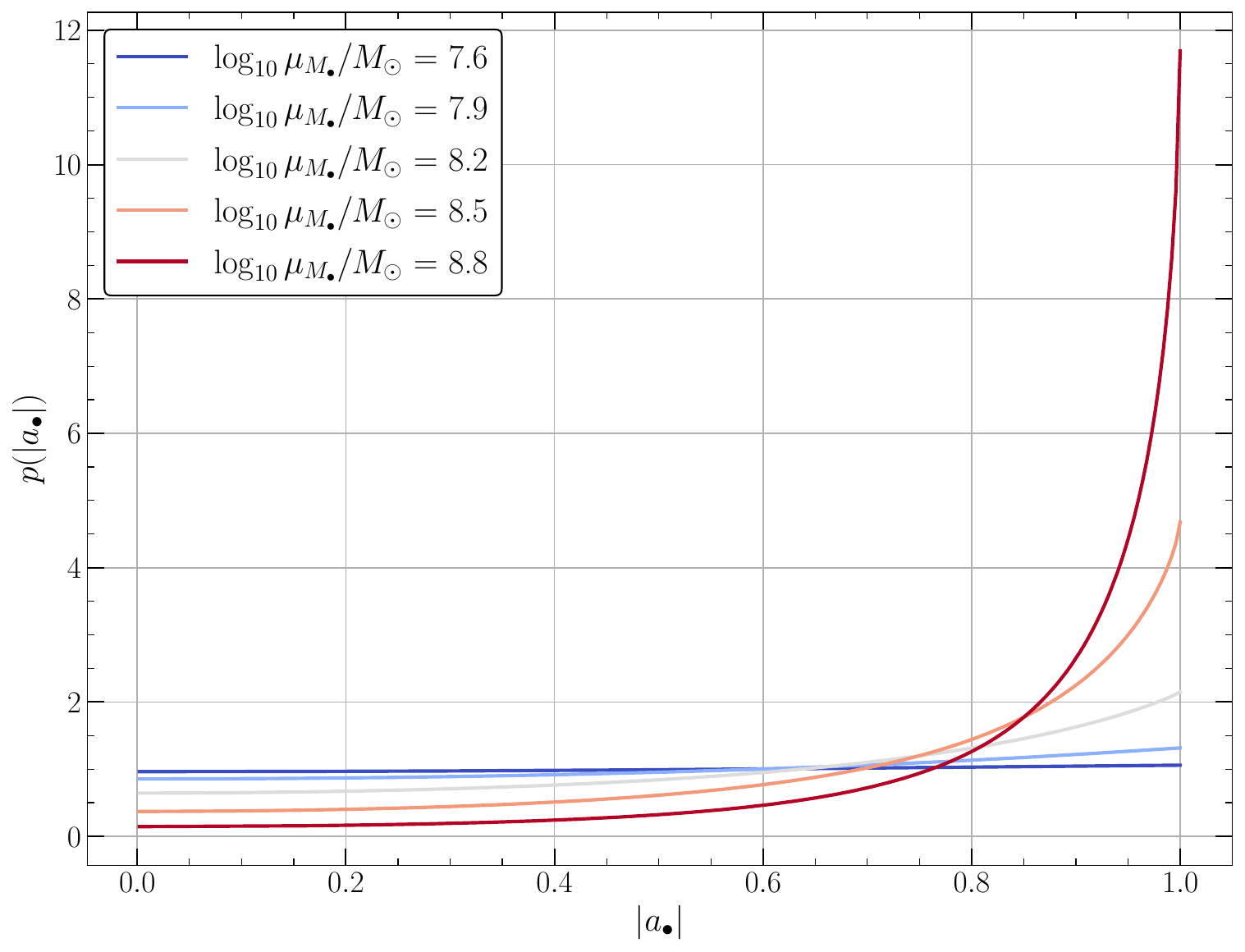}
    \includegraphics[width=0.49\textwidth]{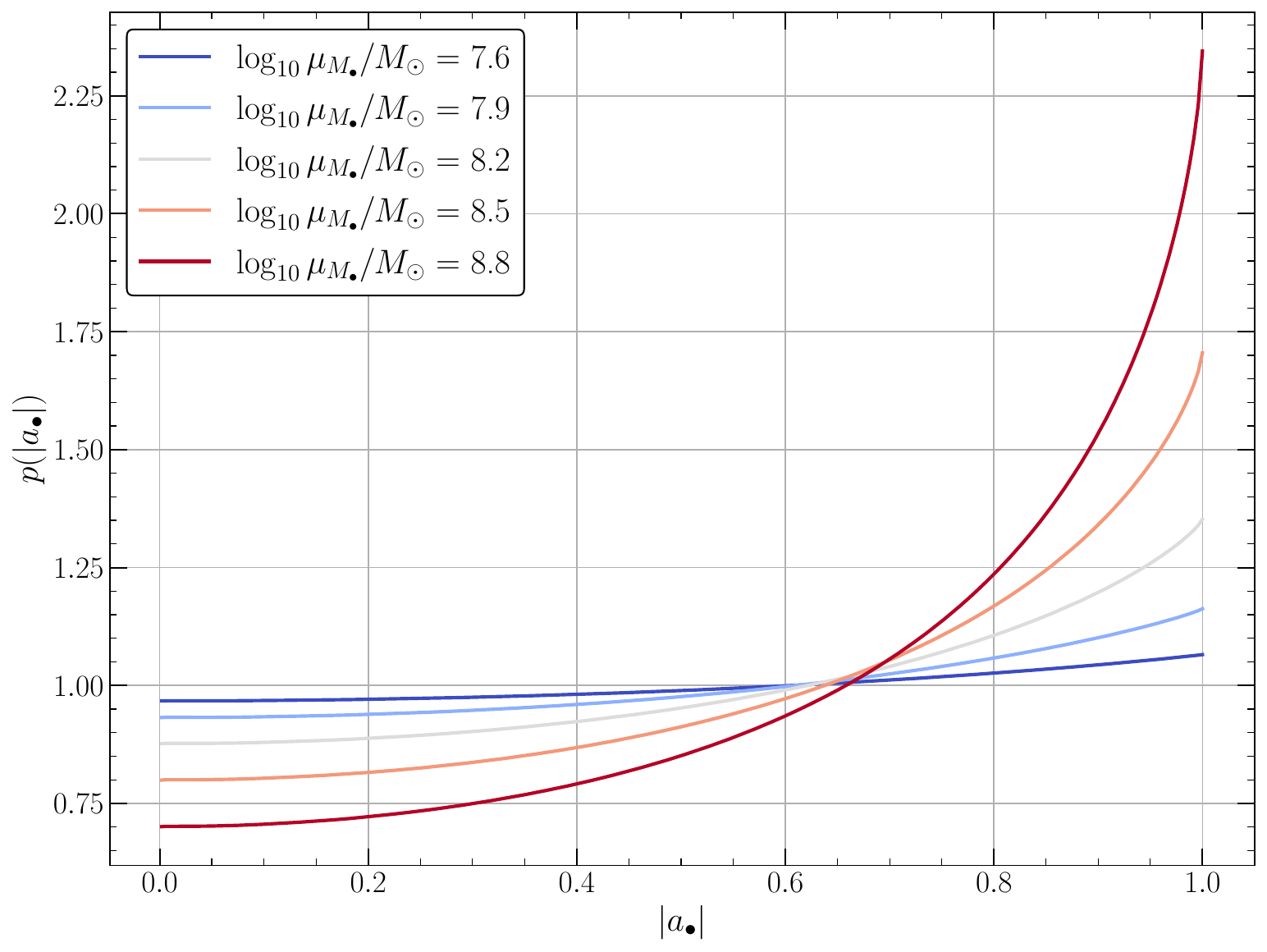}
    \includegraphics[width=0.49\textwidth]{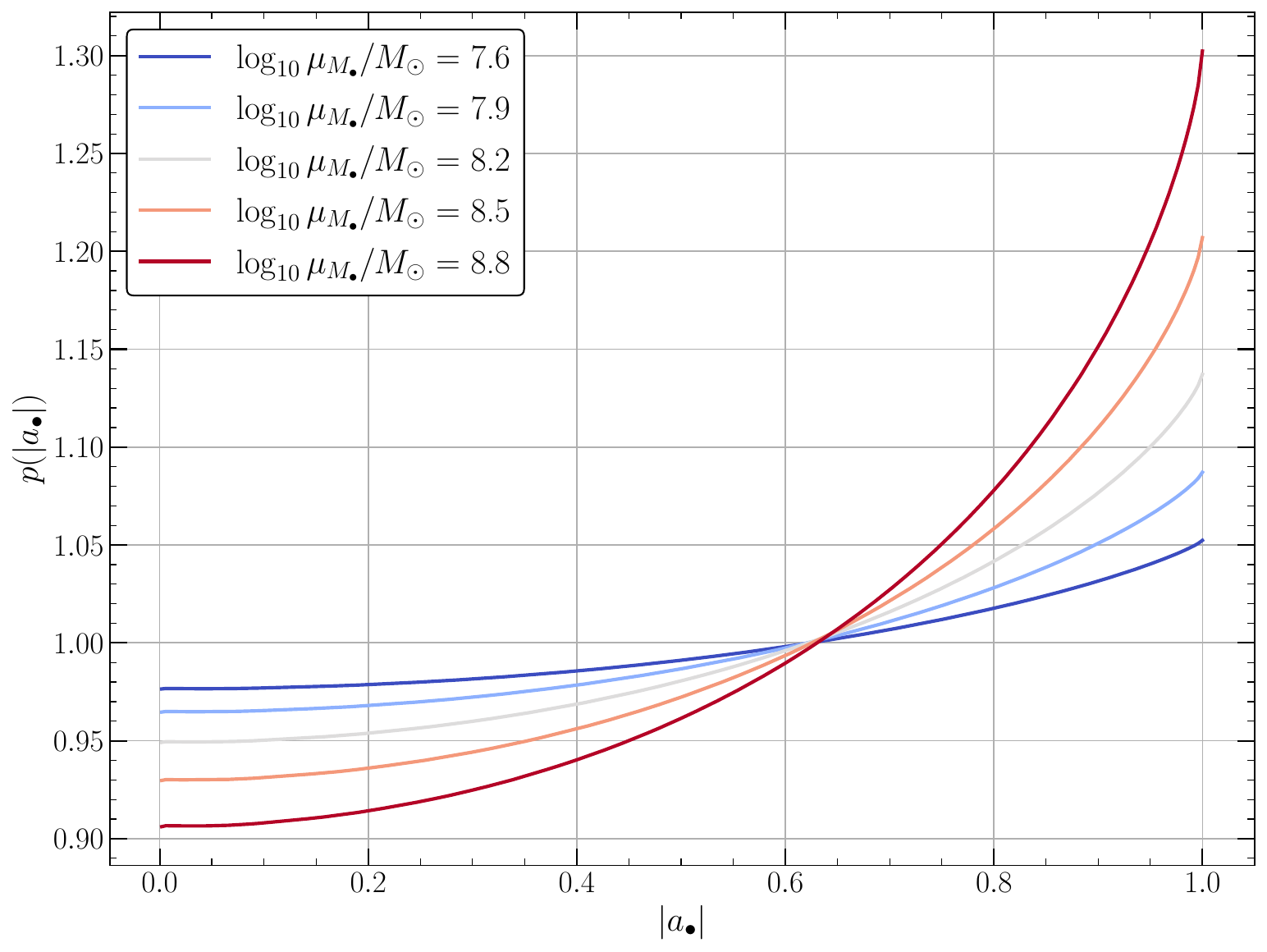}
    \caption{The inferred spin posterior distributions for tidal disruption events about black holes with median masses denoted in the figure legend, and three different values of the black hole mass uncertainty. We assume a flat prior on the black hole spin $\hat p(a_\bullet) \propto 1$. In the upper panel we use $\sigma_{M_\bullet} = 0.3$, in the middle panel $\sigma_{M_\bullet} = 0.5$ and in the lower panel $\sigma_{M_\bullet} = 0.8$. The certainty with which a black hole can be said to be rapidly rotating is a relatively strong function of the uncertainty on the black hole mass measurement.  In this plot we display the posterior distribution of the absolute value of the black hole spin parameter $p(|a_\bullet|) = p(a_\bullet) + p(-a_\bullet)$.  }
    \label{fig:spin_dists_var_mass}
\end{figure}

\begin{figure}
    \centering    
    \includegraphics[width=0.49\textwidth]{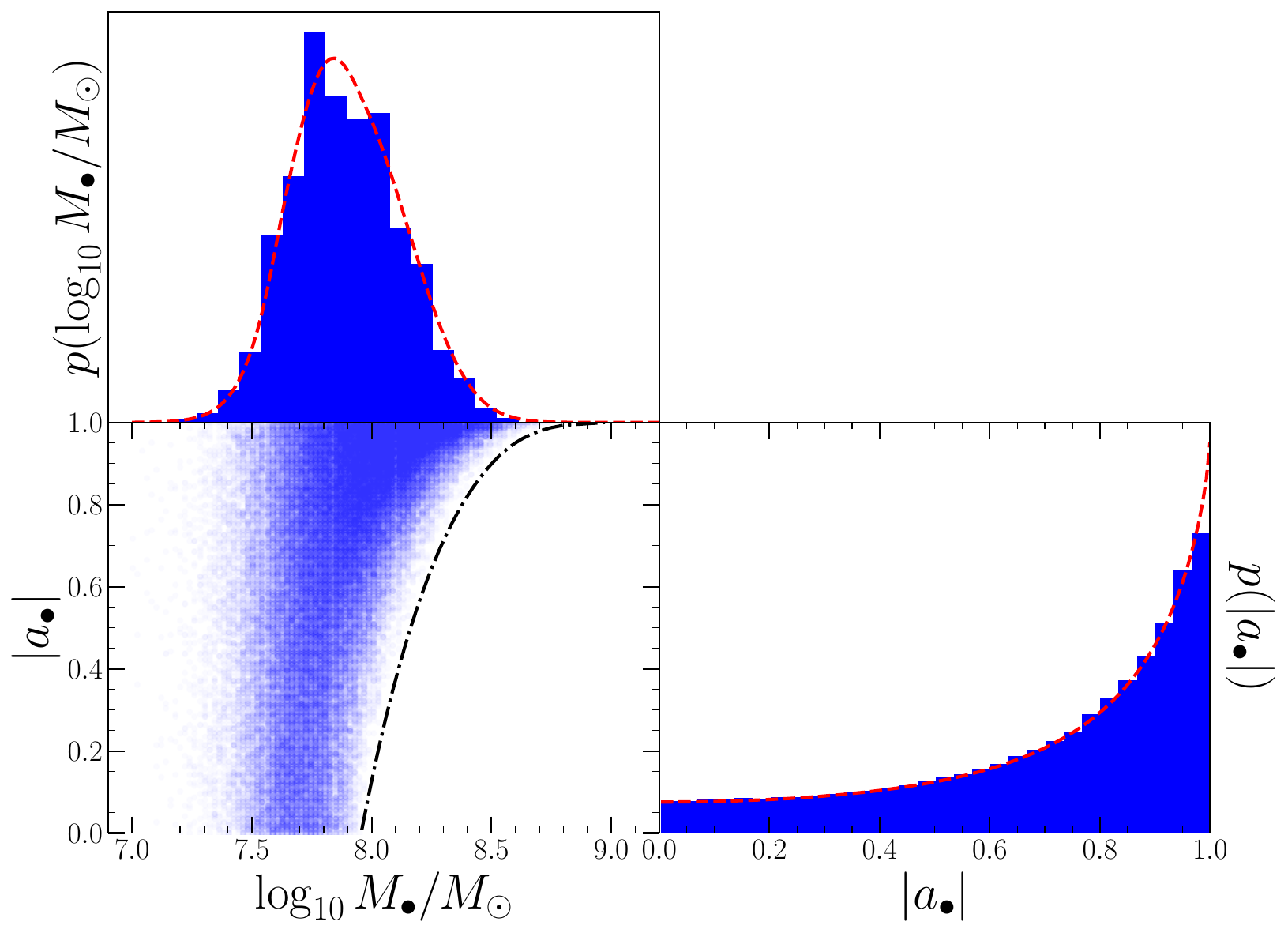}
    \caption{Two dimensional black hole mass-spin posterior distributions, for a tidal disruption event about a black hole with prior mass parameters $\log \mu_{M_\bullet}/M_\odot = 8.5$, $\sigma_{M_\bullet} = 0.3$ dex. Note how the one dimensional posterior black hole mass distribution is shifted to lower values than the prior, owing the increased probability of tidal disruption at these masses, and is no longer symmetrical (in log space) owing to the black hole mass function. The red dashed curves show the one dimensional Bayesian posteriors computed from the integrals described in the text.  The two dimensional mass-spin distribution shows a strong degeneracy with higher masses associated with rapidly rotating black holes.  The black dot-dashed curve shows the maximum Hills mass for a solar type star as a function of black hole spin.  }
    \label{fig:2d_mass_spin}
\end{figure}

\begin{equation}
    \hat p(M_\bullet) \propto {{\rm d}N_\bullet \over {\rm d}M_\bullet} \times {\cal R}(M_\bullet) \times {\cal V}(M_\bullet) \times \widetilde p(\mu_{M_\bullet}, \sigma_{M_\bullet}) .
\end{equation}
{We now construct the explicit form of this prior. The log-normal distribution has the following functional form}
\begin{equation}
    \widetilde p(\mu_{M_\bullet}, \sigma_{M_\bullet}) \propto {1 \over M_\bullet} \exp\left(-\left({\log(M_\bullet) - \log(\mu_{M_\bullet}) \over \sqrt{2}\, \sigma_{M_\bullet}}\right)^2 \right) .
\end{equation}
{The differential rate at which tidal disruption events occur as a function of black hole mass can be estimated from galactic properties \citep{Stone16}, or from first principles \citep[e.g.][]{Magorrian99, Wang04}. In this work we take the value found empirically in \cite{Stone16} \citep[which is similar to analytic estimates][]{Wang04}, namely }
\begin{equation}
    {\cal R}(M_\bullet) \propto M_\bullet^{-2/5} . 
\end{equation}
{In a recent paper \cite{Mummery_et_al_23} showed that the black hole mass in the centre of a tidal disruption event (measured either using theoretical disc models or galactic scaling relationships) scales positively with the peak g-band luminosity of the tidal disruption event. Inverting the \cite{Mummery_et_al_23} scaling relationship we find }
\begin{equation}\label{scaling}
    \log_{10} \left({L_{g} \over 10^{43} \, {\rm erg\, s}^{-1}} \right) = -6.65 + 1.02 \, \log_{10} \left({M_\bullet \over M_\odot} \right) .  
\end{equation}
{This positive scaling means that more massive black holes will be over-represented in a sample of sources, as they can be detected out to a larger volume. In a simplified Newtonian treatment,  a flux limited sample (which cannot detect sources below $F_{\rm min}$) can observe a source with luminosity $L$ out to a total volume proportional to }
\begin{equation}
    {\cal V} \propto D_{\rm max}^3 \propto \left(\sqrt{L\over 4\pi F_{\rm min}}\right)^{3} \propto M_\bullet^{1.53} ,
\end{equation}
{where in the final proportionality we have substituted the \cite{Mummery_et_al_23} scaling (eq. \ref{scaling}).   }

{Our final step is to specify a model for the black hole mass function ${\rm d}N_\bullet/{\rm d}M_\bullet$. We use the \cite{Shankar04} parameterisation, which takes the form }
\begin{equation}
    {{\rm d}N_\bullet \over {\rm d}M_\bullet} \propto M_\bullet^{-1.1} \, \exp\left( - \left({M_\bullet \over 6.4 \times 10^7  M_\odot} \right)^{0.49} \right) , 
\end{equation}
{which is consistent with other estimates \citep[see e.g.][]{Stone16}. Bringing these various factors together we have a tidal disruption event black hole mass prior of }
\begin{multline}
    \hat p(M_\bullet) \, {\rm d} \log M_\bullet \propto M_\bullet^{0.03} \, \exp\left( - \left({M_\bullet \over 6.4 \times 10^7  M_\odot} \right)^{0.49}\right) \\ \exp\left( - \left({\log(M_\bullet) - \log(\mu_{M_\bullet}) \over \sqrt{2}\, \sigma_{M_\bullet}}\right)^2 \right) {\rm d} \log M_\bullet ,
\end{multline}
{The near perfect cancellation of the power law exponents in this expression is a coincidence, and in general the leading power law index has the value $\alpha + 3\beta/2  + \gamma$, where $\alpha$ and $\gamma$ are the exponents from the tidal disruption event rate and black hole mass functions respectively, and $\beta$ is the exponent from the $L_{\rm peak}(M_\bullet)$ scaling law.  }


Some example spin posteriors {(absolute magnitude)} with differing mass uncertainties $\sigma_{M_\bullet}$ are displayed in Fig. \ref{fig:spin_dists_var_mass}. Again, we assume a flat prior on the black hole spin $\hat p(a_\bullet) \propto 1$. We note that the certainty with which a black hole can be said to be rapidly rotating is a relatively strong function of the uncertainty on the black hole mass measurement.

Of course, it is not just the black hole spin for which we can construct a posterior distribution which (potentially) differs markedly from the prior. Indeed any posterior distribution can be obtained from marginalising over the other system parameters.  To understand degeneracy's between different system parameters it is simplest to Monte Carlo sample the underlying prior distributions of each parameter, and record only those systems with masses below the maximum Hills mass.   We show an example of such a sampling procedure in Fig. \ref{fig:2d_mass_spin}, where we plot the two dimensional black hole mass-spin posterior distributions, for a tidal disruption event about a black hole with prior mass parameters $\log \mu_{M_\bullet}/M_\odot = 8.5$, $\sigma_{M_\bullet} = 0.3$ dex. Note how the one dimensional posterior black hole mass distribution is shifted to lower values than the prior, owing the increased probability of tidal disruption at these masses. The two dimensional mass-spin distribution shows a strong degeneracy with higher masses associated with rapidly rotating black holes.

\subsection{Black hole spin constraints of real tidal disruption events}
This Bayesian approach can be applied to the population of tidal disruption events already discovered. We use the most recent compilation of optically discovered tidal disruption events published in \cite{Mummery_et_al_23}. Of those tidal disruption events, nine have masses inferred from a galactic scaling relationship to be $\log_{10}M_\bullet/M_\odot > 7.8$, and for which the methods developed here produce non-trivial results.

We compute the black hole masses from either the galactic mass scaling relationship \citep{Greene20} 
\begin{equation}\label{galmass_scale}
\log_{10} \left[M_\bullet/M_\odot \right]= 7.43 + 1.61 \log_{10} \left[ M_{\rm gal}\big/(3\times 10^{10} M_\odot) \right],
\end{equation}
or the $M_\bullet-\sigma$ relationship \citep{Greene20}
\begin{equation}\label{sig_scale}
\log_{10} \left[M_\bullet/M_\odot \right] = 7.87 + 4.38 \log_{10} \left[\sigma\big/(160\, {\rm km\,s^{-1}}) \right].
\end{equation}
The intrinsic scatter in the $M_\bullet - M_{\rm gal}$ relationship is 0.8 dex, while the intrinsic scatter in the $M_\bullet -\sigma$ relationship is 0.3 dex. We list the tidal disruption event properties in Table \ref{mass_table} (See Tables \ref{tab:sigmarefs}, \ref{tab:tderefs} for literature references).

    \renewcommand{\arraystretch}{1.25}
    \begin{table}
    \centering
    \begin{tabular}{|p{2.0cm} p{.7cm} p{1.cm} p{1.cm} p{1.cm}|  }
    \hline
    Event name  & $ \sigma $ & $M_\bullet$ & $M_{\rm gal}$ & $M_\bullet$ \\
    
    & km/s & $  M_\odot$  & $  M_\odot$ & $  M_\odot$ \\
    \hline
 ASASSN-15lh &  $225$ & $8.52$  & $10.85$ & $8.02$ \\ 
 
 AT2018fyk &  $158$ & $7.85$  & $10.61$ & $7.65$ \\ 
 
 AT2020qhs &  $215$ & $8.44$  & $11.33$ & $8.81$ \\ 
 
 AT2020ysg & -- & --  & $10.90$ & $8.12$ \\
 
 AT2020riz &  -- & --  & $10.84$ & $8.02$ \\ 
 
 AT2020acka &  $174$ & $8.03$  & $11.00$ & $8.28$ \\ 
 
 AT2019cmw &  -- & --  & $11.09$ & $8.41$ \\ 
 
 AT2018iih &  $149$ & $7.73$  & $10.84$ & $8.02$ \\ 
 
 AT2021yzv & -- & --  & $10.86$ & $8.05$ \\ \hline
 
 \end{tabular}

    \caption{The black hole masses of the tidal disruption events used in this paper, computed from the galactic mass and velocity dispersion different scaling relationships. 
    The quoted error ranges correspond to $1\sigma$ uncertainties. All quantities other than the velocity dispersion are presented as logarithms $\log_{10}$. 
    The column directly to the right of each observed quantity corresponds to the black hole mass computed from that quantity. 
    We do not quote uncertainties on the black hole masses computed from the velocity dispersion and galactic mass scaling relationships. 
    The intrinsic scatter in each relationship is $\sim 0.3$ dex and $\sim 0.8$ dex respectively.   }

    \label{mass_table}
\end{table}

\begin{figure}
    \centering
    \includegraphics[width=0.49\textwidth]{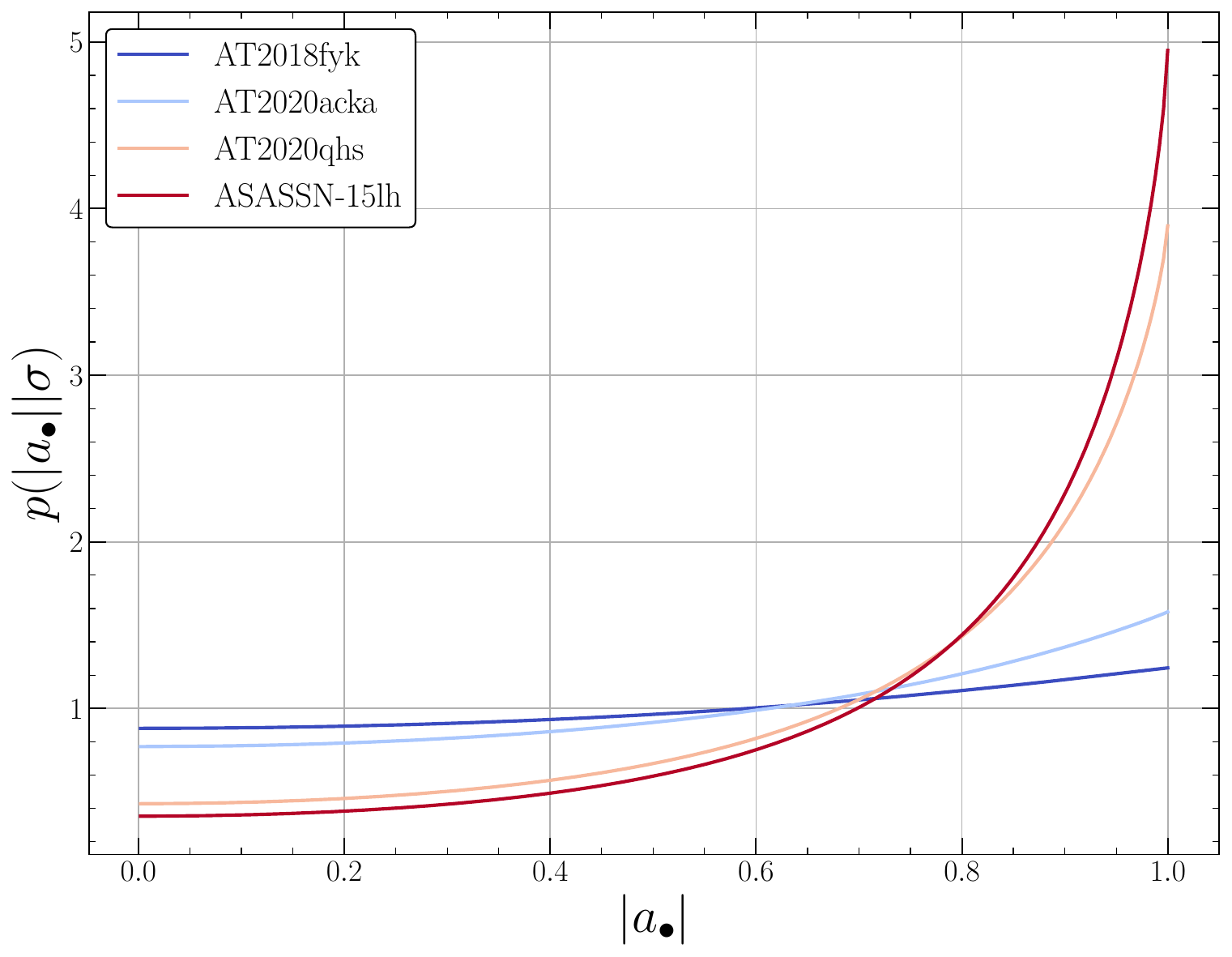}
    \caption{Black hole spin constraints {(absolute magnitude $|a_\bullet|$)} resulting from tidal disruption events with large inferred black hole masses, with masses inferred from the $M_\bullet -\sigma$ relationship. Strong spin constraints can be placed on tidal disruption events with large velocity dispersion $\sigma$ measurements.  }
    \label{fig:spin_m_sigma}
\end{figure}

\begin{figure}
    \centering
    \includegraphics[width=0.49\textwidth]{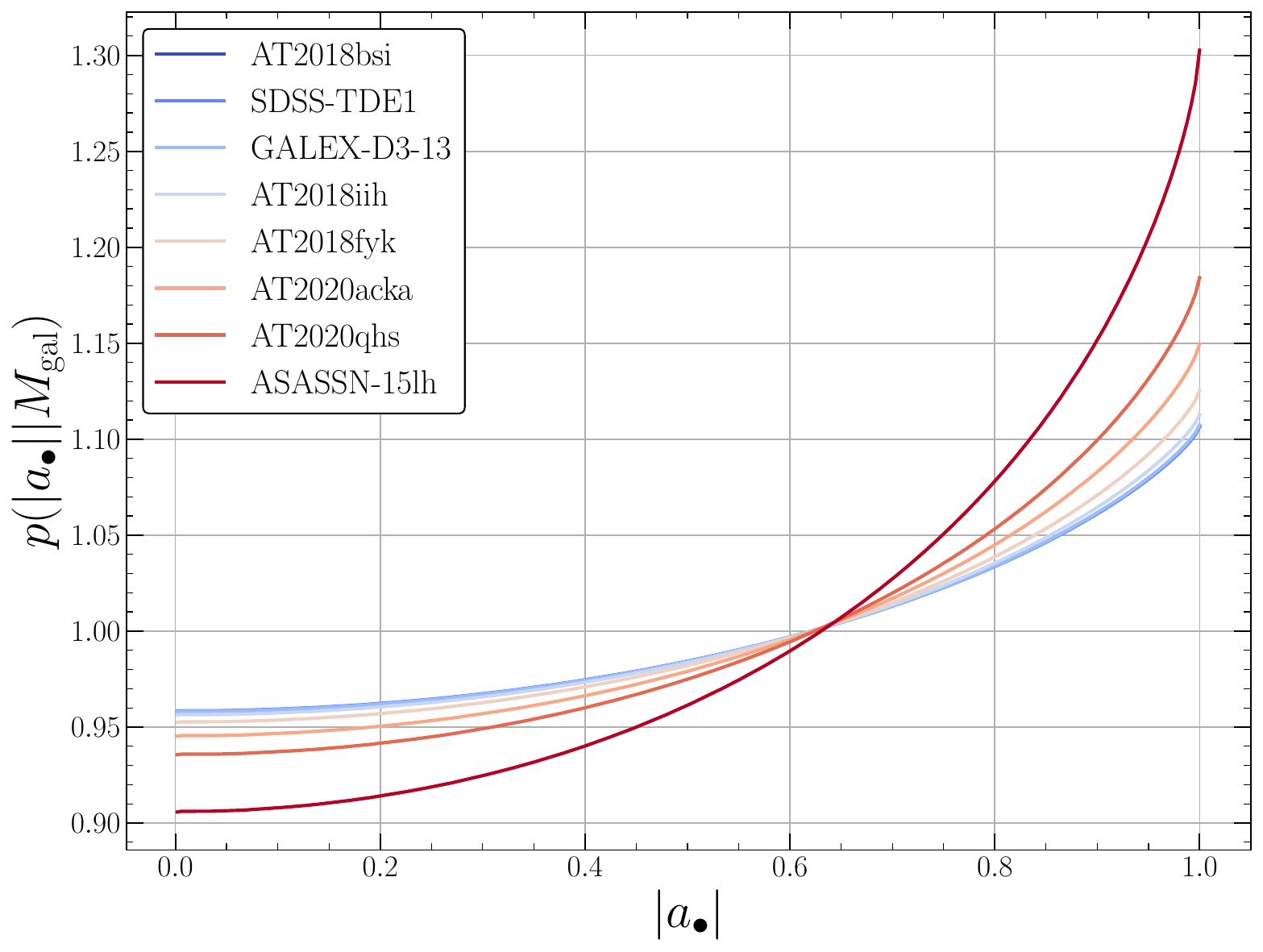}
    \caption{Black hole spin constraints {(absolute magnitude $|a_\bullet|$)} resulting from tidal disruption events with large inferred black hole masses, measured from the $M_\bullet -M_{\rm gal}$ relationship. The large intrinsic scatter in the galaxy mass-black hole mass scaling relationship reduces the ability of tidal acceleration based Bayesian inference to constrain  black hole spins in tidal disruption events. }
    \label{fig:spin_m_gal}
\end{figure}

In Figures \ref{fig:spin_m_sigma} and \ref{fig:spin_m_gal} we display the posterior spin constraints of each event in Table \ref{mass_table}, for those events with $\log_{10} M_\bullet/M_\odot > 7.8$. In Fig. \ref{fig:spin_m_sigma} we use the $M_\bullet-\sigma$ relationship, while in Fig. \ref{fig:spin_m_gal} we use the $M_\bullet-M_{\rm gal}$ relationship. An important result highlighted by these two figures is that the larger intrinsic scatter in the galaxy mass scaling relationship severely hampers the ability of Bayesian inference to constrain black hole spins in tidal disruption events. On the contrary, the relatively small intrinsic scatter in the $M_\bullet - \sigma$ relationship allows tighter constrains to be derived.  For the two most massive black holes (using the $M_\bullet-\sigma$) relationship we infer spins 
\begin{align}
    |\bar a_\bullet| &=  0.77^{+0.18}_{-0.43}, \quad {\rm AT2020qhs} \\
    |\bar a_\bullet| &=  0.81^{+0.15}_{-0.41}, \quad {\rm ASASSN \,\,15lh} ,
\end{align}
where $|\bar a_\bullet|$ is the median of the posterior spin-magnitude distribution, and the error range corresponds to the spin magnitude range associated with a 1$\sigma$ interval about the median. These spin constraints compliment the analysis in \cite{Mummery_et_al_23}, where spin constraints were found based on physical models of the tidal disruption event emission. 

\section{ Population constraints }\label{sec:population}
{In this paper so far we have considered the constraints that can be placed on the black hole at the centre of an individual tidal disruption event, however the formalism described in the previous section can be equally well applied to populations of tidal disruption events. For example, by writing the posterior black hole mass distribution in terms of a marginalisation over all other parameters }
\begin{align}
    & p(M_\bullet) \propto \hat p(M_\bullet)  \left\{ \int_{-1}^{+1} \hat p(a_\bullet) \, \left[ \int\limits_{M_{\star, {\rm min}}}^{M_{\star, {\rm max}}} \hat p(M_\star) \nonumber \right. \right.\\ & \left. \left. \Bigg( \int\limits_0^{\pi/2}  \,{\cal L}(a_\bullet, \psi, M_\star, M_\bullet) \cos \psi \, {\rm d}\psi \Bigg) {\rm d}M_\star \right] {\rm d} a_\bullet \right\} ,
\end{align}
{the effects of varying the assumed spin prior $\hat p(a_\bullet)$ on the observed black hole mass distribution of tidal disruption events can be determined.  For this analysis we take the prior on the tidal disruption event black hole mass function to be  }
\begin{align}
    \hat p(M_\bullet)   & \, {\rm d}M_\bullet \propto {{\rm d}N_\bullet \over {\rm d}M_\bullet} \times {\cal R}(M_\bullet) \times {\cal V}(M_\bullet) \,  {\rm d}M_\bullet, \\
    & \propto  M_\bullet^{0.03} \, \exp\left( - \left({M_\bullet \over 6.4 \times 10^7  M_\odot} \right)^{0.49}\right) \,  {\rm d}M_\bullet .
\end{align}
{With the prior defined in this manner the posterior is to be interpreted as the prediction for the {\it observed} distribution of black hole masses in a population of tidal disruption events (for a given spin prior $\hat p(a_\bullet)$).  }

\begin{figure}
    \centering
    \includegraphics[width=\linewidth]{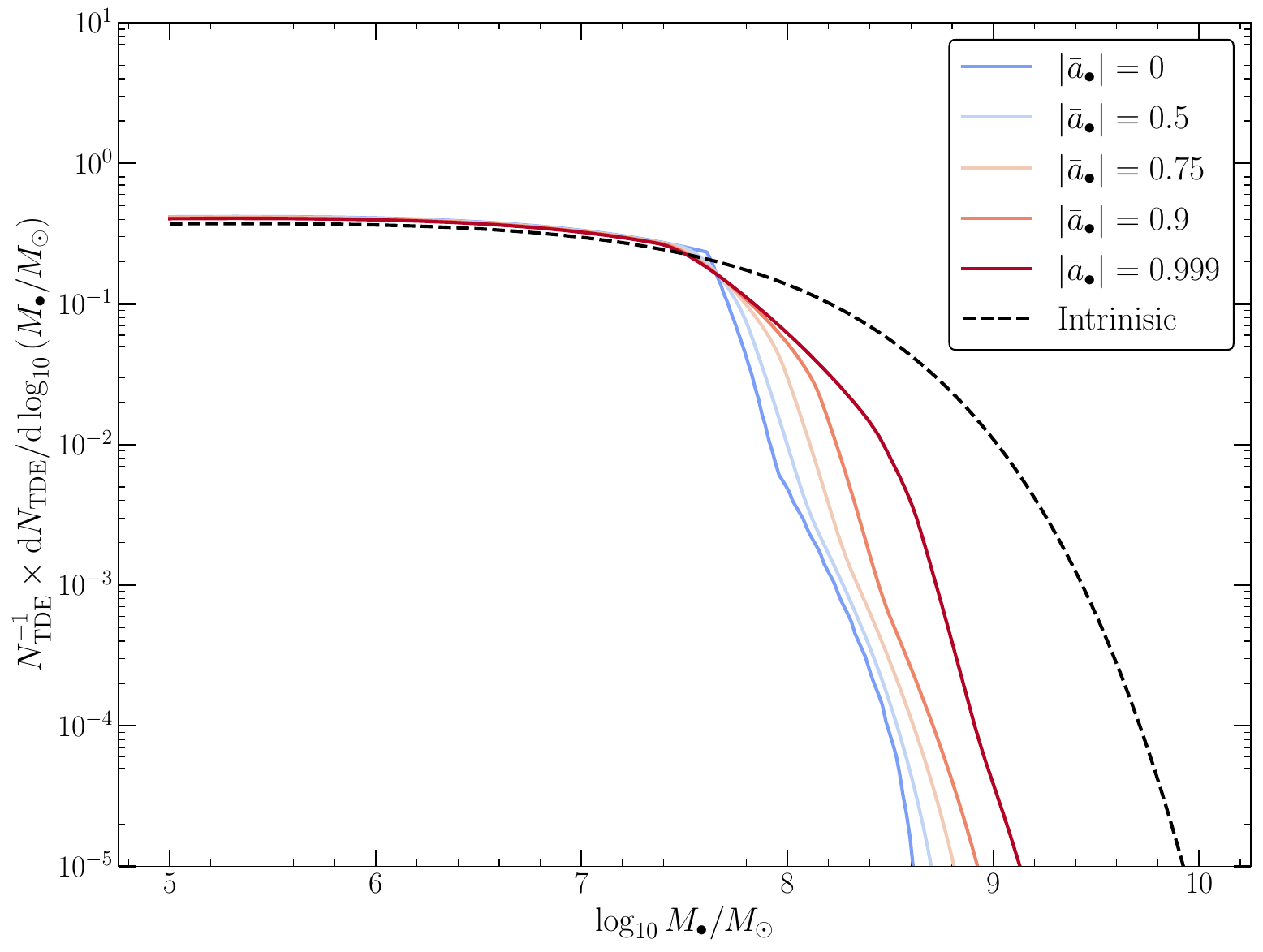}
    \includegraphics[width=\linewidth]{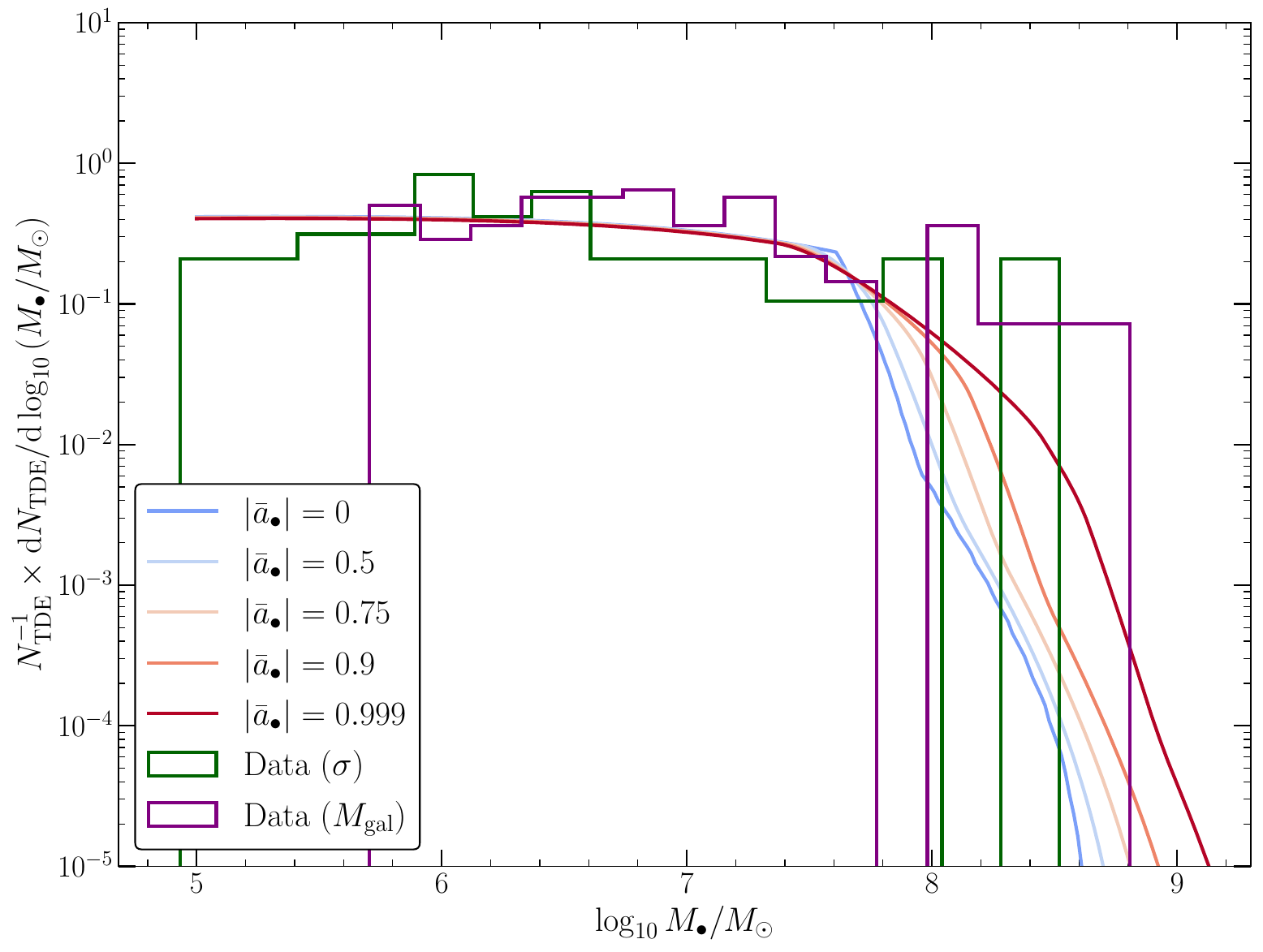}
    \caption{Upper: the effect of black hole spin on the expected observed tidal disruption event mass function. As a black dashed curve we display the prior model, while the coloured curves show the effects of tidal forces, with all black holes assumed to have the same spin value (displayed in legend; note that  both $a_\bullet = \pm \bar a_\bullet$ appear in our integrals, due to the isotropic stellar population).  Lower:  We display the observed mass distribution inferred from the $M_\bullet-\sigma$ relationship (green), and the $M_\bullet-M_{\rm gal}$ relationship (purple). The predicted flat distribution at lower black hole masses is recovered by the data, while no tidal disruption event has been observed with black hole mass (as inferred from galactic scaling relationships) greater than $10^9M_\odot$, as expected from the properties of relativistic tidal forces. }
    \label{fig:masspop}
\end{figure}

{In Figure \ref{fig:masspop} we plot the predicted theoretical black hole mass distributions (coloured curves) and the observed tidal disruption event distributions implied by galactic scaling relationships  \citep[data from][displayed as coloured histograms]{Mummery_et_al_23}.  We assume simplistic monochromatic spin distributions, with all black hole spins given by a single value $|a_\bullet| = \bar a_\bullet$ (note that this means both $a_\bullet = \pm \bar a_\bullet$ appear in our integrals, due to the isotropic stellar population). This is done purely for simplicity and to highlight the effects of black hole spin. We do not expected monochromatic spin distributions to be an accurate physical model. By a black dashed curve (upper panel) we display the distribution in the absence of tidal effects (i.e., this curve is the prior $\hat p(M_\bullet)$). As expected \citep[and discussed previously in e.g.,][]{Kesden12}, higher spin values extend the predicted mass function upwards towards the ``intrinsic'' distribution.  

In green (lower panel) we display the observed mass distribution inferred from the $M_\bullet-\sigma$ relationship (eq. \ref{sig_scale}), while in purple we  display the observed mass distribution inferred from the $M_\bullet-M_{\rm gal}$ relationship (eq. \ref{galmass_scale}). We see that the predicted flat distribution at lower black hole masses is recovered by the data, while no tidal disruption event has been observed with black hole mass (as inferred from galactic scaling relationships) greater than $10^9M_\odot$, as expected from the properties of relativistic tidal forces.  }

{Extending this analysis to determine predicted populations of different observables is possible if we have a model which links an observable ${\cal O}$ to the black hole mass $M_\bullet$ (for example the peak $g$-band luminosity described above).  This is done simply by marginalising over black hole mass (which we shall assume is the principle physical parameter which determines the observable, e.g. equation \ref{scaling}) to determine the observable posterior $p({\cal O})$ }
\begin{equation}
    p({\cal O}) = \int_0^\infty p\left({\cal O} | M_\bullet \right) \, p(M_\bullet) \, {\rm d}M_\bullet . 
\end{equation}
{In this expression $p(M_\bullet)$ is the posterior distribution constructed above. Specifying the marginal probability density $p({\cal O} | M_\bullet)$ requires some model (either theoretical or empirical) relating the observable to the black hole mass. Taking the peak $g$-band luminosity as an example, we consider models of the form   }
\begin{equation}
    \log_{10}\left({\cal O} \right) = \alpha + \beta \log_{10} (M_\bullet/M_\odot) ,
\end{equation}
{and shall assume some intrinsic scatter in this relationship denoted $\epsilon$. We can then approximate the probability density }
\begin{equation}
    p({\cal O} | M_\bullet) \propto \exp\left( - {\big[ \log {\cal O} - \alpha - \beta \log M_\bullet/M_\odot \big]^2 \over 2 \epsilon ^ 2 }\right) , 
\end{equation}
{which allows us to construct predicted population posteriors. }

\begin{figure}
    \centering
    \includegraphics[width=\linewidth]{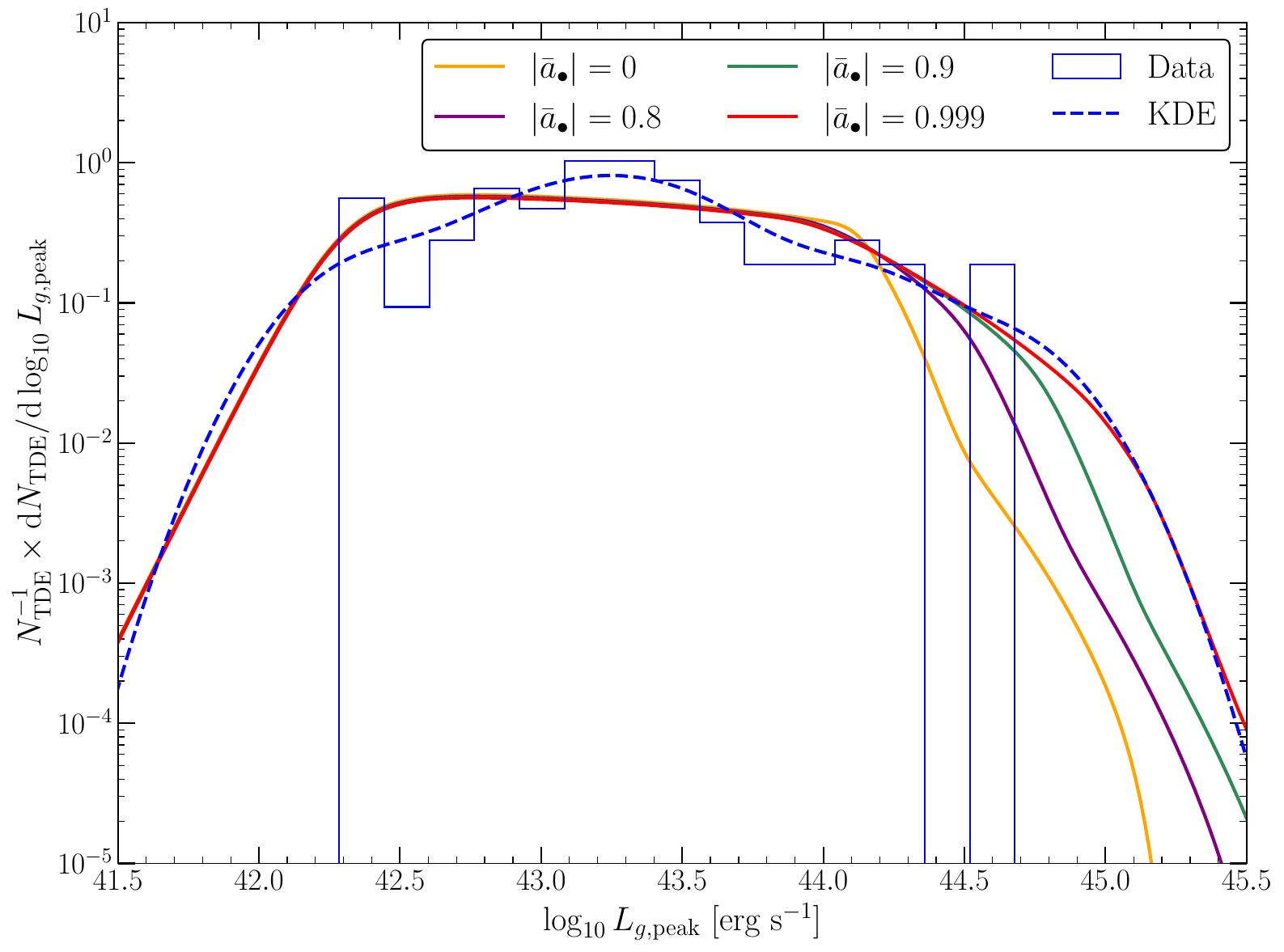}
    \includegraphics[width=\linewidth]{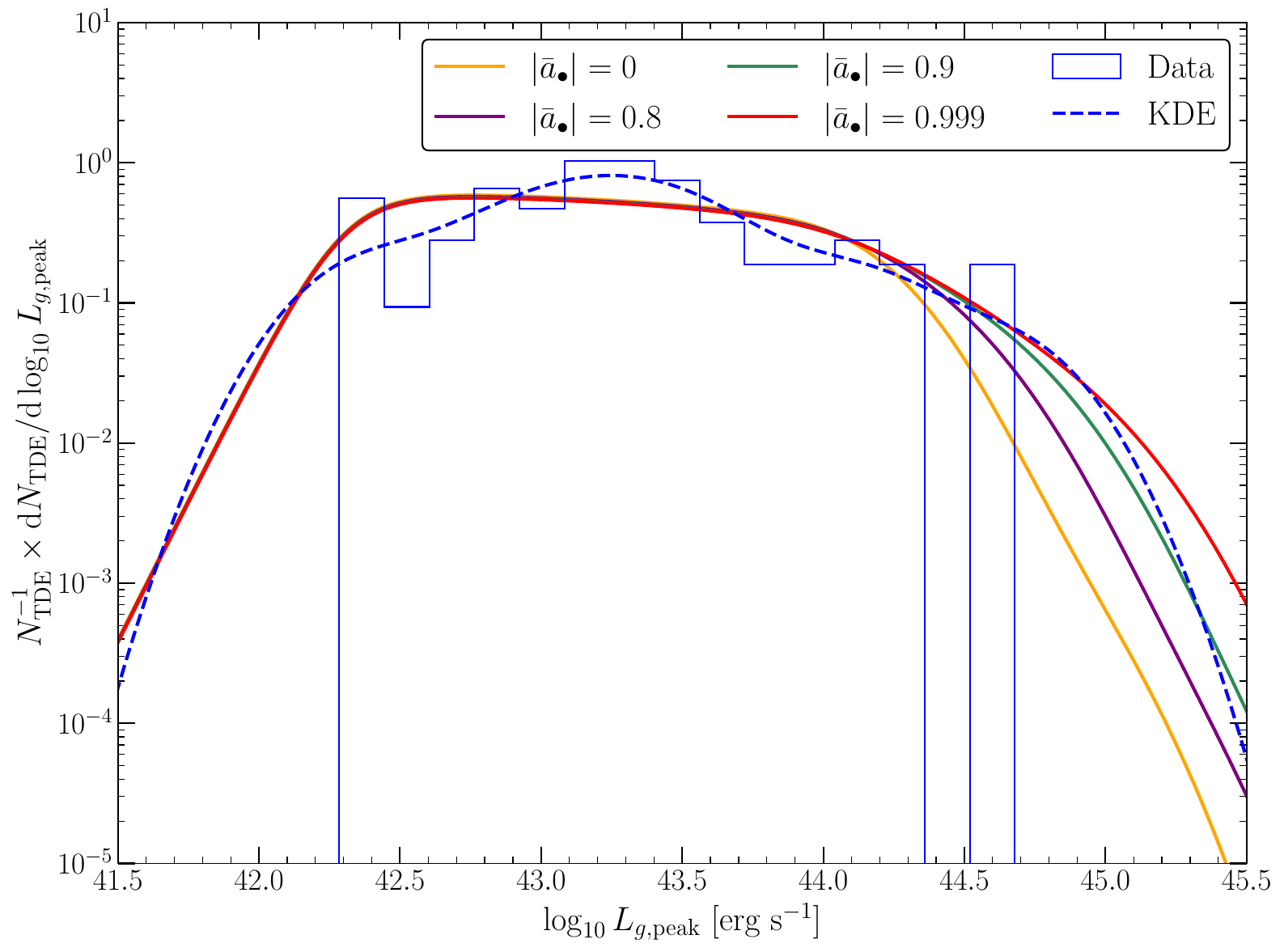}
    \includegraphics[width=\linewidth]{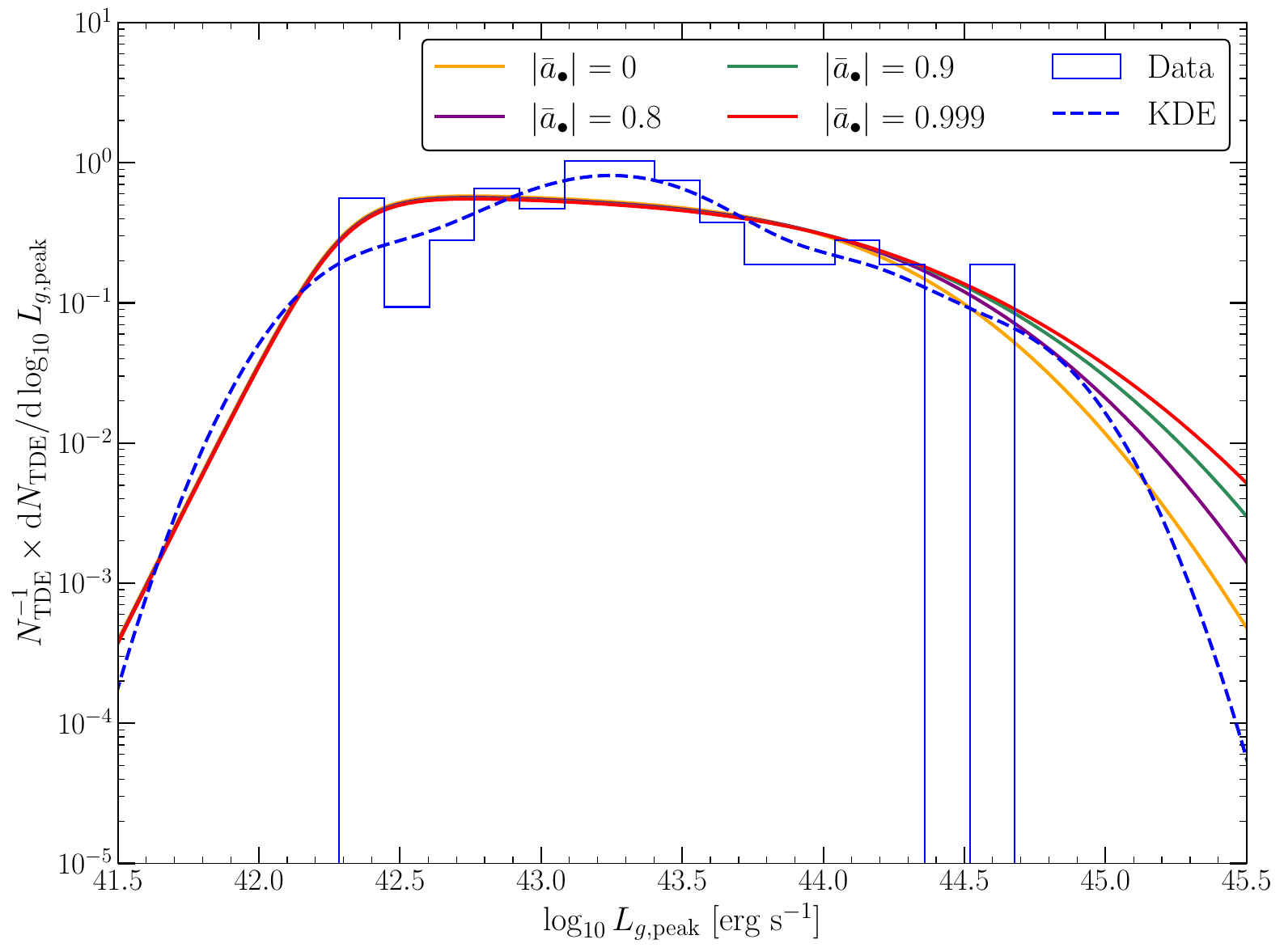}
    \caption{ The predicted (coloured curves) and observed (blue histogram and dashed curve) population density functions of the logarithm of the peak $g$-band luminosity observed from tidal disruption events. In the upper panel we assume an intrinsic scatter between mass and observable of $\epsilon = 0.05$ dex, in the middle panel $\epsilon = 0.2$ dex, and in the lower panel $\epsilon = 0.4$ dex. We note two results, firstly a large scatter in the observable - black hole mass relation washes out most of the signature of black hole spins on the population level (contrast the upper and lower panels). Secondly, the observed low luminosity $\log_{10} L_g \lesssim 44$ tidal disruption event population is in good accord with the theoretical models developed here.   }
    \label{fig:lumpop}
\end{figure}

{As an example of the power of this procedure we plot (in Fig \ref{fig:lumpop}) the posterior $g$-band luminosity distributions for the different spin priors discussed above. In blue we also plot the (normalised) histogram of the current tidal disruption event population \citep[taken from][]{Mummery_et_al_23}, and as a dashed blue line plot a Gaussian kernel density estimation of the observed probability density function. 

As different coloured curves we display the {\it predicted}  distributions of the observed peak $g$-band luminosity, for three different values of the intrinsic scatter $\epsilon$. In the upper panel we take a scatter $\epsilon = 0.05$ dex, for the middle panel $\epsilon = 0.2$ dex, and for the lower panel $\epsilon = 0.4$ dex. We also impose a low luminosity cut-off in the marginal probability density function, of the form  }
\begin{equation}
    p(L_g | M_\bullet) \propto {\exp\left( - {\big[ \log L_g - \alpha - \beta \log M_\bullet/M_\odot \big]^2 / 2 \epsilon ^ 2 }\right)\over 1 + \left(L_{\rm cut}/L_g\right)^4 }  , 
\end{equation}
{to model the fact that we cannot observe arbitrarily low peak luminosities. We take the following parameter values (derived from equation \ref{scaling}) }
\begin{align}
    &\alpha = 36.35 , \\
    &\beta = 1.02 , \\ 
    \log_{10}&L_{\rm cut} = 42.3  .
\end{align}
{Note that the value of $L_{\rm cut}$ was chosen by visual comparison to the data, and is not based on any fundamental analysis. All luminosities are measured in erg s$^{-1}$.     }

{In Figure (\ref{fig:lumpop}) we see that the observed distribution of peak $g$-band luminosities is consistent with the model predictions developed in this paper. This is an important result. We also see that increasing the scatter in the observable-black hole mass relationship washes out much of the signature of the black hole spin (contrast the upper and lower panels). A larger scatter in an observable relationship would dramatically increase the number of observations of tidal disruption event systems required to distinguish different black hole spin distributions.   }

\begin{figure}
    \centering
    \includegraphics[width=\linewidth]{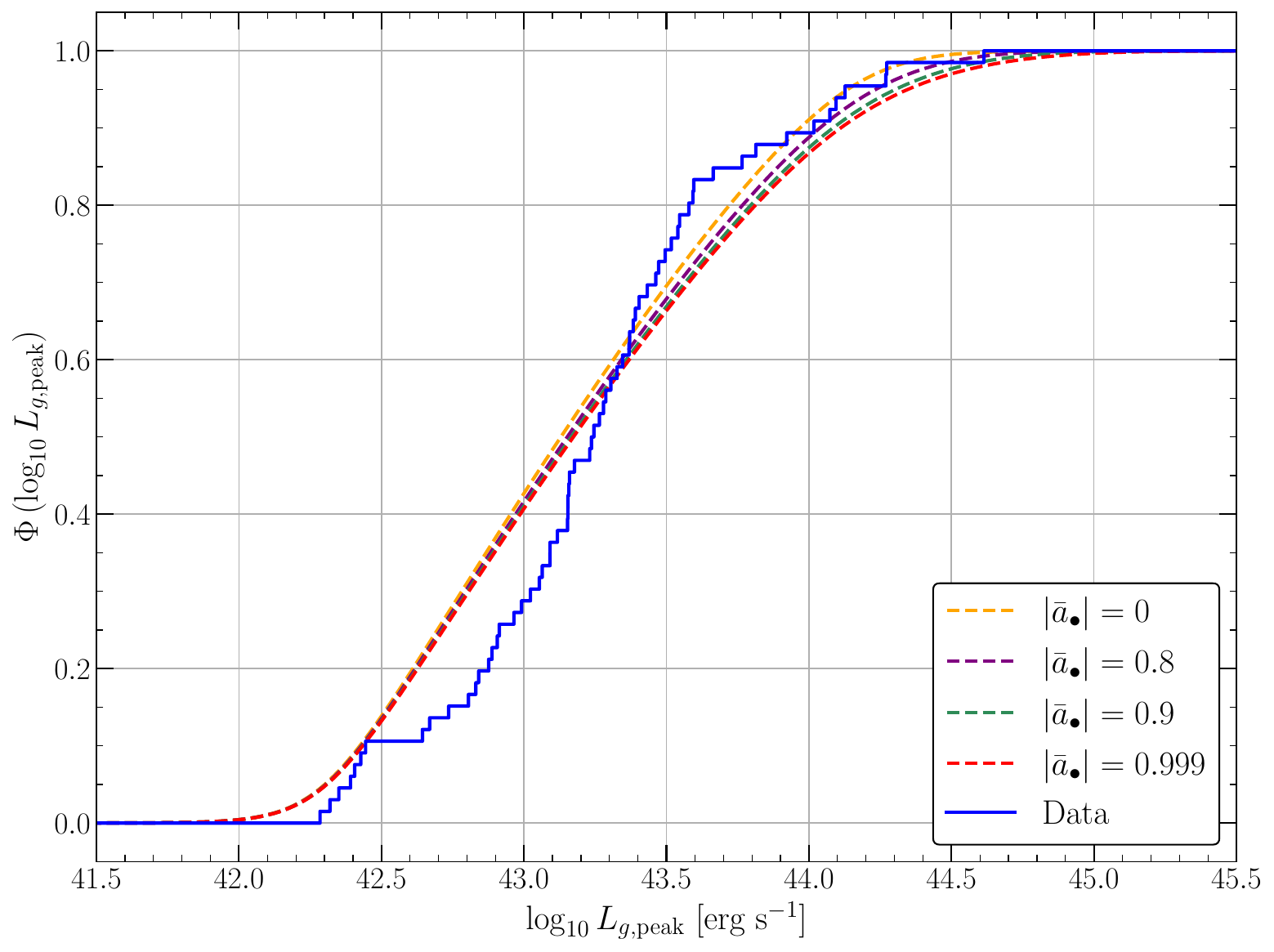}
    \caption{ The cumulative distribution function of the peak $g$-band luminosity of the tidal disruption event population (blue solid curve) and the theoretical distributions developed in this paper for different spin priors (dashed coloured curves). This curve is for $\epsilon = 0.2$ dex. The differences between the theoretical cumulative distribution functions only become apparent at luminosities  above $\log_{10} L_g \gtrsim 43.5$, where the current population contains too few sources to be constraining.  }
    \label{fig:cdf}
\end{figure}

{We can also compare the cumulative distribution function $\Phi({\cal O})$, defined by}
\begin{equation}
    \Phi({\cal O}) \equiv \int_0^{\cal O} p({\cal O'}) \, {\rm d}{\cal O'} ,
\end{equation}
{to the observed cumulative distribution function of the data. We do this in Figure \ref{fig:cdf}, for the scatter parameter $\epsilon = 0.2$ dex. With a sample of 63 tidal disruption events we cannot currently distinguish  the different cumulative distribution functions determined by differing spin priors.  }

\section{Conclusions} \label{conc}
In this paper we have determined a general analytical expression for the maximum tidal acceleration experienced by a test particle on an innermost bound spherical orbit about a Kerr black hole.  This special orbit is the critical parabolic orbit which separates plunging states from those which escape to infinity as $t\to\infty$. As such, if an object cannot be tidally disrupted on this orbit then it will never be able to produce observable emission from the tidal disruption process.  Using this insight we derive the maximum mass, for given stellar properties, of a Kerr black hole which can produce an observable tidal disruption event as a function of black hole spin $a_\bullet$ and incoming orbit inclination $\psi$ (eq. \ref{HillsMass}). 

Utilising this analytical expression we have determined some interesting, and potentially unexpected, properties of the Hills mass. For high ($|a_\bullet| \gtrsim 0.92$) black hole spins the maximum Hills mass at fixed black hole spin occurs for incoming orbital inclinations outside of the black hole's equatorial plane $\psi \neq \pi/2$. This result 
means that the maximum black hole mass which can tidally disrupt a given star can be $\sqrt{11/5} \simeq 1.48$ times higher than previously estimated.  A maximally rotating $10^9 M_\odot$ black hole can tidally disrupt a solar type star, providing the inclination of the incoming stellar orbit satisfies $ \sin\psi \simeq \sqrt{2/3} $.   

As only certain sets of system parameters $(M_\bullet, a_\bullet, \psi, M_\star)$ can possibly result in an observable tidal disruption event, we then demonstrate how Bayesian inference can place constrains on the posterior black hole spin distributions of black holes in the centre of tidal disruption events. If the black hole mass at the centre of a tidal disruption event can be estimated by independent means (for example by using a galactic scaling relationship, or some property of the tidal disruption event emission itself), and this mass is large $M_\bullet \gtrsim 10^8 M_\odot$ then non-trivial constraints on the black hole spin can be placed purely by considering the properties of tidal forces in the Kerr spacetime.   We provide some constraints on nine tidal disruption events previously discovered in the literature. Of these, ASASSN-15lh and AT2020qhs are found to be rapidly rotating. 

{This analysis can be extended from the individual source level to populations of tidal disruption events. In section \ref{sec:population} we demonstrate how this may be done, and demonstrate that the current observed population of tidal disruption event peak $g$-band luminosities is consistent with theoretical expectations of direct capture of stars above the Hills mass.  }

To aid in the measurement of black hole spins from the larger population of tidal disruption events discovered in future optical surveys, a Python package `{\tt tidalspin}' which computes black hole spin distributions for a given tidal disruption event black hole mass (Figs. \ref{fig:spin_dists}, \ref{fig:spin_dists_var_mass}, \ref{fig:2d_mass_spin}, \ref{fig:spin_m_sigma}, \ref{fig:spin_m_gal}) and population inference (Figs. \ref{fig:masspop}, \ref{fig:lumpop}, \ref{fig:cdf}) is publicly available at the following link \url{https://github.com/andymummeryastro/tidalspin}. 

\section*{Acknowledgments} 
This work was supported by a Leverhulme Trust International Professorship grant [number LIP-202-014]. For the purpose of Open Access, AM has applied a CC BY public copyright licence to any Author Accepted Manuscript version arising from this submission. I would like to thank the reviewer for detailed reports which improved the paper in a number of places. 

\section*{Data availability }
The tidal disruption event data used in this manuscript is publically available at \href{https://github.com/sjoertvv/manyTDE}{https://github.com/sjoertvv/manyTDE}. We release a Python package {\tt tidalspin} which computes black hole parameter posteriors for a given tidal disruption event black hole mass estimate. This package is available at the following github repository: \url{https://github.com/andymummeryastro/tidalspin} 

\bibliographystyle{mnras}
\bibliography{andy}

\appendix{}

\section{A more numerically stable equation for $\chi$}\label{poly_app}
Starting from the governing definition of $\chi$
\begin{multline}
    \chi^4 - 4\chi^3 - a_\bullet^2(1 - 3 \cos^2\psi)\chi^2 + a_\bullet^4\cos^2\psi \\ + 4a_\bullet \sin \psi \sqrt{\chi^5 - a_\bullet^2\chi^3\cos^2\psi} = 0 ,
\end{multline}
it is prudent numerically to isolate the final square root, square both sides, and then rearrange to leave the octic polynomial 
\begin{equation}
    \sum_{n=0}^8 c_n \chi^n = 0,
\end{equation}
with 
\begin{align}
    c_0 &= a_\bullet^8 \cos^4\psi , \\
    c_1 &= 0 , \\
    c_2 &= 6a_\bullet^6 \cos^4\psi - 2a_\bullet^6 \cos^2\psi , \\
    c_3 &= 16a_\bullet^4 \cos^2\psi \sin^2\psi - 8a_\bullet^4 \cos^2\psi , \\
    c_4 &= a_\bullet^4 - 4a_\bullet^4 \cos^2\psi + 9a_\bullet^4 \cos^4\psi , \\
    c_5 &= 8a_\bullet^2 - 24 a_\bullet^2 \cos^2\psi - 16a_\bullet^2 \sin^2\psi , \\
    c_6 &= 16 - 2a_\bullet^2 + 6a_\bullet^2 \cos^2\psi , \\
    c_7 &= -8 , \\
    c_8 &= 1. 
\end{align}
This equation proves more stable to solve numerically as it no longer contains a square root which may become complex during the implementation of a numerical root finding algorithm.  

By squaring this expression we have lost the intrinsic degeneracy between $a_\bullet < 0$ and $-\pi/2 < \psi < 0$. This physical degeneracy is contained in the fact that there are generically two real roots of the octic polynomial which are larger than the event horizon of the Kerr black hole. The larger of these two roots corresponds to the retrograde $a_\bullet < 0, \psi >0$ (or equivalently $a_\bullet >0, \psi < 0$) orbit. 

\section{Numerical integration of the orbital equations}\label{integrator_app}
Rather than use the orbital equations in their Carter form (eqs. \ref{rad_mot}, \ref{thet_mot} and \ref{phi_mot}) we solve for the radial and theta evolution using the equations of motion in geodesic form 
 \beq\label{req}
   \frac{\text{d}^2 r}{\text{d}\tau^2} = - \Gamma^r_{\mu\nu}   \frac{\text{d}x^\mu}{\text{d}\tau}   \frac{\text{d}x^\nu}{\text{d}\tau} ,
 \eeq
 and
  \beq\label{thetaeq}
   \frac{\text{d}^2 \theta}{\text{d}\tau^2} = - \Gamma^\theta_{\mu\nu}   \frac{\text{d}x^\mu}{\text{d}\tau}   \frac{\text{d}x^\nu}{\text{d}\tau} ,
 \eeq
where $\Gamma^{\mu}_{\nu \kappa} = \Gamma^{\mu}_{\kappa\nu}$ are the Christoffel coefficients for the Boyer-Lindquist Kerr metric. We solve these expressions as opposed to eqs. \ref{rad_mot} \& \ref{thet_mot} to avoid sign ambiguity which results from taking the square root of the Carter equations.   For an axi-symmetric metric the non-zero coefficients are $\Gamma^\star_{00}$, $\Gamma^\star_{rr}$, $\Gamma^\star_{\phi\phi}$, $\Gamma^\star_{\theta\theta}$, $\Gamma^\star_{\phi 0}$ and $\Gamma^\star_{\theta r}$, where $\star$ takes the place of $r$ and $\theta$ in equations (\ref{req}) and (\ref{thetaeq}) respectively. 

For the $\phi$ and $t$ evolution equations we utilise the conserved quantities $l_z$ and $\epsilon$. These constants of motion are related to the particles 4-velocity $u^\mu$ by 
 \beq\label{L}
 l_z = p_\phi/M_\star = \left(g_{\phi\phi} u^{\phi} + g_{\phi 0} u^0\right)/M_\star  ,
 \eeq
 and
 \beq\label{E}
 \epsilon = -p_0/M_\star = - \left(g_{00}u^0 + g_{0\phi}u^{\phi}\right)/M_\star .
 \eeq
  These two conservation laws can be re-written as equations of motion for the coordinates $t$ and $\phi$, explicitly: 
 \beq\label{teq}
 {{u^0}} = \frac{\text{d}t}{\text{d}\tau} = - \left(\frac{l_z g_{\phi0} + \epsilon g_{\phi\phi}}{g_{\phi\phi}g_{00} - g_{\phi0}^2} \right),
 \eeq
and
 \beq\label{phieq}
 {{u^\phi}} = \frac{\text{d}\phi}{\text{d}\tau} = \frac{l_z g_{00} + \epsilon g_{0\phi}}{g_{\phi\phi}g_{00} - g_{\phi0}^2} .
 \eeq

By writing 
\beq
  \frac{\text{d}^2 r}{\text{d}\tau^2} =   \frac{\text{d} u^r}{\text{d}\tau}, 
\eeq
and 
\beq
  \frac{\text{d}^2 \theta}{\text{d}\tau^2} =   \frac{\text{d} u^\theta}{\text{d}\tau},
\eeq
equations (\ref{req}, \ref{thetaeq}, \ref{teq}, \ref{phieq}) can be expressed as four coupled first order differential equations for the variables $(t, \,\phi,\, u^r,\, u^\theta)$. These four equations, together with the definitions
\beq\label{ureq}
\frac{\text{d} r}{\text{d}\tau} =  u^r ,
 \eeq
 and
 \beq\label{uteq}
 \frac{\text{d} \theta}{\text{d}\tau} =  u^\theta,
 \eeq
 completely specify the photons trajectory. We solve these  six (\ref{req}, \ref{thetaeq}, \ref{teq}, \ref{phieq}, \ref{ureq}, \ref{uteq})  coupled first order differential equations using a fourth order Runge-Kutta integrator. 
 
 For the initial condition of the particle we note that the asymptotic properties of the particular innermost bound spherical orbit are well defined.  As the particle approaches a large distance from the black hole (which we shall denote $r_\infty$), the inclination of the particle satisfies 
 \begin{equation}
     \theta_\infty \to \psi, \quad \dot \theta_\infty \to 0 ,
 \end{equation}
where $\dot x$ denotes the derivative of $x$ with respect to proper time. As $l_z$ is completely specified by $a_\bullet$ and $\psi$ 
 \begin{equation}
    l_z = \sqrt{4 G M_\bullet r_g \chi^3    \over \chi^2 - a_\bullet^2 \cos^2\psi } \sin \psi ,  
\end{equation}
and $\epsilon = 1$, $\dot \phi_\infty$ and $\dot t_\infty$ are given by eqs. \ref{phieq} and \ref{teq}, evaluated at ($r_\infty, \theta_\infty$). The initial radial component of the 4-velocity is then determined from 
\begin{equation}
    g_{\mu\nu} u^\mu u^\nu = -1,
\end{equation}
which can be solved for $\dot r_\infty$ at $(r_\infty, \theta_\infty)$. Note that as the Kerr metric posses axial symmetry, all orbits are independent of the value of $\phi_\infty$, which we set to zero. The initial condition of the particle is then fully specified. For the figures in this paper we use $r_\infty = 10^4 r_g$. 

 We employ a variable time step $\delta \tau$, set as a fixed fraction $h$ of the fastest changing variable
 \beq
 \delta \tau = h \times \text{min} \left[ r  \left(\frac{\text{d}r}{\text{d}\tau}\right)^{-1}, \,  \left(\frac{\text{d}\theta}{\text{d}\tau}\right)^{-1},\,   \left(\frac{\text{d}\phi}{\text{d}\tau}\right)^{-1} \right]  .
 \eeq

To determine the appropriate size of the fixed step size $h$ we require a measure of the accuracy of the algorithm. The final integral of motion $g_{\mu\nu}u^\mu u^\nu = -1$ is useful for this purpose. Errors propagating throughout the particles trajectory will cause the norm of the particles 4-velocity to deviate from $-1$. We therefore define the parameter $\Delta$ by 
 \begin{multline}
 \Delta = 1 + \Bigg[g_{rr} \left(\frac{\text{d}r}{\text{d}\tau}\right)^2 + g_{\theta\theta} \left(\frac{\text{d}\theta}{\text{d}\tau}\right)^2 + g_{\phi\phi} \left(\frac{\text{d}\phi}{\text{d}\tau}\right)^2   \\ 
 + 2 g_{\phi 0} \left(\frac{\text{d}t}{\text{d}\tau}\right)\left(\frac{\text{d}\phi}{\text{d}\tau}\right) + {g_{00} \left(\frac{\text{d}t}{\text{d}\tau}\right)^2} \Bigg] ,
 \end{multline}
which would satisfy  $\Delta = 0$ for an error free integration. We set the fixed step size $h$ by requiring that 
\beq
 \Delta < 10^{-7},
 \eeq
for all particle trajectories. This was found empirically to be satisfied by 
\beq
h = 2 \times 10^{-5}. 
\eeq

\section{The full tidal tensor}\label{tidal_tensor_app}
In this Appendix we present the full tidal tensor $C_{ij}$ as derived by \cite{Marck83}.  As discussed in section \ref{gr_tides} the tidal tensor is a 3x3 symmetric tensor, and so has six independent components. These are\footnote{Note that we correct a typo in the \cite{Kesden12} paper here, equation (8.c) of \cite{Kesden12} should read $T = K - a^2 \cos^2\theta$ \citep[using the notation of][]{Kesden12}. }
\begin{align}
    C_{11} &= \Bigg( 1 - {3 (r^2 + k)(k - a^2 \cos^2\theta)(r^2 - a^2 \cos^2\theta) \cos^2\gamma \over k (r^2 + a^2\cos^2\theta)^2 } \Bigg) I_1 \nonumber \\ 
    &+ 6ar\cos\theta { (r^2 + k)(k - a^2 \cos^2\theta) \cos^2\gamma \over k (r^2 + a^2\cos^2\theta)^2 } I_2 , \\
    C_{12} &= -{3\sqrt{(r^2 + k)(k - a^2 \cos^2\theta) } \cos\gamma \over k (r^2 + a^2\cos^2\theta)^2  } \nonumber \\ & \Bigg[   a r \cos\theta (r^2 -a^2\cos^2\theta + 2k) I_1 \nonumber \\ & \,\,\, - (a^2\cos^2\theta (r^2 + k) - r^2(k-a^2\cos^2\theta))I_2 \Bigg] ,
\end{align}
\begin{align}
    C_{13} &= {3{(r^2 + k)(k - a^2 \cos^2\theta) } \cos\gamma \sin\gamma \over k (r^2 + a^2\cos^2\theta)^2  } \nonumber \\ & \Bigg[ (a^2\cos^2\theta - r^2)I_1 + 2ar\cos\theta I_2 \Bigg], \\
    C_{22} &= \Bigg( 1 + {3r^2(k - a^2 \cos^2\theta)^2  - 3a^2\cos^2\theta (r^2+k)^2 \over k (r^2 + a^2\cos^2\theta)^2 }  \Bigg) I_1 \nonumber \\ & -  6ar\cos\theta { (r^2 + k)(k - a^2 \cos^2\theta) \over k (r^2 + a^2\cos^2\theta)^2 } I_2 , \\
    C_{23} &= - {3\sqrt{(r^2 + k)(k - a^2 \cos^2\theta) } \sin\gamma \over k (r^2 + a^2\cos^2\theta)^2  } \nonumber \\ & \Bigg[    a r \cos\theta (r^2 -a^2\cos^2\theta + 2k) I_1 \nonumber \\ & \,\,\, - (a^2\cos^2\theta (r^2 + k) - r^2(k-a^2\cos^2\theta))I_2 \Bigg] , 
\end{align}
and finally
\begin{multline}
    C_{33} = \Bigg( 1 - {3 (r^2 + k)(k - a^2 \cos^2\theta)(r^2 - a^2 \cos^2\theta) \sin^2\gamma \over k (r^2 + a^2\cos^2\theta)^2 } \Bigg) I_1  \\ 
    + 6ar\cos\theta { (r^2 + k)(k - a^2 \cos^2\theta) \sin^2\gamma \over k (r^2 + a^2\cos^2\theta)^2 } I_2 , 
\end{multline}
where 
\begin{equation}
    I_1 \equiv {GM_\bullet r \over (r^2 + a^2\cos^2\theta)^3}(r^2 - 3a^2 \cos^2\theta), 
\end{equation}
and 
\begin{equation}
    I_2 \equiv {GM_\bullet a \cos\theta \over (r^2 + a^2\cos^2\theta)^3}(3r^2 - a^2 \cos^2\theta) .
\end{equation}
The equatorial plane ($\theta = \pi/2$) limit of this tensor is then found by setting $\cos\theta = 0 = I_2$.

\section{Modified likelihood functions and their (lack of) effect on spin estimates } \label{app:like}
\begin{figure}
    \centering
    \includegraphics[width=\linewidth]{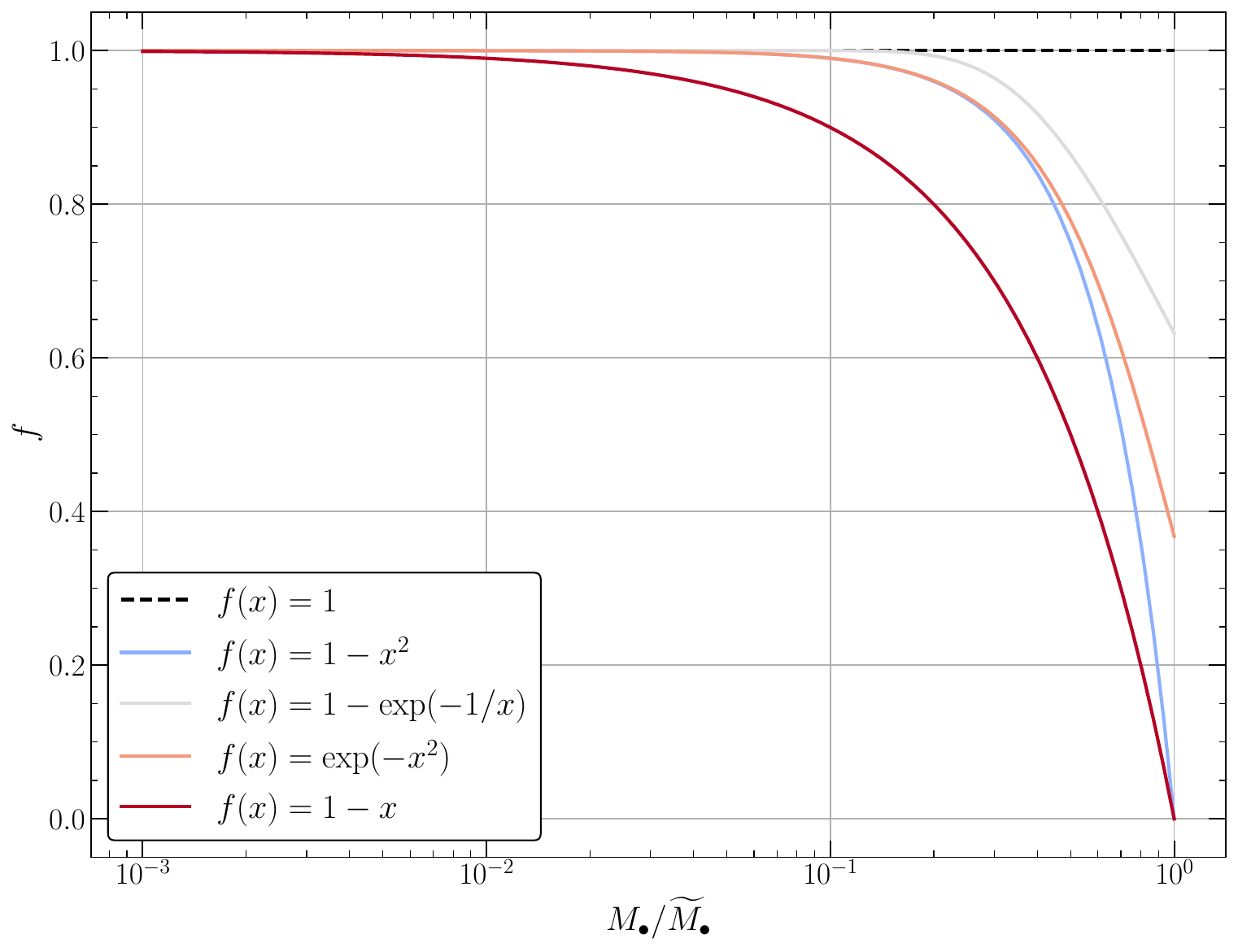}
    \includegraphics[width=\linewidth]{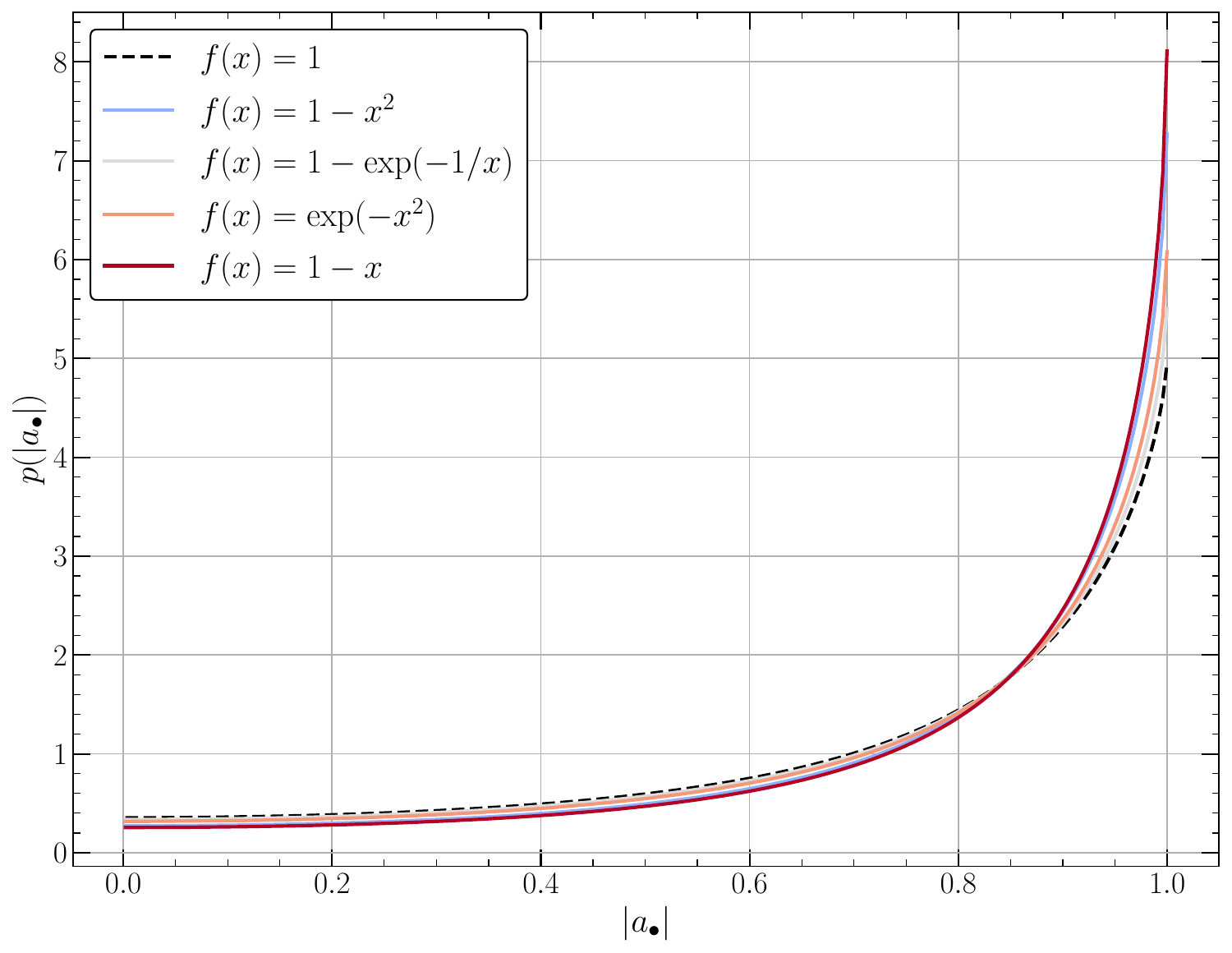}
    \caption{Upper: the different functional forms of the likelihood functions considered in this Appendix. Lower: the posterior spin constraints for the tidal disruption event ASASSN-15lh (from its velocity dispersion measurement; see main body of the paper) under these different assumptions. The black dashed line shows the likelihood used in the paper, and returns the lowest spin constraints. The spin constraints reported in the paper can therefore be considered conservative.  }
    \label{fig:diffL}
\end{figure}

Consider the likelihood function 
\begin{equation}
    {\cal L} = \theta\left(\widetilde M_\bullet - M_\bullet \right) \, f\left(M_\bullet / \widetilde M_\bullet \right) ,
\end{equation}
which describes the probability of a tidal disruption event occurring around a black hole with mass $M_\bullet$ when the Hills mass of the stellar-black hole system is $\widetilde M_\bullet$. In this Appendix we shall examine the effects of various parameterisations of the function $f(x)$ on the inferred spin posterior distributions of tidal disruption events. We have the following constraints on the likelihood 
\begin{equation}
    f(x) \leq 1 \quad \forall \quad x \leq 1, \quad \lim_{x\to 0} f(x) \to 1 . 
\end{equation}
In Figure \ref{fig:diffL} we examine the effects of the following likelihood functions 
\begin{align}
    f(x) &= 1 , \quad {\rm (as \, in \, paper)}, \\
    f(x) &= 1-x^2, \\
    f(x) &= 1 - \exp(-1/x), \\
    f(x) &= \exp(-x^2) , \\
    f(x) &= 1 - x.
\end{align}
Of course, these functions are only valid for $x \leq 1$. None of these likelihoods are intended to be based on a physical model, they are merely chosen to show the lack of sensitivity of the spin constraints to assumptions about $f(x)$ and how the choice used in the paper results in lower bounds on the spin parameter. 

In the upper panel of Fig. \ref{fig:diffL} we display the different functional forms listed above, while in the lower panel we show the spin inference for the tidal disruption event ASASSN-15lh under the different likelihood functions. The black dashed line shows the likelihood used in the paper, and returns the lowest spin constraints. 

{At first this result may seem counter intuitive, as it (potentially) appears that likelihoods which favour lower black hole masses have resulted in a favouring of higher spins, which appears to contradict  the Hills mass mechanism spelled out in this paper. However, what is vital to remember is that this modified likelihood is only one part of the Bayesian inference which ultimately results in the spin constraints presented in Figure \ref{fig:diffL}. These modified likelihoods are also competing with the {\it prior knowledge} we have about the black hole mass. Each  likelihood function reduces the probability that all black holes (of differing spins) with masses close to the star's Hills mass caused the tidal disruption, but importantly this effect also reduces the probability for (e.g.) Schwarzschild black holes.  Therefore, if the tidal disruption event really did occur about a Schwarzschild black hole, then the mass of that black hole would have to be {\it even lower} than the usual Hills mass.   If the prior mass estimate heavily disfavours masses at this lower end (which is the case for ASASSN-15lh), then this likelihood  actually has a net effect of favouring  higher spins, as low spin black holes have been pushed even further into the tail of the prior black hole mass function, an effect which ultimately beats the additional likelihood function.  

To see this most clearly, imagine an extreme likelihood which states that a TDE can only occur if the central black hole mass is an order of magnitude lower than the Hills mass (this is not supposed to be reasonable, only to highlight the point). Then, Schwarzschild black holes could only disrupt (e.g.) a solar type star up to masses $\sim 8 \times 10^6M_\odot$. If the prior information about the black hole mass is some log-normal about $\mu_{M_\bullet} = 5 \times 10^8M_\odot$ (for example), then this upper Schwarzschild mass is pushed deep into the exponential tail of the prior, and a Schwarzschild black hole is even less likely to have caused the TDE than for the Heaviside likelihood function used in the main body of the paper, favouring higher spins. 

For this reason, the Heaviside likelihood used in this paper provides {\it conservative} spin constraints on the black holes involved in tidal disruption events.}

\section{Literature references for tidal disruption events used in this work}
See Tables \ref{tab:sigmarefs} and \ref{tab:tderefs} for the references for the velocity dispersion measurements, and original discovery papers,  of the tidal disruption events used in this paper.  
\begin{table}
    \centering
    \begin{tabular}{l p{140pt}}
    Reference & Event name   \\
    \hline\hline

\citet{Yao23} & AT2018iih, \mbox{AT2020acka} \\
\citet{Hammerstein23b} &  \mbox{AT2020qhs} \\
\citet{Wevers20} & \mbox{AT2018fyk} \\
\citet{Kruhler18} & ASASSN-15lh \\
   
    \end{tabular}
\caption{Origin for velocity dispersion measurements.}
\label{tab:sigmarefs}
\end{table}

\begin{table}
    \centering
    \begin{tabular}{p{90pt} p{140pt}}
    Reference & Event name   \\
    \hline\hline

\citet{Dong16,Leloudas16} & ASASSN-15lh \\
\citet{Wevers19} & AT2018fyk \\
\citet{vanVelzen20} & \mbox{AT2018iih} \\
\citet{Hammerstein23} &  \mbox{AT2020qhs}, \mbox{AT2020riz}, \mbox{AT2020ysg} \\
\citet{Yao23} & AT2019cmw, \mbox{AT2020acka},  \mbox{AT2021yzv} \\
   
    \end{tabular}
\caption{Literature references (``discovery papers") for the TDEs used in work.}
\label{tab:tderefs}
\end{table}

\label{lastpage}
\end{document}